\documentclass[11pt]{article}
\usepackage{verbatim,amsmath,amssymb}
\usepackage{epsfig,float,color}
\usepackage{geometry}
\usepackage{setspace}
\usepackage{natbib}
\usepackage{amsmath,amssymb,amsthm,mathrsfs,amsfonts,dsfont} 
\usepackage{longtable}
\usepackage{hyperref}
\usepackage{multirow}
\usepackage{array}
\usepackage{titlesec}
\usepackage[bottom]{footmisc}
\usepackage{tabularx}
\usepackage{tabulary}
\usepackage{tabu}
\usepackage{booktabs}
\usepackage{accents}
\usepackage{xfrac}

\definecolor{darkblue}{rgb}{0,0,1}
\hypersetup{colorlinks=true, breaklinks=true, linkcolor=darkblue, menucolor=darkblue, citecolor=darkblue, urlcolor=darkblue}

\newcommand{\bitm}{\begin{itemize}}
\newcommand{\eitm}{\end{itemize}}
\newcommand{\bnumr}{\begin{enumerate}}
\newcommand{\enumr}{\end{enumerate}}

\newcommand {\sigab}{\sigma^{\alpha\beta}}

\newcommand {\aab}{a^{\alpha\beta}}
\newcommand {\auab}{a_{\alpha\beta}}
\newcommand {\agd}{a^{\gamma\delta}}

\newcommand {\augd}{a_{\gamma\delta}}

\newcommand {\Aab}{A^{\alpha\beta}}
\newcommand {\Auab}{A_{\alpha\beta}}
\newcommand {\Agd}{A^{\gamma\delta}}

\newcommand {\Mab}{M^{\alpha\beta}}

\newcommand {\bab}{b^{\alpha\beta}}
\newcommand {\Bab}{B^{\alpha\beta}}
\newcommand {\Bgd}{B^{\gamma\delta}}

\newcommand {\buab}{b_{\alpha\beta}}

\newcommand {\bgd}{b^{\gamma\delta}}

\newcommand {\bugd}{b_{\gamma\delta}}

\newcommand {\tauab}{\tau^{\alpha\beta}}

\newcommand{\mrB}{\mathrm{B}}
\newcommand{\mrM}{\mathrm{M}}

\newcommand {\eqb}[1]{\begin{equation}\begin{array}{#1}}
\newcommand {\eqe}{\end{array}\end{equation}}

\newcommand {\esb}[1]{\begin{equation*}\begin{array}{#1}}
\newcommand {\ese}{\end{array}\end{equation*}}
\newcommand {\ds}{\displaystyle}
\newcommand {\ts}{\textstyle}

\newcommand {\pa}[2]{\frac{\partial{#1}}{\partial{#2}}}

\newcommand {\paqq}[3]{\frac{\partial^2{#1}}{\partial{#2}\,\partial{#3}}}
\newcommand {\back}{\! \! \!}
\newcommand {\is}{\back &=& \back}
\newcommand {\dis}{\back &:=& \back}

\newcommand {\plus}{\back &+& \back}
\newcommand {\mi}{\back &-& \back}

\newcommand {\norm}[1]{\|#1\|}
\newcommand {\tr}{\mathrm{tr}\,}

\newcommand {\dif}{\mathrm{d}}

\newcommand {\II}{{I\kern-.3em I}}
\newcommand {\III}{{I\kern-.3em I\kern-.3em I}}

\newcommand {\mra}{\mathrm{a}}
\newcommand {\mrb}{\mathrm{b}}
\newcommand {\mrc}{\mathrm{c}}

\newcommand {\mrf}{\mathrm{f}}

\newcommand {\mrm}{\mathrm{m}}

\newcommand {\mrp}{\mathrm{p}}

\newcommand {\ba}{\boldsymbol{a}}
\newcommand {\bb}{\boldsymbol{b}}

\newcommand {\be}{\boldsymbol{e}}
\newcommand {\bff}{\boldsymbol{f}}
\newcommand {\bg}{\boldsymbol{g}}

\newcommand {\bn}{\boldsymbol{n}}

\newcommand {\bt}{\boldsymbol{t}}

\newcommand {\bx}{\boldsymbol{x}}

\newcommand {\bnu}{\mbox{\boldmath$\nu$}}

\newcommand {\bA}{\boldsymbol{A}}

\newcommand {\bC}{\boldsymbol{C}}

\newcommand {\bE}{\boldsymbol{E}}
\newcommand {\bF}{\boldsymbol{F}}
\newcommand {\bG}{\boldsymbol{G}}

\newcommand {\bI}{\boldsymbol{I}}

\newcommand {\bL}{\boldsymbol{L}}
\newcommand {\bM}{\boldsymbol{M}}
\newcommand {\bN}{\boldsymbol{N}}

\newcommand {\bone}{\mathbf{1}}

\newcommand {\bbR}{\mathbb{R}}

\newcommand {\IR}{{\rm\kern.24em
   \vrule width.02em height1.53ex depth-.05ex
   \kern-.3em R}}
\newcommand {\ic}{{\rm\kern.20em
   \vrule width.02em height1.0ex depth-.05ex
   \kern-.22em c}}
\newcommand {\ia}{{\rm\kern.20em
   \vrule width.02em height1.05ex depth-.0ex
   \kern-.25em a}}
\newcommand {\IC}{{\rm\kern.24em
   \vrule width.02em height1.4ex depth-.05ex
   \kern-.26em C}}
\newcommand {\ID}{{\rm\kern.34em
   \vrule width.02em height1.5ex depth-.05ex
   \kern-.36em D}}
\newcommand {\IS}{{\rm\kern.24em
   \vrule width.02em height1.6ex depth.05ex
   \kern-.26em S}}
\newcommand {\IT}{{\rm\kern.50em
   \vrule width.02em height1.55ex depth-.05ex
   \kern-.52em T}}

\newcommand {\IE}{{\rm\kern.24em
   \vrule width.02em height1.55ex depth-.05ex
   \kern-.33em E}}
\newcommand {\IEa}{{\rm\kern.24em
   \vrule width.02em height1.55ex depth-.05ex
   \kern-.33em E}^{1}_{ijkl}}
\newcommand {\IEb}{{\rm\kern.24em
   \vrule width.02em height1.55ex depth-.05ex
   \kern-.33em E}^{2}_{ijkl}}

\newcommand {\sS}{\mathcal{S}}

\newcommand {\sV}{\mathcal{V}}

\newcommand {\vaua}{\ba_\alpha}

\newcommand {\Ass}[2]{\kern 0.9ex \vrule width0.45em height0.2ex depth0ex \kern -2.1ex \bigwedge_{#1}^{#2}}
\newcommand {\ASS}[2]{\kern 1.45ex \vrule width0.5em height0.2ex depth0ex \kern -2.65ex \bigwedge_{#1}^{#2}}

\newcommand {\hc}{\hat{c}}
\newcommand {\hd}{\hat{d}}

\newcommand {\hg}{\hat{g}}

\newcommand {\hD}{\hat{D}}
\newcommand {\hE}{\hat{E}}
\newcommand {\hF}{\hat{F}}

\newcommand {\hI}{\hat{I}}
\newcommand {\hJ}{\hat{J}}

\newcommand {\hL}{\hat{L}}

\newcommand {\hR}{\hat{R}}

\newcommand {\hW}{\hat{W}}

\newcommand {\htau}{\hat{\tau}}

\newcommand {\tc}{\tilde{c}}

\newcommand {\tg}{\tilde{g}}

\newcommand {\tk}{\tilde{k}}

\newcommand {\tp}{\tilde{p}}

\newcommand {\tD}{\tilde{D}}
\newcommand {\tE}{\tilde{E}}

\newcommand {\tI}{\tilde{I}}
\newcommand {\tJ}{\tilde{J}}

\newcommand {\tW}{\tilde{W}}

\newcommand {\tbx}{\tilde{\boldsymbol{x}}}

\newcommand {\tbC}{\tilde{\boldsymbol{C}}}

\newcommand {\tbF}{\tilde{\boldsymbol{F}}}

\newcommand {\tbH}{\tilde{\boldsymbol{H}}}

\newcommand {\tbL}{\tilde{\boldsymbol{L}}}
\newcommand {\tbM}{\tilde{\boldsymbol{M}}}

\newcommand {\tbX}{\tilde{\boldsymbol{X}}}

\newcommand {\ttau}{\tilde{\tau}}
\newcommand {\tmu}{\tilde{\mu}}

\newcommand {\aE}{\accentset{*}{E}}

\newcommand {\aH}{\accentset{*}{H}}
\newcommand {\aI}{\accentset{*}{I}}
\newcommand {\aJ}{\accentset{*}{J}}

\newcommand {\aL}{\accentset{*}{L}}

\newcommand {\abC}{\accentset{*}{\boldsymbol{C}}}

\newcommand {\abE}{\accentset{*}{\boldsymbol{E}}}
\newcommand {\abF}{\accentset{*}{\boldsymbol{F}}}

\newcommand {\abI}{\accentset{*}{\boldsymbol{I}}}

\newcommand {\abL}{\accentset{*}{\boldsymbol{L}}}
\newcommand {\abM}{\accentset{*}{\boldsymbol{M}}}

\newcommand {\asS}{\accentset{*}{\sS}}

\newcommand{\lam}{\lambda_3}

\newcommand {\gab}{g^{\alpha\beta}}
\newcommand {\guab}{g_{\alpha\beta}}

\newcommand {\Guab}{G_{\alpha\beta}}

\newcommand{\tmod}[1]{{\color{black}#1}}  

\titleformat{\paragraph}{\normalfont\normalsize\itshape}{\theparagraph}{1em}{}
\titlespacing*{\paragraph}{\parindent}{3.25ex plus 1ex minus .2ex}{.75ex plus .1ex}

\newcommand\tst{\rule{0pt}{3.0ex}}         
\newcommand\bst{\rule[-1.0ex]{0pt}{2.0ex}} 

\begin{document}

\theoremstyle{definition}
\newtheorem{thm}{Theorem}[section]
\newtheorem{defn}[thm]{Definition}
\newtheorem{lem}[thm]{Lemma}
\newtheorem{cor}[thm]{Corollary}
\newtheorem{prop}[thm]{Proposition}
\newtheorem{rem}[thm]{Remark}
\newtheorem{note}[thm]{Note}

%
%

\begin{center}
\Large{\bf{Efficient isogeometric thin shell formulations for soft biological materials}}\\

\end{center}

\begin{center}
\large{Farshad Roohbakhshan and Roger A. Sauer
\footnote{corresponding author, email: sauer@aices.rwth-aachen.de}}\\
\vspace{4mm}

\small{\textit{Aachen Institute for Advanced Study in Computational Engineering Science (AICES), RWTH Aachen
University, Templergraben 55, 52056 Aachen, Germany}}

\vspace{4mm}

Published\footnote{This pdf is the personal version of an article whose final publication is available at \href{http://dx.doi.org/10.1007/s10237-017-0906-6}{http://www.springer.com/}} 
in \textit{Biomechanics and Modeling in Mechanobiology}, \\
\href{http://dx.doi.org/10.1007/s10237-017-0906-6}{DOI: 10.1007/s10237-017-0906-6} \\
Submitted on 23.~December 2016 2016, Accepted on 27.~March 2017, Published online on 12.~April 2017

\end{center}

%
%

\rule{\linewidth}{.15mm}
{\bf Abstract:}
This paper presents three different constitutive approaches to model thin rotation-free shells based on the Kirchhoff--Love hypothesis. One approach is based on numerical integration through the shell thickness while the other two approaches do not need any numerical integration and so they are computationally more efficient. The formulation is designed for large deformations and allows for geometrical and material nonlinearities, which makes it very suitable for the modeling of soft tissues. Furthermore, six different isotropic and anisotropic material models, which are commonly used to model soft biological materials, are examined for the three proposed constitutive approaches. Following an isogeometric approach, NURBS-based finite elements are used for the discretization of the shell surface. Several numerical examples are investigated to demonstrate the capabilities of the formulation. Those include the contact simulation during balloon angioplasty.

{\bf Keywords:}
Angioplasty, contact modeling, isogeometric analysis, Kirchhoff--Love shell, soft biological materials, thin rotation-free shells

\vspace{-4mm}
\rule{\linewidth}{.15mm}
%
%
%
\section{Introduction}\label{s:intro}
Many biological systems are thin structures, composed of nonlinear soft materials, which can easily undergo large deformations. In many cases, such structures do not resist any bending moments \citep{humphrey98}; thus, a membrane formulation \citep[e.g.][]{biomembrane} is efficient and robust to predict the mechanical response. However, if the bending effects are not negligible, a shell formulation is required. For thin structures, where the transverse shear strains can be neglected, rotation-free formulations based on the Kirchhoff--Love hypothesis are the best choice. Here, we introduce a new approach to model thin biological shells\footnote{Here, we distinguish between membranes and shells as two thin structures with different mechanical characteristics. Membranes bear only in-plane stresses but shells bear bending moments as well. However, the term ``(bio)membrane'' is also used for structures that are mechanical shells \citep[cf.][]{tepole15}.} constructed from nonlinear constitutive laws, without any need for numerical integration. Following \citet{shelltheo} and \citet{solidshell}, the model is formulated in a curvilinear coordinate system, without resorting to a transformation from/to the Cartesian coordinate system. Furthermore, the geometry, kinematic variables and weak form of the governing equation are discretized within the framework of isogeometric analysis (IGA) in order to take advantage of the $C^1$-continuity {NURBS}-based interpolation, which is a necessary condition for the Kirchhoff--Love shells.

The finite element modeling and analysis of thin soft tissues has been the subject of extensive research although only the membrane forces are considered in general \citep[e.g.][]{humphrey92,humphrey98,prot07,kroon09,abdessalem11,rausch13,rausch14,biomembrane} and mostly planar tissues are studied \citep[e.g.][]{flynn98,sun05tissue,holzapfel08,jacobs13,fan14}. The first isogeometric Kirchhoff--Love shell, specially formulated for soft tissues, was introduced by \citet{tepole15}. It is based on numerical integration through the shell thickness. Furthermore, \citet{kiendl15} and \citet{solidshell} have suggested two different isogeometric formulations for the  modeling of the rotation-free thin shells with arbitrary nonlinear hyperelastic materials. Both approaches can be used for the modeling of biological shells. The former requires numerical integration through the shell thickness. In contrast, the latter allows for both projected shell models that are extracted from existing 3D material models using numerical integration and shell models that are directly formulated on a 2D manifold, like the Koiter model or the Canham model.

In the present work, we extend the earlier work of \citet{solidshell}, which allows an arbitrary choice of the membrane and bending strain energies. For specific applications, like biological shells, a physically well-defined link between the bending and membrane parts is needed. Here, for any given {3D} material model, a systematic approach is introduced to derive the corresponding 2D shell formulation, which (1) requires no numerical integration through the thickness, (2) provides a natural link between membrane and bending parts and (3) admits many isotropic and anisotropic material models. To show the accuracy of the new approach, a simplified version of the projected shell formulation of \citet{solidshell} is used for reference.  

The presented work adds novelties to the existing literature on the computational modeling of soft biological shells:
\vspace{-\topsep}
\bitm
\setlength{\itemsep}{0pt}
\setlength{\parskip}{0pt}
\setlength{\parsep}{0pt}
\item First and foremost, it provides two new approaches that do not require any numerical thickness integration and that are therefore computationally more efficient. 
\item Second, the resultant stresses and bending moments are expressed in terms of the first and second fundamental forms of the shell mid-surface, which allows flexible coupling of the bending and membrane modes of the shell (see Sec.~\ref{s:DD}). 
\item Third, an efficient and accurate treatment of the compression/extension switch, to exclude compressed fibers from the constitutive law, is introduced. Such a switch, which is used for the anisotropic material models like the Gasser--Ogden--Holzapfel (GOH) model \citep{gasser06}, is needed to guarantee the polyconvexity of the strain energy density function in order to avoid non-physical responses \citep{balzani06}.       
\eitm
\vspace{-\topsep}

The remaining part of this paper is organized as follows: Sec.~\ref{s:theory} provides a short summary of the rotation-free thin shell theory, including the kinematics and weak form of the governing equations. Sec.~\ref{s:shells}  discusses the three constitutive approaches to model thin shells in detail. Those are the \emph{numerically-projected} (NP) shell model, which is based on numerical integration through the shell thickness, and the \emph{analytically-projected} (AP) and \emph{directly-decoupled} (DD) shell models, which need no numerical integration. In Sec.~\ref{s:material}, those three shell models are specifically derived for different isotropic and anisotropic material models, which are commonly used for soft tissues. Several numerical experiments are presented in Sec.~\ref{s:examples} to illustrate the capabilities of the new model. Sec.~\ref{s:con} concludes the paper.
%
%
\section{Thin shell theory}\label{s:theory}
This section summarizes the nonlinear theory of rotation-free thin shells. Further details can be found in e.g.~\citet{naghdi82}, \citet{steigmann99} and \citet{shelltheo}. Here, first the kinematics of thin shells based on the Kirchhoff--Love hypothesis is reviewed. Those kinematics can either be derived from 3D kinematics or be formulated directly on the shell mid-surface. This is followed by a brief discussion of the weak form of the governing equation. Last, the weak form is linearized in order to be solved by the Newton-Raphson method. Upon this theoretical foundation, three different shell models are constructed in the next section.

\subsection{Kinematics of Kirchoff--Love shells}\label{s:kin}
A thin shell is a structure that can be presented as a 2D manifold defined by the shell mid-surface. Alternatively, the shell can be described as a thin 3D continuum, which is confined by an upper and a lower surface. Here, a framework is presented that can capture both approaches. First, the shell mid-surface is described and then the description is extended to the other shell layers according to the Kirchhoff--Love hypothesis. 

In the deformed configuration, the shell mid-surface $\sS$ is described by the mapping 
\eqb{l}
\bx = \bx(\xi^\alpha)~,\quad(\alpha=1,2)~,
\eqe
where $\xi^\alpha$ are the convective coordinates defined in a parametric domain. According to this surface description, the co-variant tangent vectors are $\ba_\alpha=\partial{\bx}/\partial\xi^\alpha$ and the contra-variant tangent vectors are defined by $\ba^\alpha=a^{\alpha\beta}\ba_\beta$, where the co-variant components of the metric tensor are $a_{\alpha\beta}=\ba_\alpha\cdot\ba_\beta$ and the contra-variant components are $[a^{\alpha\beta}] = [a_{\alpha\beta}]^{-1}$. Then, from the tangent vectors $\ba_\alpha$, the normal vector of the surface $\sS$ is given by $\bn = (\ba_1 \times \ba_2)/\norm{\ba_1 \times \ba_2}$. Another important object associated with a surface is the curvature tensor $\bb = b_{\alpha\beta}\,\ba^\alpha\otimes\ba^\beta$, where $b_{\alpha\beta} := \bn\cdot\ba_{\alpha,\beta}$ are the co-variant components of the curvature tensor. The mean curvature of the deformed surface is $H:=\tfrac{1}{2}\,a^{\alpha\beta}\,b_{\alpha\beta}$. Likewise, the shell mid-surface can be described in its reference configuration, denoted by $\sS_0$.

Such a surface description can be extended to any shell layer $\asS$ within the shell thickness. Based on the Kirchhoff--Love assumptions, the position $ \tbx $ of any material point in the deformed shell body is related to a corresponding point $ \bx $ on the shell mid-surface $\sS$ as
\eqb{l}
\tbx(\xi^\alpha,\xi)  = \bx(\xi^\alpha) + \xi\,\bn ~,
\label{e:tx}
\eqe
where $ \xi \in [-\tfrac{T}{2},\tfrac{T}{2}] $ is the out-of-plane coordinate and $ T $ is the initial shell thickness.

\begin{rem} 
Henceforth, the variables of 3D continua are distinguished by a tilde and the corresponding variables of a shell layer $\asS$, located at $\xi$ within the shell thickness, are distinguished by an asterisk. The variables of the shell mid-surface $\sS$, located at $\xi = 0$, have no mark. All the variables in the reference and current configurations are denoted by uppercase and lowercase letters, respectively. 
\end{rem} 

\begin{rem} 
Greek indices take values in $\{1, 2\}$, where the Einstein summation convention is assumed. Further, one needs to distinguish between variables with Greek upper and lower indices as contra-variant and co-variant objects, respectively. Latin indices that appear later can be any positive integer and may be arbitrarily used in upper or lower positions. 
\end{rem} 

Likewise to the shell mid-surface, the tangent vectors on a shell layer $\asS$ can be expressed as
\eqb{l}
\bg_\alpha := \ds\pa{\tbx}{\xi^\alpha} = \ba_\alpha - \xi\,b_\alpha^\gamma\,\ba_\gamma ~,
\label{e:bg_i}\eqe
which give the co-variant components of the metric tensor of a shell layer 
\eqb{l}
g_{\alpha\beta} := \bg_\alpha\cdot\bg_\beta = a_{\alpha\beta} - 2\,\xi\,b_{\alpha\beta}
\label{e:g_ab}\eqe
if the second and higher order terms are neglected. The contra-variant components are then $[g^{\alpha\beta}]=[g_{\alpha\beta}]^{-1}$, which give the contra-variant tangent vectors $\bg^\alpha=\gab\bg_\beta$. Similarly, corresponding variables on $\asS_0$ are defined in the same fashion. 

The mapping between the reference configuration and the current configuration is characterized by the deformation tensor $\tbF := \partial\tbx/\partial\tbX$. On the shell layer $\asS$, the deformation gradient can be decomposed into in-plane and out-of-plane components as $\tbF = \abF + \tmod{\sqrt{g_{33}}} \, \bn \otimes \bN$ and correspondingly as $\tbF :=  \bF + \lam \, \bn \otimes \bN $ on $\sS$, where the surface deformation tensors are $ \bF = \ba_\alpha \otimes \bA^\alpha $ and $ \abF = \bg_\alpha \otimes \bG^\alpha$.
Here, $\tmod{\sqrt{g_{33}}}$ and $\lam$ measure the out-of-plane stretches at $\xi\in[-T/2,T/2]$ and $\xi=0$, respectively. In  general, $\lam$ is the average of $\sqrt{g_{33}}$ over the thickness. Such a layer-wise decomposition is also applied to the other kinematical variables. For instance, the volume change, measured by the determinant of deformation gradient, is given by
\eqb{l}
\tJ := \det \tbF = J \, \lam ~,\quad \tJ := \det \tbF = \aJ \, \tmod{\sqrt{g_{33}}} ~,
\label{e:tJ}\eqe
for $\sS$ and $\asS$, respectively, where the surface changes are determined by 
\eqb{l}
J := \det \bF = \sqrt{a/A} ~, \quad \aJ := \det \abF = \sqrt{g/G} ~.
\label{e:J}\eqe
Here, $g := \det[g_{\alpha\beta}]$ and $a := \det[a_{\alpha\beta}]$ are defined in the current configuration and they are correspondingly denoted by $G$ and $A$ in the reference configuration. The other important tensors to describe the deformation are the  Cauchy--Green deformation tensors and the Green--Lagrange strain tensor. The right Cauchy--Green deformation tensors for a Kirchhoff--Love shell are
\eqb{l}
\tbC = \bC + \lam^2 \, \bN \otimes \bN ~,\quad \tbC = \abC + \tmod{g_{33}} \, \bN \otimes \bN ~,
\label{e:tbC}\eqe
where the right Cauchy--Green deformation tensors of the shell mid-surface and a layer within the shell thickness are, respectively,
\eqb{lllll}
\bC \dis \bF^\mathrm{T}\bF \is a_{\alpha\beta} \, \bA^{\alpha} \otimes \bA^{\beta} ~, \\
\abC \dis \abF^\mathrm{T}\abF \is g_{\alpha\beta} \, \bG^{\alpha} \otimes \bG^{\beta} ~.
\label{e:bC}\eqe
Accordingly, the first three invariants of $\tbC$ are
\eqb{l}
\tI_1 := \tr \tbC = I_1 + \lam^2~,\quad \tI_1 := \tr \tbC = \aI_1 + \tmod{g_{33}} ~,
\label{e:tI1}\eqe
with
\eqb{l}
I_1 := \tr \bC = A^{\alpha\beta} a_{\alpha\beta} ~, \quad \aI_1 := \tr \abC = G^{\alpha\beta} g_{\alpha\beta} ~,
\label{e:I1}\eqe
\eqb{l}
\tI_2 := \ds\frac{1}{2}\left[\big(\tr \tilde\bC \big)^2 - \tr \tilde\bC^2  \right] = \lam^2 \, I_1 + J^2 ~,\quad \tI_2 = \tmod{g_{33}} \, \aI_1 + \aJ^2
\label{e:tI2}\eqe
and
\eqb{l}
\tI_3 := \det \tbC = \tJ^2 = \lam^2 \, J^2~,\quad \tI_3 := \det \tbC = \tmod{g_{33}} \, \aJ^2 ~.
\label{e:tI3}\eqe

The Green--Lagrange strain tensors then are 
\eqb{l}
\tilde\bE := \ds\frac{1}{2}\,\big(\tilde\bC - \bone\big) = \bE + E_{33}\,\bN\otimes\bN ~,\quad
\tilde\bE := \ds\frac{1}{2}\,\big(\tilde\bC - \bone\big)= \abE + \aE_{33}\,\bN\otimes\bN ~,
\label{e:tE}\eqe
where $\bone$ is the usual identity tensor in $\bbR^3$. Further, $\abE$ and $\bE$ are the in-plane components of the Green--Lagrange strain tensor given by
\eqb{lllll}
\bE \dis \ds\frac{1}{2}\,\big(\bC - \bI\big) \is E_{\alpha\beta}\,\bA^\alpha\otimes\bA^\beta ~,\\[2mm] 
\abE \dis \ds\frac{1}{2}\,\big(\abC - \abI\big) \is \aE_{\alpha\beta}\,\bG^\alpha\otimes\bG^\beta ~,
\label{e:E}\eqe 
where $\bI = \Auab\,\bA‍‍‍‍‍‍‍^\alpha\otimes\bA^\beta$ and $\abI = \Guab\,\bG^\alpha\otimes\bG^\beta$ are the surface identity tensors on $\sS$ and $\asS$, respectively, and thus
\eqb{l}
E_{\alpha\beta} = \ds\frac{1}{2}\,\big(a_{\alpha\beta} - A_{\alpha\beta}) ~,\quad \aE_{\alpha\beta} = \ds\frac{1}{2}\,\big(g_{\alpha\beta} - G_{\alpha\beta}) ~.
\label{e:Eab}\eqe
The out-of-plane components of the Green--Lagrange strain tensor \eqref{e:tE} are then
\eqb{l}
E_{33} = \ds\frac{1}{2}\,\big(\lam^2 - 1\big) ~, \quad 
\aE_{33} = \ds\frac{1}{2}\,\big(\tmod{g_{33}} - 1\big)~.
\eqe

Moreover, Eqs.~\eqref{e:g_ab},~\eqref{e:E}~and~\eqref{e:Eab} imply that the distribution of the in-plane strain across the shell thickness is linear, i.e.
\eqb{l}
\aE_{\alpha\beta} = E_{\alpha\beta} - \xi\,K_{\alpha\beta} ~,
\label{e:linstr}\eqe
where we have defined 
\eqb{l}
K_{\alpha\beta} = b_{\alpha\beta} - B_{\alpha\beta}~.
\eqe

\subsection{Governing weak form}\label{s:gov}
Having described the kinematics of thin shells, now the weak form of the governing equation of a thin shell is introduced. Here, a brief review is presented. Further details of the stress and moment tensors as well as the balance laws and  strong form can be found in \citet{shelltheo}.

Neglecting the inertial effects, for any admissible variation $\delta\bx\in\sV$, the weak form of the governing equation is formulated in terms of the internal and external virtual work contributions as
\eqb{l}
G_\mathrm{int} - G_\mathrm{ext} = 0 \quad\forall\,\delta\bx\in\sV~.
\label{e:wf1}\eqe
The internal virtual work is
\eqb{l}
G_\mathrm{int} =  \ds\int_{\sS_0} \frac{1}{2}\,\delta a_{\alpha\beta} \, \tau^{\alpha\beta} \, \dif A  + \int_{\sS_0} \delta b_{\alpha\beta} \, M_0^{\alpha\beta} \, \dif A ~,
\eqe
where $\tau^{\alpha\beta}$ is the Kirchhoff stress and $M_0^{\alpha\beta}$ is the moment tensor defined in the reference configuration. They are associated with their counterparts in the current configuration by $\tau^{\alpha\beta} = J\,\sigab$ and $M^{\alpha\beta}_0 = J\,M^{\alpha\beta}$. The external virtual work is 
\eqb{l}
G_\mathrm{ext} = \ds\int_{\sS}\delta\bx\cdot\bff\,\dif a 
+ \ds\int_{\partial_t\sS} \delta\bx\cdot\bt\,\dif s +  \ds\int_{\partial_m\sS}\delta\bn\cdot m_\tau\,\bnu\,\dif s + [\delta\bx\cdot m_\nu\,\bn\big]~,
\label{e:wf2}\eqe
where $\bff = f^\alpha\,\vaua + p_\mathrm{ext}\,\bn$ is a prescribed body force on $\sS$ with $p_\mathrm{ext}$ as the external pressure. Further, $\bt$, $m_\tau$ and $m_\nu$ are distributed forces and moments prescribed on the boundary and $\bnu = \nu_\alpha\,\ba^\alpha$ is the normal to $\partial_m\sS$, where the bending moment $m_\tau$ is applied.

\subsection{Linearization of the weak form}\label{s:lin}
As the weak form \eqref{e:wf1} is nonlinear, it needs to be linearized in order to be solved by the Newton--Raphson method. The linearized internal virtual work contribution is \citep{shelltheo} 
\eqb{lll}
\Delta G_\mathrm{int}
\is \ds\int_{\sS_0}\ts \ds\frac{1}{2}\delta\auab\,\big(c^{\alpha\beta\gamma\delta}\,\frac{1}{2}\Delta\augd 
+ d^{\alpha\beta\gamma\delta}\,\Delta\bugd\big)\,\dif A \\ [4mm]
\plus \ds\int_{\sS_0}\ts~~\delta\buab\,\big(e^{\alpha\beta\gamma\delta}\,\ds\frac{1}{2}\Delta\augd 
+ f^{\alpha\beta\gamma\delta}\,\Delta\bugd \big)\,\dif A \\ [4mm] 
\plus \ds\int_{\sS_0}\ts \big(\tauab\,\ds\frac{1}{2}\Delta\delta\auab + \Mab_0\,\Delta\delta\buab \big)\,\dif A ~,
\label{e:dgint}\eqe
where 
\eqb{l}
c^{\alpha\beta\gamma\delta} := 2\,\ds\frac{\partial \tau^{\alpha\beta}}{\partial a_{\gamma\delta}}~, \quad
d^{\alpha\beta\gamma\delta} :=    \ds\frac{\partial \tau^{\alpha\beta}}{\partial b_{\gamma\delta}}~, \quad
e^{\alpha\beta\gamma\delta} := 2\,\ds\frac{\partial M_0^{\alpha\beta}}{\partial a_{\gamma\delta}}~,  \quad
f^{\alpha\beta\gamma\delta} :=    \ds\frac{\partial M_0^{\alpha\beta}}{\partial b_{\gamma\delta}}
\label{e:cdef}\eqe
are the material tangent tensors. The linearized external virtual work contribution, $\Delta G_\mathrm{ext}$, is given in Appendix~\ref{s:DGext}. The discretized weak form, which gives the FE force vectors and the FE tangent matrices, can be found in \citet{solidshell} for a NURBS-based FE implementation.

\section{Shell constitution: Three modeling approaches}\label{s:shells}
As shown in Tab.~\ref{t:models}, in general, there are two different structural modeling approaches in shell theory \citep{bischoff04}. In the \emph{projection approach}, a shell is assumed to be a 3D continuum, thus the stress resultants are derived rarely analytically and mostly by numerical integration through the shell thickness \citep[e.g.][]{solidshell}. In the \emph{direct surface approach}, a shell is considered as a 2D manifold, defined on the mid-surface of the shell continuum and thus the stresses and moments can be directly derived from a well-postulated 2D strain energy density function \citep[e.g.][]{shelltheo}. Furthermore, the \emph{degenerated solid approach} can also be used, which is in fact not based on a shell theory but rather is a method to reduce the dimension of 3D finite elements \citep{bischoff04}.

\begin{table}[ht]
\centering
\unitlength1cm
\tabulinesep = 7pt
\begin{tabu} to 0.95\textwidth { p{10pt} | X[c,m] | X[c,m] | }
\cline{2-3}
& \textit{Projection approach} & \textit{Direct surface approach}
\end{tabu}\par\vskip-1.4pt

\tabulinesep = 0pt
\begin{tabu} to 0.95\textwidth {| p{10pt} | X[c,m] | X[c,m] | X[c,m] | }
\hline
\multirow{2}{*}[2.5pt]{\parbox[c]{\textwidth}{\rotatebox[origin=c]{90}{Model}}} & \tst Numerically-projected \bst & \tst Analytically-projected \bst & \tst Directly-decoupled \bst \\ 
& (NP) \bst & (AP) \bst & (DD) \bst \\ \hline
\parbox[c]{\textwidth}{\rotatebox[origin=c]{90}{Geometry}} &\raisebox{0\height}{\includegraphics[height=30mm]{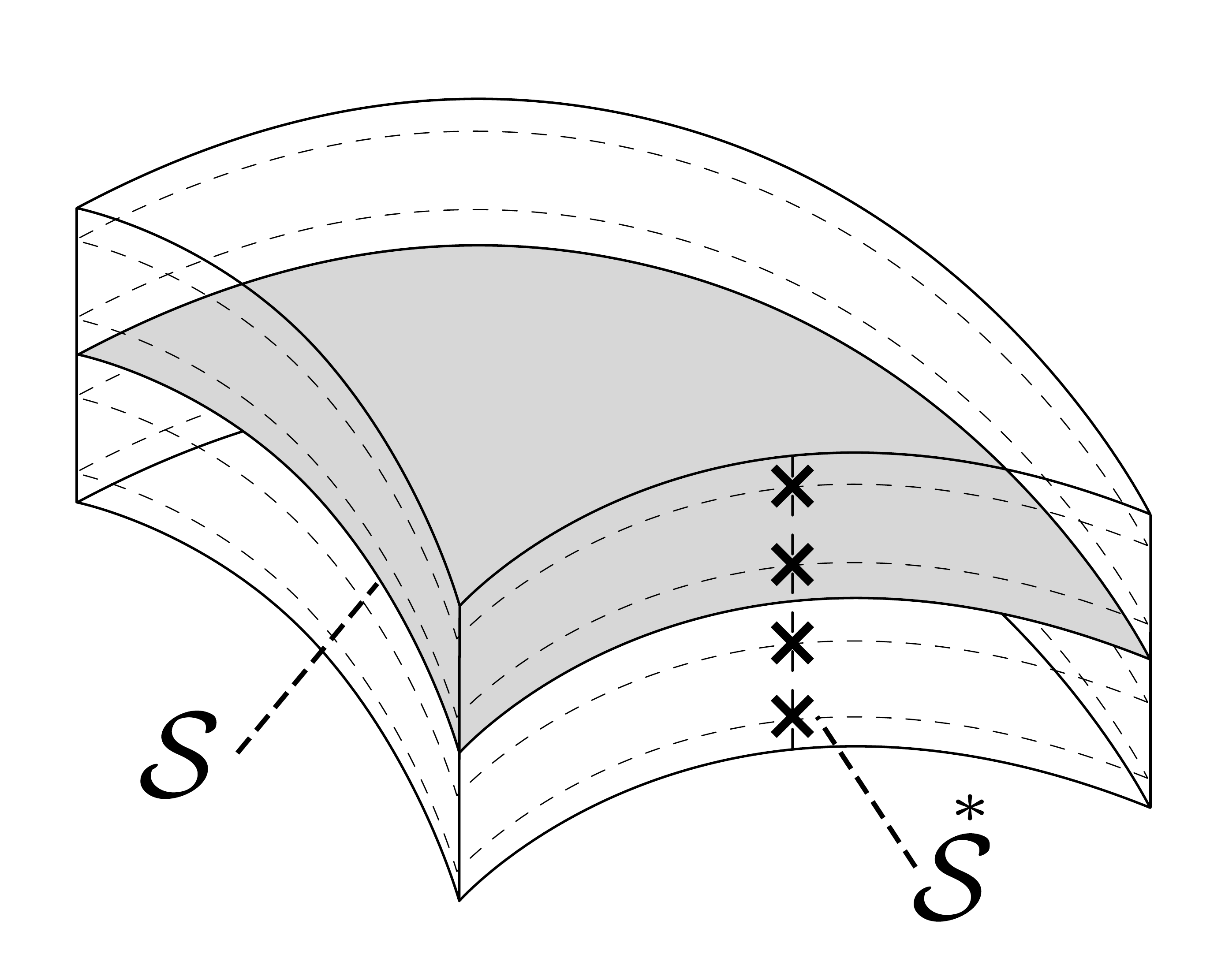}} & 
\raisebox{0\height}{\includegraphics[height=30mm]{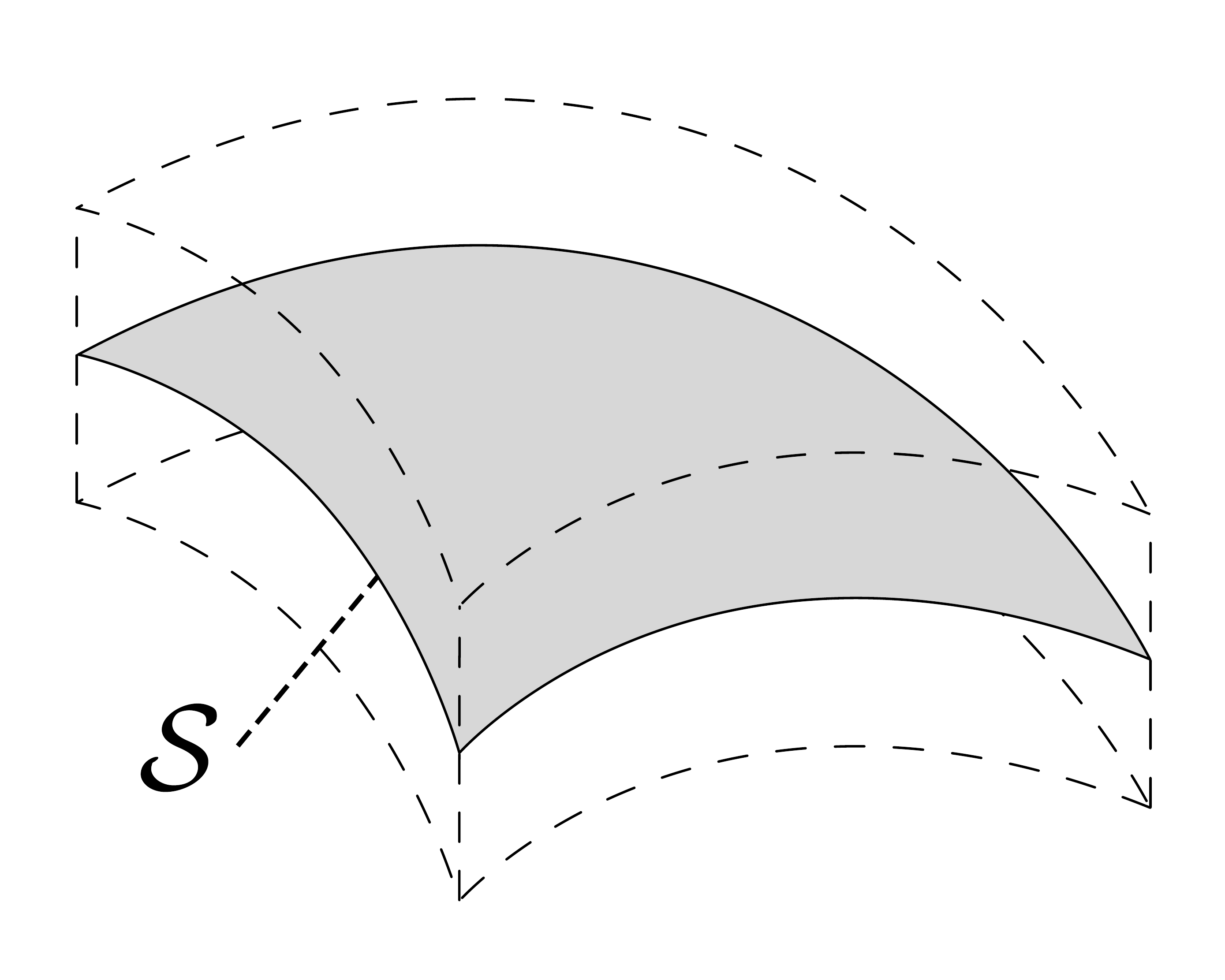}} &
\raisebox{0\height}{\includegraphics[height=30mm]{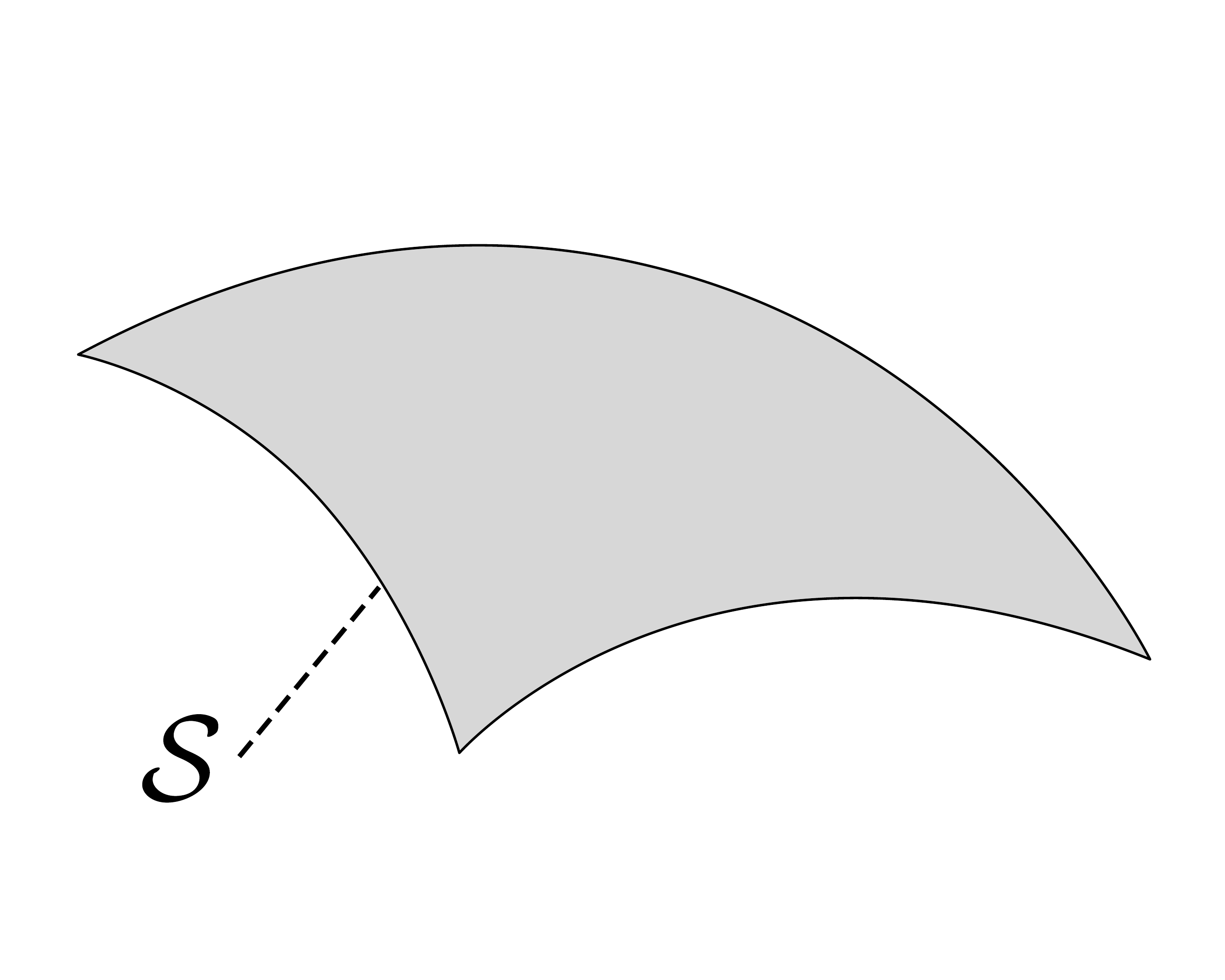}} \\ \hline
\parbox[c]{\textwidth}{\rotatebox[origin=c]{90}{Strain}} &
\raisebox{0\height}{\includegraphics[height=20mm]{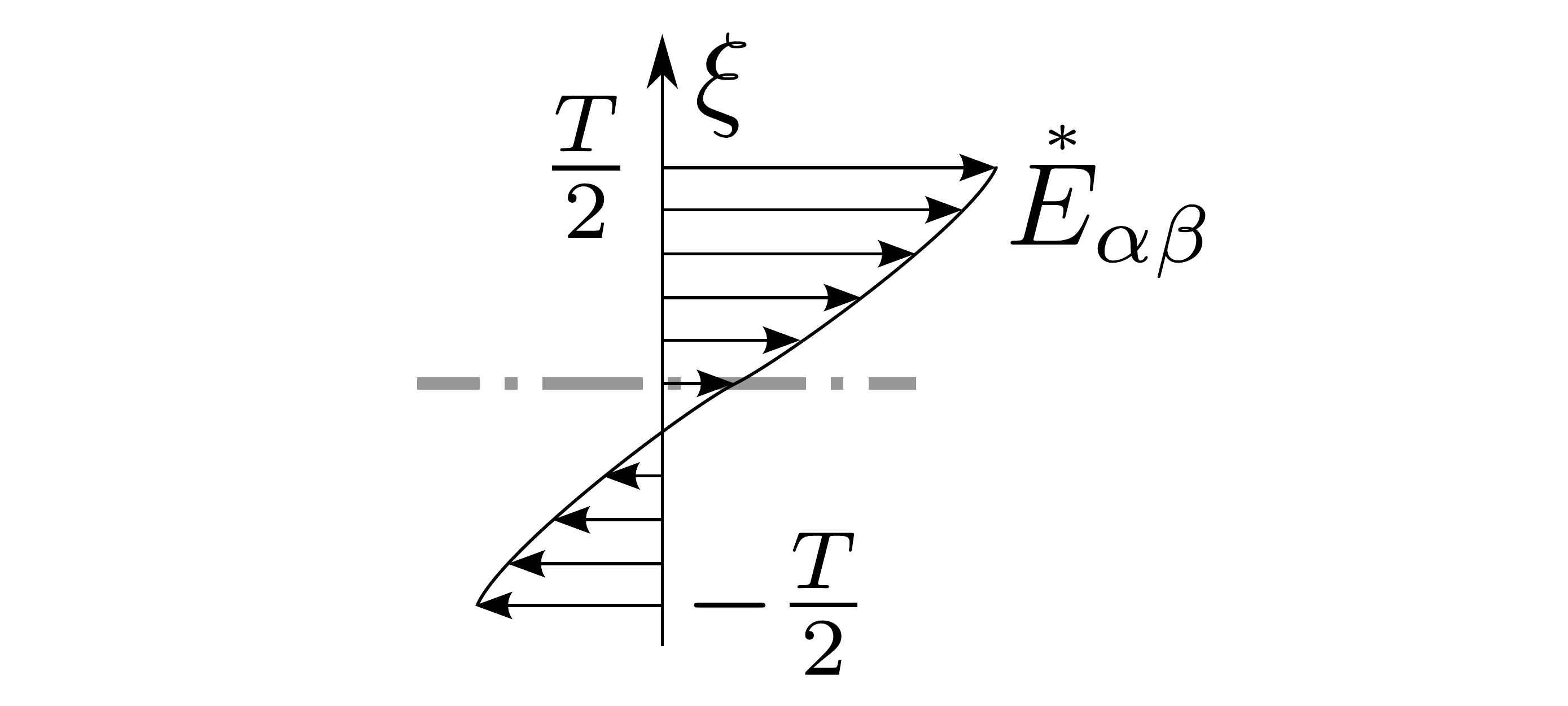}} & 
\raisebox{0\height}{\includegraphics[height=20mm]{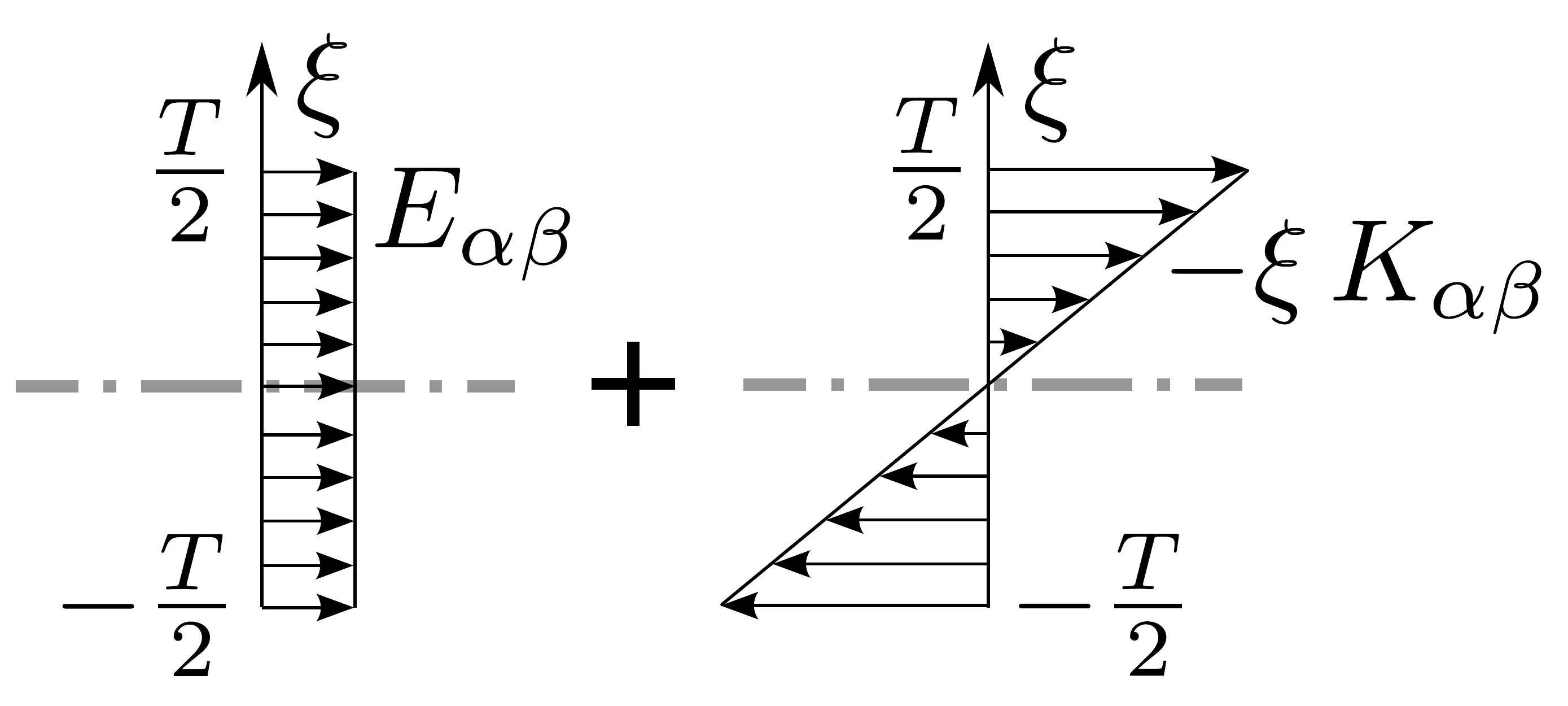}} &
\raisebox{0\height}{\includegraphics[height=20mm]{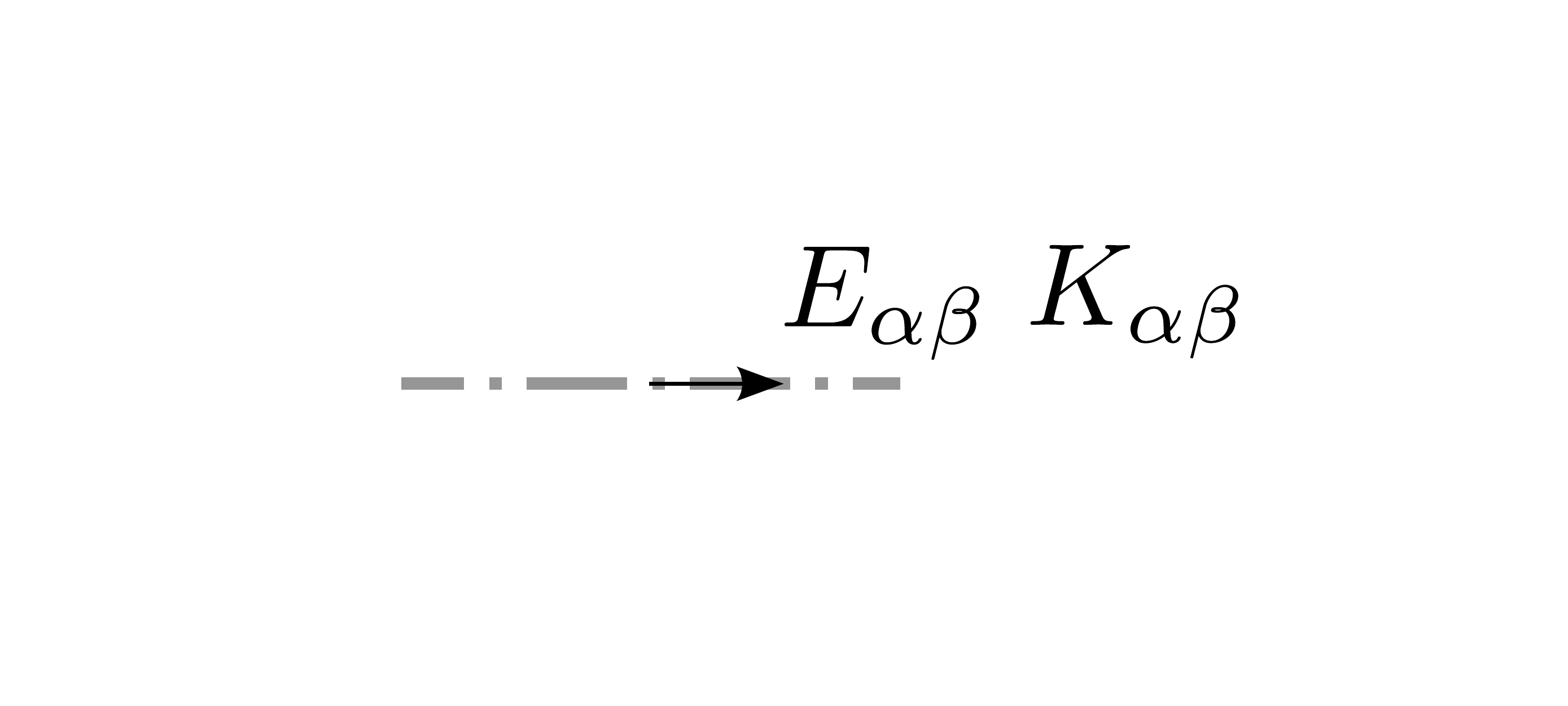}} \\ \hline
\parbox[c]{\textwidth}{\rotatebox[origin=c]{90}{Stress}} &
\raisebox{0\height}{\includegraphics[height=20mm]{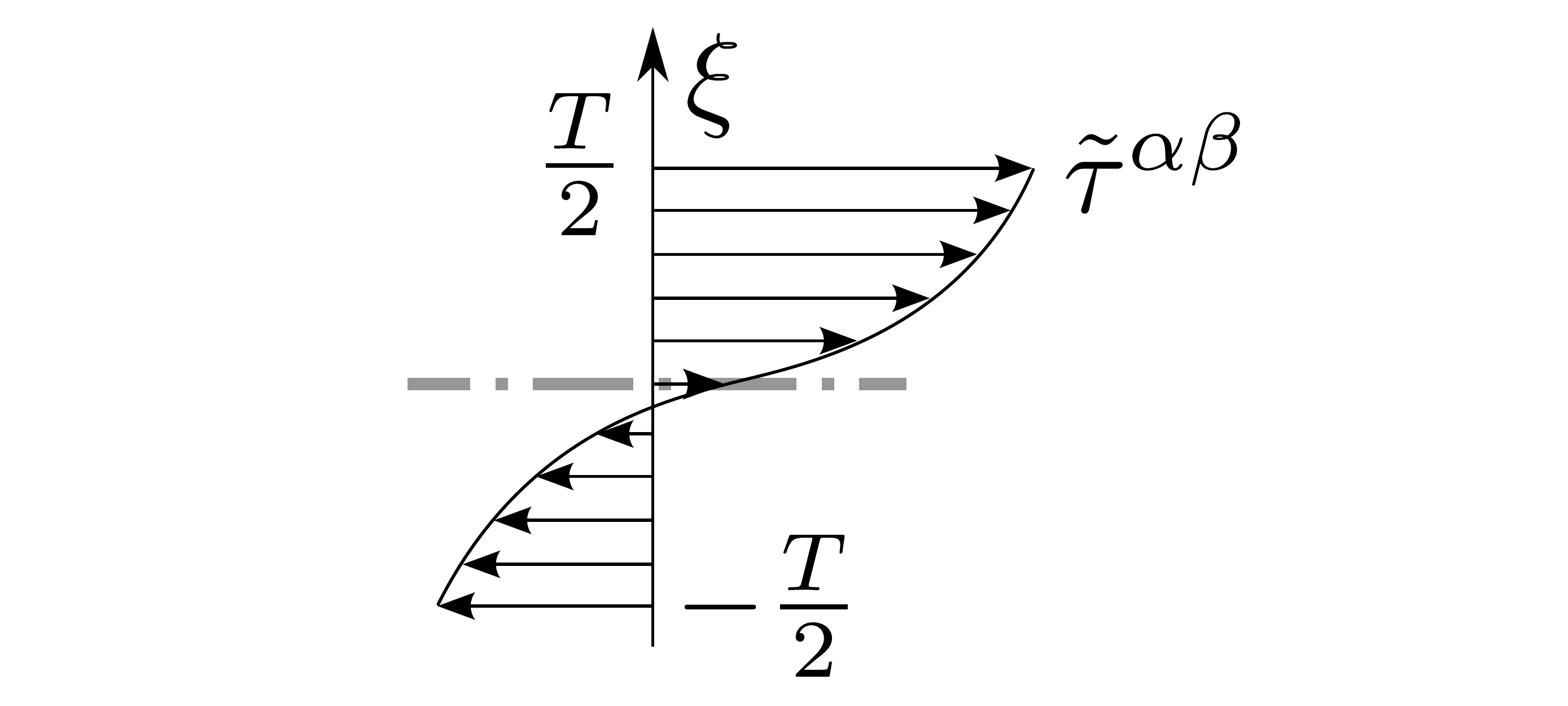}} & 
\raisebox{0\height}{\includegraphics[height=20mm]{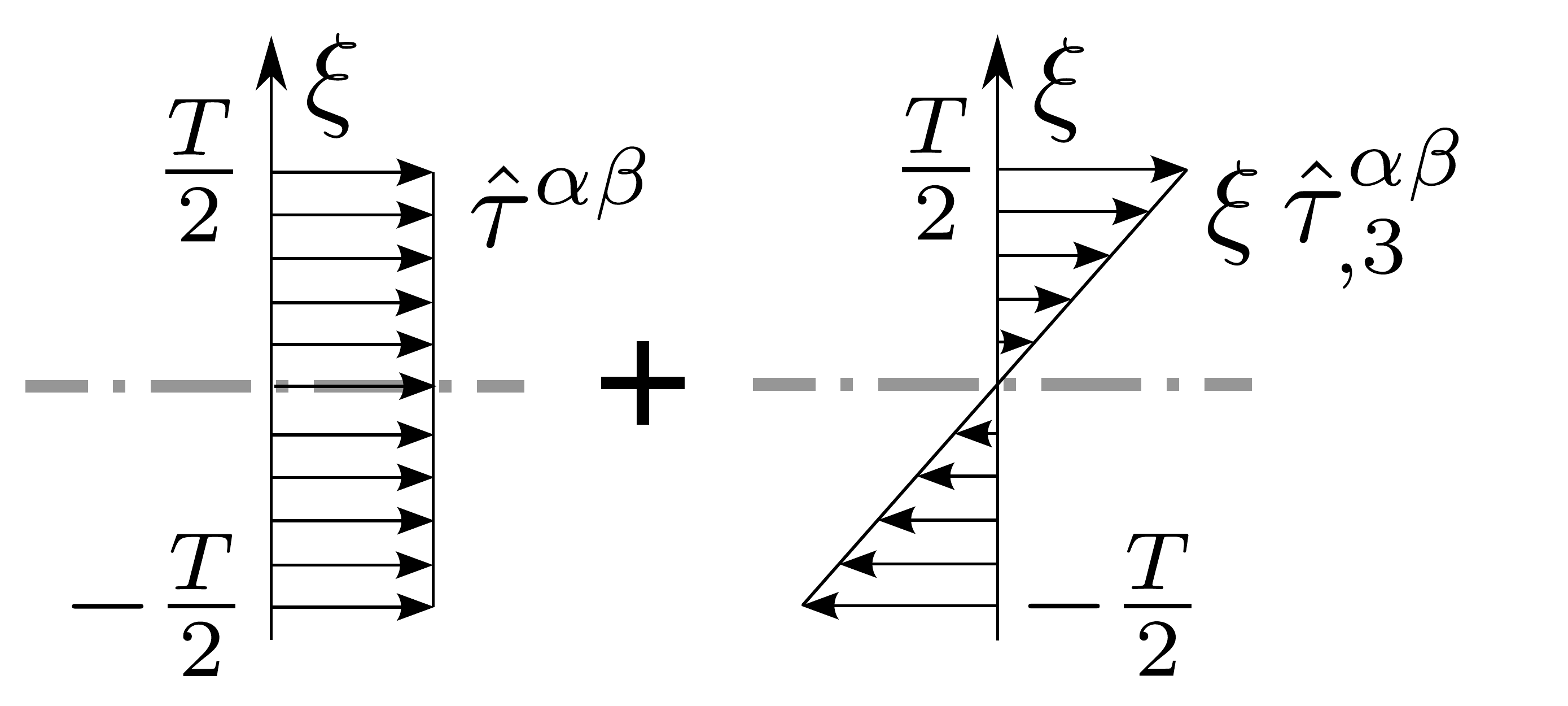}} &
\raisebox{0\height}{\includegraphics[height=20mm]{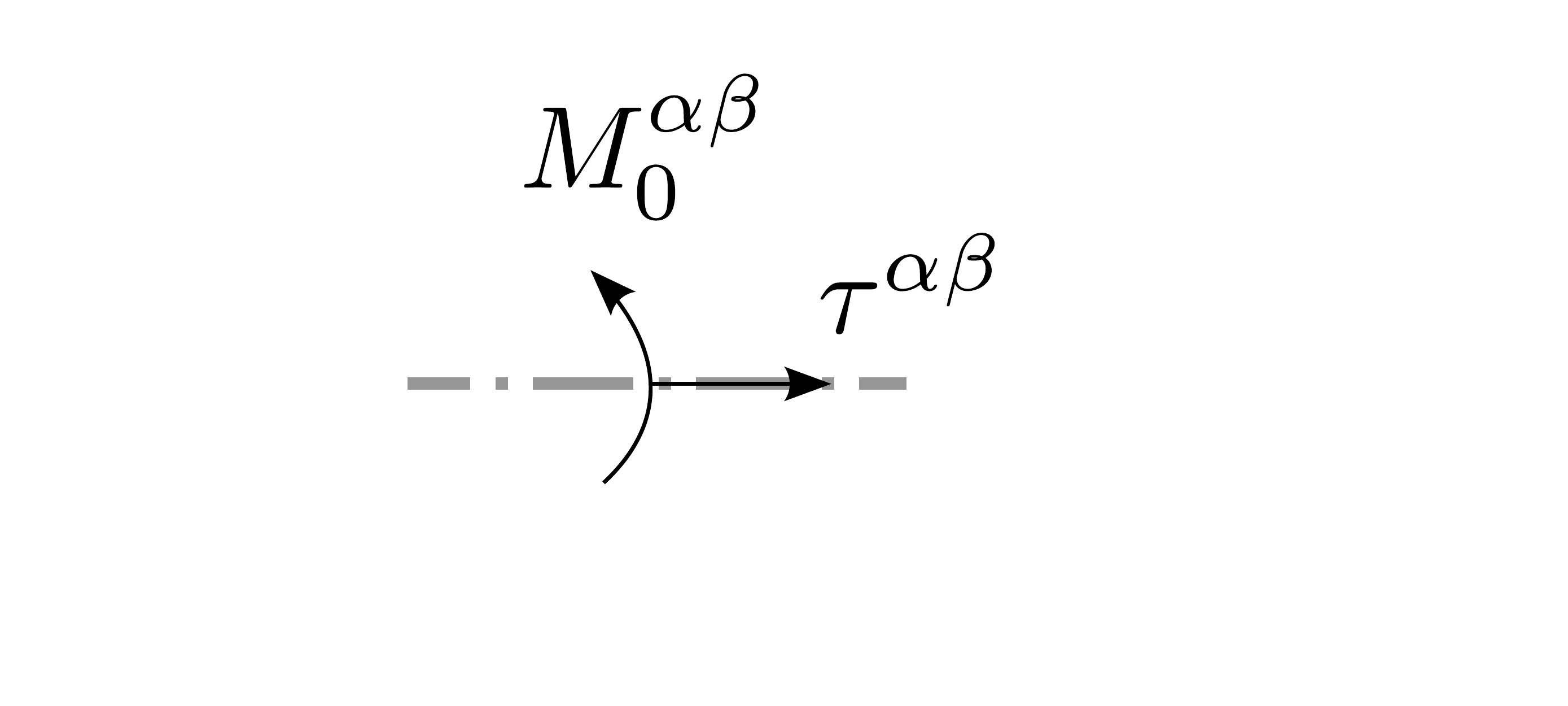}} \\ \hline
\end{tabu}

\caption{Three different modeling approaches for defining the constitution of thin shells}
\label{t:models}
\end{table}

Here, a model of the first approach, namely the \emph{numerically-projected} (NP) shell model, and a model of the second approach, namely the \emph{directly-decoupled} (DD) shell model are introduced. Further, the \emph{analytically-projected} (AP) shell model is presented, which combines elements of both approaches and provides an algorithm to analytically evaluate the integration through the shell thickness. For the NP shell model, it is assumed that the in-plane strains vary linearly across the shell thickness, which considerably simplifies the formulation yet it is accurate only for thin shells. For the DD shell model, by extending the formulation of \citet{solidshell} for a combined Koiter/Neo-Hooke shell, a systematic algorithm is introduced to consistently find the bending counterparts for any given membrane formulation. 

As shown in detail in the next sessions, the NP shell model is a fully nonlinear shell formulation. The AP shell model is a first-order approximation of the NP shell model, in which the membrane and bending forces are still coupled. The DD shell model combines a fully nonlinear membrane with a linear elastic bending model.

Further theoretical background and implementation details for the DD and the NP shell models can be found in the earlier works of the authors. Here, the principal concepts and the new extensions are introduced.  

\subsection{Numerically-projected (NP) shell model}\label{s:NP}    
In this approach, the strain energy density function of a {3D} shell continuum is projected onto the mid-surface of the shell so that the stress and bending moment resultants can be found by an appropriate integration through the shell thickness, which has to be evaluated numerically in general. The projected strain energy density function is
\eqb{l}
W = W(a_{\alpha\beta},\,b_{\alpha\beta}) = \ds\int_{-\frac {T}{2}}^{\frac{T}{2}} \, \tW(g_{\alpha\beta},g_{33})\, \dif\xi ~,
\label{e:Wproj}\eqe
where $\tW  = \tW (g_{\alpha\beta},g_{33})$ is the {3D} strain energy density function defined on a shell layer $\asS$ at $\xi\in[-T/2,T/2]$. Then, the Kirchhoff stress on $\asS$ is derived by the variation $\delta \tW = \tfrac{1}{2}\tilde\tau^{\alpha\beta}\, \delta g_{\alpha\beta}  +  \tmod{\tfrac{1}{2}}\tilde\tau^{33}\,\delta g_{33}$, which gives
\eqb{l}
\ttau^{\alpha\beta} := \ds 2\pa{\tW}{g_{\alpha\beta}}, \quad \ttau^{33} := \tmod{2}\ds\pa{\tW}{g_{33}} ~.
\label{e:ttau}\eqe
Writing the variation of the projected strain energy $W$ as
\eqb{l}
\delta W = \ds\int_{-\frac {T}{2}}^{\frac{T}{2}} \, \delta\tW(g_{\alpha\beta},g_{33})\,\dif\xi = \ds\frac{1}{2}\tau^{\alpha\beta}\, \delta a_{\alpha\beta} + M_0^{\alpha\beta}\, \delta b_{\alpha\beta}~,
\eqe
the resultant stress and moment tensors are
\eqb{l}
\tau^{\alpha\beta} := \ds 2\pa{W}{a_{\alpha\beta}}~, \quad M_0^{\alpha\beta}:= \ds \pa{W}{b_{\alpha\beta}} ~.
\label{e:tau1}\eqe
Plugging Eq.~\eqref{e:vg_ab} into Eqs.~\eqref{e:Wproj},~\eqref{e:ttau}~and~\eqref{e:tau1}, one can find the resultant stresses and bending moments as
\eqb{lll}
\tau^{\alpha\beta} \is \ds\int_{-\frac{T}{2}}^{\frac{T}{2}}\,\ttau^{\alpha\beta}\, \dif\xi ~,\\[5mm]
M^{\alpha\beta}_0 \is -\ds\int_{-\frac{T}{2}}^{\frac{T}{2}}\,\xi\,\ttau^{\alpha\beta} \,\dif\xi ~.
\label{e:tau2}\eqe

\begin{rem}\label{r:pstress}
Assuming the plane-stress condition, i.e.~$ \ttau^{33} = 0$, the out-of-plane \tmod{squared} stretch $g_{33}$ is eliminated by static condensation \citep{bischoff04,echter13t,kiendl15,solidshell}. 
\end{rem}

From Eqs.~\eqref{e:tau2},~\eqref{e:cdef}~and~\eqref{e:g_ab}, the material tangents follow as
\eqb{lll}
c^{\alpha\beta\gamma\delta} \is \ds\int_{-\frac{T}{2}}^{\frac{T}{2}}\,\tc^{\alpha\beta\gamma\delta}\,\dif\xi ~,\\[5mm]
d^{\alpha\beta\gamma\delta} \is e^{\alpha\beta\gamma\delta} = -\ds\int_{-\frac{T}{2}}^{\frac{T}{2}}\,\xi\,\tc^{\alpha\beta\gamma\delta}\,\dif\xi ~,\\[5mm]
f^{\alpha\beta\gamma\delta} \is \ds\int_{-\frac{T}{2}}^{\frac{T}{2}}\,\xi^2\,\tc^{\alpha\beta\gamma\delta}\,\dif\xi ~,
\label{e:tcabgd}\eqe 
where, on $\asS$, we have introduced the elasticity tensor
\eqb{l}
\tc^{\alpha\beta\gamma\delta}:=2\,\ds\pa{\ttau^{\alpha\beta}}{g_{\gamma\delta}}~.
\label{e:tcabgd2}\eqe

\subsection{Analytically-projected (AP) shell model}\label{s:AP}
If the shell thickness is small enough compared to the curvature radii of the shell, one can use a first order Taylor expansion to analytically evaluate the integrals in Eqs.~\eqref{e:tau2}~and~\eqref{e:tcabgd}. For this purpose, all the kinematical objects, stresses and bending moments are linearized w.r.t.~the out-of-plane coordinate $\xi$. The linearization of the kinematical parameters are given in Appendix~\ref{s:vxi}. In this section, the stresses, bending moments and material tangents are derived for two cases: (1) The whole thickness of the shell contributes to the strain energy density function and (2) only a portion of the shell thickness, e.g.~$[T_1~T_2] \subset [-T/2~T/2]$, is active. The former case is the typical condition of thin shells while the latter happens for instance if the material bears only compression (e.g.~concrete) or only tension (e.g.~collagen fibers).  

Using a Taylor expansion of $\ttau^{\alpha\beta}$ about $\xi=0$, we have
\eqb{l}
\ttau^{\alpha\beta} = \htau^{\alpha\beta} + \xi\,\htau^{\alpha\beta}_{,3} + O(\xi^2)~,
\label{e:ltau}\eqe
where we have defined
\eqb{l}
\hat\tau^{\alpha\beta} := \Big(\tilde\tau^{\alpha\beta}\Big)_{\xi=0}~, \quad\quad
\hat\tau^{\alpha\beta}_{,3} := \Big(\ds\pa{\tilde\tau^{\alpha\beta}}{\xi}\Big)_{\xi=0} ~.
\label{e:ltau1}\eqe

\begin{rem}
Henceforth, a hat is used to denote the quantities calculated at $\xi=0$, i.e.~$\hat{\bullet}=(\accentset{*}{\bullet})_{\xi=0}$. In general, such quantities can be defined for each shell layer. In particular, they can be dimensionally linked to a counterpart in the membrane theory (e.g.~$ \auab = \hg_{\alpha\beta} $ and $\tau^{\alpha\beta} = T\,\htau^{\alpha\beta}$) or there might be no corresponding quantity in the membrane theory (e.g.~for $\htau^{\alpha\beta}_{,3}$).
\end{rem} 

\subsubsection{Fully-stressed cross-section}\label{s:fullsh}
Plugging Eq.~\eqref{e:ltau} into Eq.~\eqref{e:tau2} and integrating analytically, the resultant stresses and bending moments are 
\eqb{lll}
\tau^{\alpha\beta} \is T\,\hat\tau^{\alpha\beta}~,\\[2mm]
M_0^{\alpha\beta} \is -\ds\frac{T^3}{12}\,\hat\tau^{\alpha\beta}_{,3}~. 
\label{e:ltau2}\eqe

The tangent matrices are derived from Eq.~\eqref{e:ltau2} as
\eqb{lllllllll}
c^{\alpha\beta\gamma\delta} \dis 2\,\ds\pa{\tau^{\alpha\beta}}{a_{\gamma\delta}} \is T\,\hat{c}^{\alpha\beta\gamma\delta}~,\quad\quad
d^{\alpha\beta\gamma\delta} \dis \ds\pa{\tau^{\alpha\beta}}{b_{\gamma\delta}} \is T\,\hat{d}^{\alpha\beta\gamma\delta}~,\\[2mm]
e^{\alpha\beta\gamma\delta} \dis 2\,\ds\pa{M_0^{\alpha\beta}}{a_{\gamma\delta}} \is -\ds\frac{T^3}{12}\,\hat{c}^{\alpha\beta\gamma\delta}_{,3}~,\hfill
f^{\alpha\beta\gamma\delta} \dis \ds\pa{M_0^{\alpha\beta}}{b_{\gamma\delta}} \is -\ds\frac{T^3}{12}\,\hat{d}^{\alpha\beta\gamma\delta}_{,3}~,
\label{e:lcabgd1}\eqe
where we have defined
\eqb{l}
\hat{c}^{\alpha\beta\gamma\delta} := 2\,\ds\pa{\hat\tau^{\alpha\beta}}{a_{\gamma\delta}}~,\quad\quad
\hat{d}^{\alpha\beta\gamma\delta} := \ds\pa{\hat\tau^{\alpha\beta}}{b_{\gamma\delta}}~,\quad\quad
\hat{c}^{\alpha\beta\gamma\delta}_{,3} := 2\,\ds\pa{\hat\tau^{\alpha\beta}_{,3}}{a_{\gamma\delta}}~,\quad\quad
\hat{d}^{\alpha\beta\gamma\delta}_{,3} := \ds\pa{\hat\tau^{\alpha\beta}_{,3}}{b_{\gamma\delta}}~.
\label{e:lcabgd2}\eqe

\subsubsection{Partially-stressed cross-section}\label{s:parsh}
As already mentioned, in many applications, the strain energy density function and accordingly the in-plane stresses are nonzero only in a portion of the shell thickness, i.e.
\eqb{lll}
\tau^{\alpha\beta} \is \ds\int_{T_1}^{T_2}\,\ttau^{\alpha\beta}\, \dif\xi ~,\\[5mm]
M^{\alpha\beta}_0 \is -\ds\int_{T_1}^{T_2}\,\xi\,\ttau^{\alpha\beta} \,\dif\xi ~.
\label{e:ltau3}
\eqe

Plugging Eq.~\eqref{e:ltau} into Eq.~\eqref{e:ltau3}, one can analytically calculate the stress and the bending moment resultants as
\eqb{rcrcl}
\tau^{\alpha\beta} \is \big(T_2 -T_1\big)\,\hat{\tau}^{\alpha\beta} \plus \ds\frac{1}{2}\big(T_2^2 -T_1^2\big)\,\hat{\tau}^{\alpha\beta}_{,3} ~, \\[3mm]
M_0^{\alpha\beta} \is \ds\frac{1}{2}\big(T_1^2 -T_2^2\big)\,\hat{\tau}^{\alpha\beta} \plus \ds\frac{1}{3}\big(T_1^3 -T_2^3\big)\,\hat{\tau}^{\alpha\beta}_{,3}~.
\label{e:ltau4}\eqe

However, the derivation of the material tangents is not so simple since $T_1$ and $T_2$ are not considered to be generally fixed and they may vary with $\bx$, i.e.
\eqb{lllll}
T_1 \is T_1(\bx) \is T_1(a_{\alpha\beta}, b_{\alpha\beta})~,\\[2mm]
T_2 \is T_2(\bx) \is T_2(a_{\alpha\beta}, b_{\alpha\beta})
\eqe  
and they are defined based on the constitution and application. Thus, introducing
\eqb{lllllllll}
U_1^{\alpha\beta} := \ds\pa{T_1}{a_{\alpha\beta}}~,\quad\quad U_2^{\alpha\beta} := \ds\pa{T_2}{a_{\alpha\beta}}~, \quad\quad
V_1^{\alpha\beta} := \ds\pa{T_1}{b_{\alpha\beta}}~,\quad\quad V_2^{\alpha\beta} := \ds\pa{T_2}{b_{\alpha\beta}}~,
\label{e:Uab_i}\eqe
the material tangents are derived as
\eqb{lll}
c^{\alpha\beta\gamma\delta} \is 
\big(T_2 -T_1\big)\,\hat{c}^{\alpha\beta\gamma\delta} + \ds\frac{1}{2}\big(T_2^2 -T_1^2\big)\,\hat{c}^{\alpha\beta\gamma\delta}_{,3} + 2\,\tilde\tau^{\alpha\beta}_2\,U_2^{\gamma\delta} - 2\,\tilde\tau^{\alpha\beta}_1\,U_1^{\gamma\delta}~, \\[2mm]
d^{\alpha\beta\gamma\delta} \is 
\big(T_2 -T_1\big)\,\hat{d}^{\alpha\beta\gamma\delta} + \ds\frac{1}{2}\big(T_2^2 -T_1^2\big)\,\hat{d}^{\alpha\beta\gamma\delta}_{,3} + \tilde\tau^{\alpha\beta}_2\,V_2^{\gamma\delta} - \tilde\tau^{\alpha\beta}_1\,V_1^{\gamma\delta}~, \\[2mm]
e^{\alpha\beta\gamma\delta} \is 
\ds\frac{1}{2}\big(T_1^2 -T_2^2\big)\,\hat{c}^{\alpha\beta\gamma\delta} + \ds\frac{1}{3}\big(T_1^3 -T_2^3\big)\,\hat{c}^{\alpha\beta\gamma\delta}_{,3} + 2\,T_1\,\tilde\tau^{\alpha\beta}_1\,U_1^{\gamma\delta} - 2\,T_2\,\tilde\tau^{\alpha\beta}_2\,U_2^{\gamma\delta}~, \\[2mm]
f^{\alpha\beta\gamma\delta} \is 
\ds\frac{1}{2}\big(T_1^2 -T_2^2\big)\,\hat{d}^{\alpha\beta\gamma\delta} + \ds\frac{1}{3}\big(T_1^3 -T_2^3\big)\,\hat{d}^{\alpha\beta\gamma\delta}_{,3} + T_1\,\tilde\tau^{\alpha\beta}_1\,V_1^{\gamma\delta} - T_2\,\tilde\tau^{\alpha\beta}_2\,V_2^{\gamma\delta}~.
\label{e:lcabgd}\eqe
Here, $\tilde\tau^{\alpha\beta}_1$ and $\tilde\tau^{\alpha\beta}_2$ are the stresses corresponding to the lower and upper limits, respectively, which are defined as
\eqb{lllll}
\tilde\tau^{\alpha\beta}_1 \dis \left(\tilde\tau^{\alpha\beta}\right)_{\xi=T_1} \is \hat\tau^{\alpha\beta} + T_1\,\hat\tau^{\alpha\beta}_{,3}~, \\[3mm]
\tilde\tau^{\alpha\beta}_2 \dis \left(\tilde\tau^{\alpha\beta}\right)_{\xi=T_2} \is \hat\tau^{\alpha\beta} + T_2\,\hat\tau^{\alpha\beta}_{,3}~.
\label{e:ttau_12}\eqe
\begin{rem}
It should be noted that the tangent tensors given by Eqs.~\eqref{e:lcabgd1}~and~\eqref{e:lcabgd} may not be symmetric as the linearization and variation are treated differently. 
\end{rem}

Tab.~\ref{t:AP_sum} summarizes the procedure to derive a AP shell model from any given {3D} material model. 

\begin{table}[ht]
\centering
\tabulinesep = 2.0pt
\begin{tabu} to 0.9\textwidth { | X[-10,r,p] X[l,p] | }
  \hline
  1) & For any {3D} constitution, with a given strain energy function $\tW$, derive $\ttau^{\alpha\beta}$ according to Eq.~\eqref{e:ttau}. \\
  2) & Determine $\htau^{\alpha\beta}$ from Eqs.~(\ref{e:ltau1}.1) and $\hc^{\alpha\beta\gamma\delta}$ and $\hd^{\alpha\beta\gamma\delta}$ from Eq.~\eqref{e:lcabgd2}. \\
  3) & Determine $\htau^{\alpha\beta}_{,3}$ from Eq.~(\ref{e:ltau1}.2) using the linearized kinematical variables (see Appendix~\ref{s:vxi}). Compute $\hc^{\alpha\beta\gamma\delta}_{,3}$ and $\hd^{\alpha\beta\gamma\delta}_{,3}$ from Eq.~\eqref{e:lcabgd2}.\\
  4.a) & For fully-stressed shells: $\tau^{\alpha\beta}$ and $M_0^{\alpha\beta}$ and their corresponding tangents are found from Eq.~\eqref{e:ltau2}~and~\eqref{e:lcabgd1}. \\
  4.b) & For partially-stressed shells: \\
  & 4.b.1) Find the effective thickness $[T_1,T_2]$ and its corresponding tensors, given by Eq.~\eqref{e:Uab_i}, using Appendices~\ref{s:vxi}~and~\ref{s:switch_lin}. \\
  & 4.b.2) Determine $\tau^{\alpha\beta}$ and $M_0^{\alpha\beta}$ and their corresponding tangents from Eqs.~\eqref{e:ltau4}~and~\eqref{e:lcabgd}. \\
  \hline
\end{tabu}
\caption{Summary of the analytically-projected shell formulation}
\label{t:AP_sum}
\end{table}

\subsection{Directly-decoupled (DD) shell model}\label{s:DD}
In this approach, the stresses and moments are directly derived from a {2D} strain energy density function; therefore, there is no need for numerical integration thought the shell thickness. The model completely decouples the membrane and bending forces. It predicts the stretching deformation of the shell through the metric tensor while the bending of the shell depends only on the curvature tensor \citep{ciarlet05,shelltheo}. The classic shell formulations of this kind, namely Koiter shell model and Canham model, are described in detail e.g.~by \citet{ciarlet05} and \citet{shelltheo}. Recently, \citet{solidshell} have proposed a mixed formulation that combines the bending stored energy of a Koiter shell and the strain energy of a compressible Neo--Hookean membrane. Here, a systematic approach is introduced to find an appropriate and consistent bending energy for any given isotropic or anisotropic membrane formulation so that their combination results in a directly-decoupled shell model with a polyconvex {2D} strain energy density function as 
\eqb{l}
W = W(a_{\alpha\beta},b_{\alpha\beta}) = W_\mrM(a_{\alpha\beta}) + W_\mrB(b_{\alpha\beta}) ~,
\label{e:Wmb}
\eqe
where $W_\mrM$ and $W_\mrB$ are the membrane and bending parts, respectively. 

\begin{rem}
Compared to the combined Koiter/Neo-Hooke shell model of \citet{solidshell}, the presented formulation (i) provides a physically well-defined link between the bending and membrane parts, (ii) is extended to many isotropic material models and (iii) allows for anisotropic behavior, which is of great importance for the modeling of soft tissues.
\end{rem}

Many biological materials, such as soft tissues, can easily undergo large deformations while being extremely stretched. Thus, the membrane formulation should allow for large  material and geometrical nonlinearities. This implies that, for the membrane part, a nonlinear stress-strain relationship is required; however, for the bending part, a linear stress-strain relationship is sufficient for most applications even when the shell exhibits large deformations.  

For membranes, the strains are constant over the thickness, i.e. $g_{\alpha\beta} = a_{\alpha\beta}$ and $g_{33} = \tmod{\lam^2}$. Thus, a {3D} material model can be reduced to a {2D} membrane one as \citep{biomembrane}
\eqb{l}
W_\mrM(a_{\alpha\beta},\lam) = \ds\int_{-\frac{T}{2}}^{\frac{T}{2}}\tW(g_{\alpha\beta},g_{33})\,\dif\xi = T\,\hW(a_{\alpha\beta},\tmod{\lam})~,
\label{e:Wm}\eqe
where 
\eqb{l}
\hW(a_{\alpha\beta},\tmod{\lam}) := \left[\tW(\guab,g_{33})\right]_{\xi=0}
\eqe
is the {3D} strain energy density function in terms of the mid-surface metric tensor $\auab$ and normal stretch $\lam$. From Eq.~\eqref{e:Wm}, it follows that 
\eqb{l}
\tau^{\alpha\beta} := 2\,\ds\pa{W_\mrM}{a_{\alpha\beta}} = 2\,T\,\ds\pa{\hW}{a_{\alpha\beta}}
\label{e:taumb}\eqe
and
\eqb{l}
c^{\alpha\beta\gamma\delta} := 2\,\ds\pa{\tau^{\alpha\beta}}{a_{\gamma\delta}} = 4\,T\,\ds\paqq{\hW}{a_{\alpha\beta}}{a_{\gamma\delta}}
\label{e:cmb}\eqe
are the components of the membrane elasticity tensor. Likewise to the NP shell model (see Remark \ref{r:pstress}), the membrane out-of-plane stretch $\lam$ is eliminated by the plane-stress assumption, i.e.~$\ttau^{33} := \lam^{-1}\partial W/\partial\lam = 0$.

The corresponding linear stress-strain relation is derived from a linearized material model as
\eqb{l}
W_\mathrm{lin}(a_{\alpha\beta},b_{\alpha\beta}) = \ds\frac{1}{2}\,c_0^{\alpha\beta\gamma\delta}\,E_{\alpha\beta}\,E_{\gamma\delta} 
+ \ds\frac{T^2}{24}\,c_0^{\alpha\beta\gamma\delta}\,K_{\alpha\beta}\,K_{\gamma\delta} ~,
\label{e:WL}\eqe
where 
\eqb{l}
c_0^{\alpha\beta\gamma\delta} := \Big(c^{\alpha\beta\gamma\delta}\Big)_{\sS_0} 
\label{e:c_0mb}\eqe
are the components of the elasticity tensor before deformation and $c^{\alpha\beta\gamma\delta}$ is given by Eq.~\eqref{e:cmb}. Then, the rear part of Eq.~\eqref{e:WL} can be used for the {2D} strain energy density function corresponding to the bending deformations as
\eqb{l}
W_\mrB = W_\mrB(b_{\alpha\beta}) = \ds\frac{T^2}{24}\,c_0^{\alpha\beta\gamma\delta}\,K_{\alpha\beta}\,K_{\gamma\delta} ~.
\label{e:Wb}\eqe

From Eq.~\eqref{e:Wb}, the bending moments and their corresponding tangents are
\eqb{l}
M_0^{\alpha\beta} := \ds\pa{W_\mrB}{b_{\alpha\beta}} = \ds\frac{T^2}{12}\,c_0^{\alpha\beta\gamma\delta}\,K_{\gamma\delta}~,
\label{e:Mmb}
\eqe
\eqb{l}
f^{\alpha\beta\gamma\delta} := \ds\pa{M_0^{\alpha\beta}}{b_{\gamma\delta}} = \ds\frac{T^2}{12}\,c_0^{\alpha\beta\gamma\delta}~.
\label{e:fmb}
\eqe

\begin{rem}
As the bending and membrane parts are decoupled, here in contrast to the NP shell model, $d^{\alpha\beta\gamma\delta}=e^{\alpha\beta\gamma\delta}=0$ \citep{shelltheo}.   
\end{rem}
\begin{rem}
The presented decoupled membrane-bending equations are derived provided that (1) the material is symmetric w.r.t.~the shell mid-surface and (2) the shell thickness is considerably smaller than the other dimensions and the radii of curvature. If these two conditions are violated, in addition to the stretching and bending strains, other mixed terms (i.e.~strain gradients) are present in Eq.~\eqref{e:WL}; therefore, the membrane and bending parts cannot be easily decoupled. 
\end{rem}

Tab.~\ref{t:DD_sum} summarizes the procedure to formulate a DD shell model for any given {3D} material model. 

\begin{table}[ht]
\centering
\tabulinesep = 2.0pt
\begin{tabu} to 0.8\textwidth { | X[-10,r,p] X[l,p] | }
  \hline
  1) & For any {3D} constitution, with a given strain energy function $\tW$, postulate a {2D} counterpart as $W_\mrM + W_\mrB$. \\
  2) & For the membrane part $W_\mrM$, compute \\
  & 2.1) the membrane stresses $\tau^{\alpha\beta}$ from Eqs.~\eqref{e:Wm}~and~\eqref{e:taumb},\\
  & 2.1) the membrane elasticity tensor $c^{\alpha\beta\gamma\delta}$ from Eq.~\eqref{e:cmb}.\\
  3) & For the bending part $W_\mrB$, compute \\
  & 3.1) $c_0^{\alpha\beta\gamma\delta}$ from Eq.~\eqref{e:c_0mb},\\
  & 3.2) the bending moments $M_0^{\alpha\beta}$ from Eq.~\eqref{e:Mmb},\\
  & 3.3) $f^{\alpha\beta\gamma\delta}$ from Eq.~\eqref{e:fmb}.\\
  \hline
\end{tabu}
\caption{Summary of the directly-decoupled shell formulation}
\label{t:DD_sum}
\end{table}

%
%
\section{Material models}\label{s:material}
Having introduced the thin shell theory and three different approaches to model shells, various isotropic and anisotropic constitutive laws can be examined now. For each material model, the three different approaches from Sec.~\ref{s:shells} (i.e.~the NP, AP and DD shell models), are derived. All the introduced material models are considered to be incompressible since most types of soft biological materials, in particular soft tissues, are regarded as incompressible \citep{holzapfel01}. Here, the incompressibility constraint
\eqb{l}
\tg := 1 - \tJ = 0
\label{e:tg}\eqe
is enforced strictly through the Lagrange multiplier method. Thus, the incompressible {3D} stored energy $\tW_\mathrm{inc}(\tbC)$ is augmented by the contribution from the Lagrange multiplier as
§
\eqb{l}
\tW(\tbC,\tJ) = \tW_\mathrm{inc}(\tbC) + \tp\,\tg~,
\label{e:tWLag}\eqe
where the unknown Lagrange multiplier $\tp$ is a hydrostatic pressure. For shells and membranes, it is \emph{analytically} determined from the plane-stress condition as $\tp = \tmod{2\,\aJ^{-2}}\,\partial \tW_\mathrm{inc}(\tbC)/\partial g_{33}$. In addition to physical reasons, the plane-stress condition is therefore advantageous for thin shells.

In the DD shell model, the incompressibility constraint is treated analogously for the membrane part. The incompressibility constraint is added to the corresponding incompressible {2D} stored energy $W_{\mathrm{inc}}(\tbC)$ through the Lagrange multiplier method \citep{membrane,memtheo} 
\eqb{l}
W_\mrM(\tbC ,\tJ) = W_{\mathrm{inc}}(\tbC) + p\,\tg ~,
\label{e:WLag}
\eqe
which is similar to the {3D} formulation, cf.~Eq.~\eqref{e:tWLag}. The unknown Lagrange multiplier $p=T\,\tp$ can also be \emph{analytically} found from the plane-stress condition. For the bending part, the effect of incompressibility constraint is condensed into $c_0^{\alpha\beta\gamma\delta}$. 

In Secs.~\ref{s:iso}~and~\ref{s:aniso}, the stress, bending moment and tangent tensors for different isotropic and anisotropic material models are derived. To avoid repetition, the derivations are not explained in detail. As summarized in Tab.~\ref{t:shell_sum}, for the NP shell model, one needs to derive $\ttau^{\alpha\beta} $ and $\tc^{\alpha\beta\gamma\delta} $ specifically for any given material model. Then, the stress and moment tensors and their corresponding tangents are determined by plugging the specific $\ttau^{\alpha\beta} $ and $\tc^{\alpha\beta\gamma\delta} $ into Eqs.~\eqref{e:tau2}~and~\eqref{e:tcabgd}, respectively. \\
For the AP shell model, $\htau^{\alpha\beta}$, $\htau^{\alpha\beta}_{,3}$, $\hc^{\alpha\beta\gamma\delta} $, $\hd^{\alpha\beta\gamma\delta} $, $\hc^{\alpha\beta\gamma\delta}_{,3} $ and $\hd^{\alpha\beta\gamma\delta}_{,3} $ are needed. Then, the stress and moment tensors and their corresponding tangents follow from step 4 in Tab.~\ref{t:AP_sum}. \\
For the DD shell model, having derived $\tau^{\alpha\beta}$ and $c^{\alpha\beta\gamma\delta} $ specifically for each material model, one can compute $c^{\alpha\beta\gamma\delta}_0 $, $M_0^{\alpha\beta}$ and $f^{\alpha\beta\gamma\delta} $ from step 3 in  Tab.~\ref{t:DD_sum}.

\begin{table}[ht]
\centering
{\renewcommand{\arraystretch}{1.2}
\begin{tabular}{ r | l}
  \toprule
  Shell model & Required constitutive variables \\\hline
  Numerically-projected (NP) & $\ttau^{\alpha\beta} $ \eqref{e:ttau} and $\tc^{\alpha\beta\gamma\delta} $ \eqref{e:tcabgd2} \\
  Analytically-projected (AP) & $\htau^{\alpha\beta}$ (\ref{e:ltau1}.1), $\htau^{\alpha\beta}_{,3}$ (\ref{e:ltau1}.2) and $\hc^{\alpha\beta\gamma\delta} $, $\hd^{\alpha\beta\gamma\delta} $, $\hc^{\alpha\beta\gamma\delta}_{,3} $ and $\hd^{\alpha\beta\gamma\delta}_{,3} $ \eqref{e:lcabgd2} \\ 
  Directly-decoupled (DD) & $\tau^{\alpha\beta}$ \eqref{e:taumb}, $c^{\alpha\beta\gamma\delta} $ \eqref{e:cmb}, $c^{\alpha\beta\gamma\delta}_0 $  \eqref{e:c_0mb} and $f^{\alpha\beta\gamma\delta} $ \eqref{e:fmb} \\ 
  \bottomrule
\end{tabular}}
\caption{Constitutive variables of the NP, AP and DD shell models}
\label{t:shell_sum}
\end{table} 

\subsection{Isotropic models}\label{s:iso}
Soft biomaterials are commonly modeled with incompressible hyperelastic constitutive models that have been introduced for rubber-like materials. Although soft tissues are constructed from elastin and collagen fibres, the anisotropic part might be neglected and a purely isotropic model can be used. Examples are the modeling of  liver, kidney, bladder and rectum, lungs, uterus, etc.~\citep{chagnon15}. This section discusses a few isotropic constitutive models that are commonly used for biomaterials and soft tissues \citep{martins06,wex15}. Both kinds of constitutive laws, i.e.~material models with polynomial and exponential forms of strain energy functions, are included in the presented examples.   

\subsubsection{Incompressible Neo--Hooke (NH)}\label{s:incNH}
The incompressible Neo--Hookean (NH) model is the most common hyperelastic constitution for rubber--like and soft biological materials. It is constructed from the first invariant of the right Cauchy--Green deformation tensor; therefore, it requires only one material constant to be set. 

\paragraph{NP shell model}
The strain-energy density function of a {3D} incompressible Neo-Hookean solid is
\eqb{l}
\tW(\tI_1, \tJ) =  \ds\frac{\tc_1}{2}\left(\tI_1 - 3 \right) + \tp\,\tg~,
\label{e:tWNH}\eqe
where $\tc_1 = \tilde{\mu}$ is the infinitesimal 3D shear modulus. The stress and elasticity tensors needed for the projection/integration procedure in Eqs.~\eqref{e:tau2}~and~\eqref{e:tcabgd} can be found in \citet{solidshell} as
\eqb{l}
\ttau^{\alpha\beta} := 2\,\ds\pa{\tW}{g_{\alpha\beta}} = \tc_1\,\ds\left( G^{\alpha\beta} - \frac{1}{\aJ^2} \,g^{\alpha\beta} \right) ~,
\label{e:ttauNH}\eqe
\eqb{l}
\tc^{\alpha\beta\gamma\delta} := 2\,\ds\pa{\ttau^{\alpha\beta}}{g_{\gamma\delta}} = \ds\frac{2\,\tc_1}{\aJ^2}\,\left( g^{\alpha\beta}\,g^{\gamma\delta} - g^{\alpha\beta\gamma\delta}\right) ~.
\label{e:tcabgdNH}\eqe
The fourth-order tensors $g^{\alpha\beta\gamma\delta}$, $a^{\alpha\beta\gamma\delta}$ and $b^{\alpha\beta\gamma\delta}$, which are used henceforth, are given in Appendix~\ref{s:var}.

\paragraph{AP shell model}
From Eqs.~(\ref{e:ltau1}.1)~and~\eqref{e:ttauNH}, we have
\eqb{l}
\htau^{\alpha\beta} = \tc_1\,\ds\left( A^{\alpha\beta} - \frac{1}{J^2} \,a^{\alpha\beta} \right) ~.
\label{e:htauNH}
\eqe
Thus, from Eqs.~(\ref{e:ltau1}.2)~and~\eqref{e:ttauNH}, one can obtain
\eqb{l}
\htau^{\alpha\beta}_{,3} = 
2\,\tc_1\,\ds\left( B^{\alpha\beta} - \frac{1}{J^2}\,\left[b^{\alpha\beta} + 2\,(H - H_0)\,a^{\alpha\beta} \right]\right)~,
\label{e:htauNH_3}\eqe
where $H_0:=\tfrac{1}{2}\,A^{\alpha\beta}\,B_{\alpha\beta}$ is the mean curvature on $\sS_0$. The linearization of kinematic variables w.r.t.~the through-the-thickness coordinate $\xi$ can be found in Appendix~\ref{s:vxi}.
The corresponding material tangents, defined in Sec.~\ref{s:AP}, are
\eqb{lllll}
\hc^{\alpha\beta\gamma\delta} \is \ds\frac{2\,\tc_1}{J^2}\left(a^{\alpha\beta}\,a^{\gamma\delta} -a^{\alpha\beta\gamma\delta}\right)~, \quad\hfill
\hd^{\alpha\beta\gamma\delta} \is 0~,\\[3mm]
\hc^{\alpha\beta\gamma\delta}_{,3} \is \ds\frac{4\,\tc_1}{J^2}\left(b^{\alpha\beta}\,a^{\gamma\delta} + a^{\alpha\beta}\,b^{\gamma\delta}\right) + 4\,(H - H_0)\,\hc^{\alpha\beta\gamma\delta}~,\quad\quad
\hd^{\alpha\beta\gamma\delta}_{,3} \is -\hc^{\alpha\beta\gamma\delta}~.
\label{e:hcabgdNH_1}\eqe

\paragraph{DD shell model}
The 2D incompressible Neo--Hookean strain energy \citep[e.g.~][]{membrane,memtheo} is
\eqb{l}
W_\mrM(I_1, J) =  \ds\frac{c_1}{2}\,\big(I_1 - 2 \big) + p\,\tg~,
\label{e:WNH}\eqe
where $c_1 = T\,\tc_1 = \mu/2$ is physically related to the 2D shear modulus $\mu $ as $\mu = T\,\tilde\mu$. The in-plane stress components now are
\eqb{l}
\tau^{\alpha\beta} = c_1\,\ds\left( A^{\alpha\beta} - \frac{1}{J^2} \,a^{\alpha\beta} \right) ~,
\label{e:tauNH}\eqe
with
\eqb{l}
c^{\alpha\beta\gamma\delta} := 2\,\ds\pa{\tau^{\alpha\beta}}{a_{\gamma\delta}} = \ds\frac{2\,c_1}{J^2}\,\left(a^{\alpha\beta}\,a^{\gamma\delta} -a^{\alpha\beta\gamma\delta} \right)~, 
\label{e:cabgdNH}\eqe
where $a^{\alpha\beta\gamma\delta} $ is given by Eq.~\eqref{e:aabgd}. Correspondingly, in the reference configuration,
\eqb{l}
c_0^{\alpha\beta\gamma\delta} := 2\,c_1\,A^{\alpha\beta}\,A^{\gamma\delta} + c_1\,\left(A^{\alpha\gamma}\,A^{\beta\delta} + A^{\alpha\delta}\,A^{\beta\gamma}\right)~.
\label{e:c0abgd_incNH}\eqe

Thus, the bending energy $W_\mrB$ can be found by plugging Eq.~\eqref{e:c0abgd_incNH} into Eq.~\eqref{e:Wb}, which gives the bending moment
\eqb{l}
M_0^{\alpha\beta} = f^{\alpha\beta\gamma\delta}\,\left(b_{\gamma\delta} - B_{\gamma\delta} \right) ~,
\label{e:MabKinc}\eqe
where
\eqb{l}
f^{\alpha\beta\gamma\delta} := \ds\frac{T^2}{12}\,\left[ 2\,c_1\, A^{\alpha\beta}\,A^{\gamma\delta} + c_1\,\left( A^{\alpha\gamma}\,A^{\beta\delta} + A^{\alpha\delta}\,A^{\beta\gamma}\right)  \right] ~.
\label{e:fabgdKinc}\eqe
\begin{rem}
As many material models, introduced in the following sections, are based on the first invariant of the right Cauchy--Green deformation tensor, their corresponding stress and elasticity tensors include expressions similar to an incompressible Neo-Hookean material. Thus, for the sake of simplicity, we introduce normalized $\ttau^{\alpha\beta}_\mathrm{NH}$ and $\tc^{\alpha\beta\gamma\delta}_\mathrm{NH}$ for the NP shell model by setting $\tc_1 = 1$ in Eqs.~\eqref{e:ttauNH} and \eqref{e:tcabgdNH}, which gives
\eqb{l}
\ttau^{\alpha\beta}_\mathrm{NH} := G^{\alpha\beta} - \ds\frac{1}{\aJ^2}\,g^{\alpha\beta}~,
\label{e:ttauNHRef}\eqe
\eqb{l}
\tc^{\alpha\beta\gamma\delta}_\mathrm{NH} := 2\,\ds\pa{\ttau^{\alpha\beta}_\mathrm{NH}}{g_{\gamma\delta}} = \ds\frac{2}{\aJ^2}\,\left(g^{\alpha\beta}\,g^{\gamma\delta} - g^{\alpha\beta\gamma\delta} \right)~.
\label{e:tcabgdNHRef}\eqe
For the DD shell model, the normalized $\tau^{\alpha\beta}_\mathrm{NH}$ and $c^{\alpha\beta\gamma\delta}_\mathrm{NH}$ are derived by setting $c_1 = 1$ in Eqs.~\eqref{e:tauNH} and \eqref{e:cabgdNH}, which yields
\eqb{l}
\tau^{\alpha\beta}_\mathrm{NH} := A^{\alpha\beta} - \ds\frac{1}{J^2}\,a^{\alpha\beta}~,
\label{e:tauNHRef}\eqe
\eqb{l}
c^{\alpha\beta\gamma\delta}_\mathrm{NH} := 2\,\ds\pa{\tau^{\alpha\beta}_\mathrm{NH}}{a_{\gamma\delta}} = \ds\frac{2}{J^2}\,\left(a^{\alpha\beta}\,a^{\gamma\delta} -a^{\alpha\beta\gamma\delta} \right)~.
\label{e:cabgdNHRef}\eqe
Accordingly, in the reference configuration,
\eqb{l}
c^{\alpha\beta\gamma\delta}_{\mathrm{NH}0} := \Big(c^{\alpha\beta\gamma\delta}_\mathrm{NH}\Big)_{\sS_0} = 2\,A^{\alpha\beta}\,A^{\gamma\delta} + A^{\alpha\gamma}\,A^{\beta\delta} + A^{\alpha\delta}\,A^{\beta\gamma}
\eqe
and trivially $\tau^{\alpha\beta}_{\mathrm{NH}0}:= \Big(\tau^{\alpha\beta}_\mathrm{NH}\Big)_{\sS_0} = 0$.

Similarly, the normalized stresses and tangent tensors for the AP shell model are defined according to Eqs.~\eqref{e:htauNH},~\eqref{e:htauNH_3}~and~\eqref{e:hcabgdNH_1} as 
\eqb{lll}
\htau^{\alpha\beta}_\mathrm{NH} \is \tau^{\alpha\beta}_\mathrm{NH}~, \\[2mm] 
\htau^{\alpha\beta}_{\mathrm{NH},3} \is 2\,\ds\left( B^{\alpha\beta} - \frac{1}{J^2}\,\left[b^{\alpha\beta} + 2\,(H - H_0)\,a^{\alpha\beta} \right]\right)~,
\label{e:htauNHRef}\eqe
\eqb{lllll}
\hc^{\alpha\beta\gamma\delta}_\mathrm{NH} \is c^{\alpha\beta\gamma\delta}_\mathrm{NH}~, \quad\quad
\hc^{\alpha\beta\gamma\delta}_{\mathrm{NH},3} \is \ds\frac{4}{J^2}\left(b^{\alpha\beta}\,a^{\gamma\delta} + a^{\alpha\beta}\,b^{\gamma\delta}\right) + 4\,(H - H_0)\,\hc^{\alpha\beta\gamma\delta}_\mathrm{NH}~,\\[3mm]
\hd^{\alpha\beta\gamma\delta}_\mathrm{NH} \is 0~,\hfill
\hd^{\alpha\beta\gamma\delta}_{\mathrm{NH},3} \is -\hc^{\alpha\beta\gamma\delta}_\mathrm{NH}~.
\label{e:hcabgdNH}\eqe
\end{rem}

\subsubsection{Incompressible Mooney--Rivlin (MR)}\label{s:incMR}
The Mooney--Rivlin (MR) model is one of the oldest and most accurate constitutive laws developed for large deformations of isotropic materials \citep{martins06,wex15}. It is based on the first and second invariants of the right Cauchy--Green deformation tensor, which requires two material constants to be specified.  

\paragraph{NP shell model}
The incompressible 3D strain-energy density function of the Mooney--Rivlin type is
\eqb{l}
\tW\big(\tI_1, \tI_2, \tJ\big) =  \ds\frac{\tc_1}{2}(\tI_1 - 3 ) + \ds\frac{\tc_2}{2}(\tI_2 - 3 )  + \tp\,\tg ~,
\label{e:tWMR}
\eqe
where $\tc_1$ and $\tc_2$ are stress-like parameters that should be found from experiments. The components of the Kirchhoff stress tensor thus are
\eqb{l}
\ttau^{\alpha\beta} = \tc_1\,\ttau^{\alpha\beta}_\mathrm{NH} 
+ \ds\frac{\tc_2}{\aJ^2}\left(G^{\alpha\beta} - \aI_1\,g^{\alpha\beta}\right) + \tc_2\,\aJ^2\,g^{\alpha\beta}~,
\label{e:ttauMR}\eqe
which gives
\eqb{l}
\tc^{\alpha\beta\gamma\delta} = \left(\tc_1 + \tc_2\,\aI_1\right)\,\tc^{\alpha\beta\gamma\delta}_\mathrm{NH}
- \ds\frac{2\,\tc_2}{\aJ^2}\,\big(G^{\alpha\beta}\,g^{\gamma\delta} + g^{\alpha\beta}\,G^{\gamma\delta}\big)
+ 2\,\tc_2\,\aJ^2\, \big(g^{\alpha\beta\gamma\delta} + g^{\alpha\beta}\,g^{\gamma\delta} \big)~.
\label{e:tcabgdMR}\eqe

\paragraph{AP shell model}
Following Eq.~\eqref{e:ttauMR},
\eqb{lll}
\htau^{\alpha\beta} \is \tc_1\,\tau^{\alpha\beta}_\mathrm{NH} + \ds\frac{\tc_2}{J^2}\,(\Aab - I_1\,\aab)  + \tc_2\,J^2\,\aab~,\\[2mm]
\htau^{\alpha\beta}_{,3} \is \tc_1\,\htau^{\alpha\beta}_{\mathrm{NH},3} + \tc_2\,\left(\htau^{\alpha\beta}_\mathrm{I} + \htau^{\alpha\beta}_\mathrm{II}\right) ~,
\label{e:htauMR}\eqe
where
\eqb{lll}
\htau^{\alpha\beta}_\mathrm{I} \is \ds\frac{1}{J^2}\left[4\,(H - H_0)\,(\Aab - I_1\,\aab) + 2\,(\Bab - I_1\,\bab) - \hI_{1,3}\,\aab\right] ~,\\[2mm]
\htau^{\alpha\beta}_\mathrm{II} \is 2\,J^2\left[\bab - 2\,(H - H_0)\,\aab\right]~.
\eqe

Thus, the tangent tensors are $\hd^{\alpha\beta\gamma\delta} = 0 $,
\eqb{l}
\hc^{\alpha\beta\gamma\delta} = (\tc_1 + \tc_2\,I_1)\,c^{\alpha\beta\gamma\delta}_\mathrm{NH} 
 - 2\,\ds\frac{\tc_2}{J^2}\,(\Aab\,\agd + \aab\,\Agd) + 2\,\tc_2\,J^2\,( \aab\,\agd + a^{\alpha\beta\gamma\delta} )~,
\label{e:hcabgd_MR}\eqe
\eqb{lll}
\hc^{\alpha\beta\gamma\delta}_{,3} \is 
\tc_1\,\hc^{\alpha\beta\gamma\delta}_{\mathrm{NH},3} + 2\,\tc_2\left(\htau^{\alpha\beta}_\mathrm{II} - \htau^{\alpha\beta}_\mathrm{I}\right)\agd + 4\,\tc_2\,J^2\left[b^{\alpha\beta\gamma\delta} + \aab\,\bgd - 2\,(H - H_0)\,a^{\alpha\beta\gamma\delta}\right] \\ [3mm]
\plus 4\,\ds\frac{\tc_2}{J^2}\left( I_1\,\aab\,\bgd - \Aab\,\bgd - \aab\,\Bgd 
-\left[2\,(H - H_0)\,\aab + \bab\right]\Agd\right) \\ [3mm]
\mi 4\,\ds\frac{\tc_2}{J^2}\,I_1\,b^{\alpha\beta\gamma\delta} - 2\,\ds\frac{\tc_2}{J^2}\left[4\,(H - H_0)\,I_1 + \hI_{1,3} \right]a^{\alpha\beta\gamma\delta}
\label{e:hcabgd_MR_xi}\eqe
and
\eqb{lll}
\hd^{\alpha\beta\gamma\delta}_{,3} \is 
\tc_1\,\hd^{\alpha\beta\gamma\delta}_{\mathrm{NH},3} -\tc_2\,I_1\,c^{\alpha\beta\gamma\delta}_\mathrm{NH} + 2\,\ds\frac{\tc_2}{J^2}\left(\Aab\,\agd + \aab\,\Agd\right) \\[3mm]
\mi 2\,\tc_2\,J^2\left(\aab\,\agd + a^{\alpha\beta\gamma\delta}\right) ~.
\label{e:hdabgd_MR}\eqe

\paragraph{DD shell model}
For this model, the incompressible 2D stored energy is 
\eqb{l}
W_\mrM\big(\tI_1, \tI_2, \tJ\big) =  \ds\frac{c_1}{2}(\tI_1 - 2) + \ds\frac{c_2}{2}(\tI_2 - 2)  + p\,\tg ~,
\label{e:WMR}
\eqe
where $c_1:=T\,\tc_1$ and $c_2:=T\,\tc_2$. Likewise to Eqs.~\eqref{e:ttauMR}~and~\eqref{e:tcabgdMR}, it can be shown that 
\eqb{l}
\tau^{\alpha\beta} = 
c_1\,\tau^{\alpha\beta}_\mathrm{NH}  
+ \ds\frac{c_2}{J^2}\left(A^{\alpha\beta} - I_1\,a^{\alpha\beta}\right) + c_2\,J^2\,a^{\alpha\beta}
\label{e:tauMR}\eqe 
and 
\eqb{l}
c^{\alpha\beta\gamma\delta} = 
\left(c_1 + c_2\,I_1\right)\,c^{\alpha\beta\gamma\delta}_\mathrm{NH}
- \ds\frac{2\,c_2}{J^2}\,\big(A^{\alpha\beta}\,a^{\gamma\delta} + a^{\alpha\beta}\,A^{\gamma\delta}\big)
+ 2\,c_2\,J^2\,\big(a^{\alpha\beta\gamma\delta} + a^{\alpha\beta}\,a^{\gamma\delta} \big)~.
\label{e:cabgdMR}\eqe
In the reference configuration, $I_1 = 2$, $J = 1$ and $a^{\alpha\beta} = A^{\alpha\beta}$, thus
\eqb{l}
c_0^{\alpha\beta\gamma\delta} = \big(c_1+c_2\big)\,c^{\alpha\beta\gamma\delta}_{\mathrm{NH}0}~.
\eqe

\subsubsection{Incompressbile Fung}\label{s:fung}
The strain energy function of this model has an exponential form in terms of the first invariant of the Cauchy--Green deformation tensor. This model has been proposed first by \citet{fung67} and was then further investigated by \citet{demiray72}\footnote{In literature, this model is mostly called as ``Fung'' model while it is also named ``Fung--Demiray'' \citep[e.g. by][]{wex15} or ``Demiray'' model \citep[e.g. by][]{gasser06}.}. Later, \citet{humphrey87} extended the formulation by including an anisotropic contribution of fibers to model passive cardiac tissue.  

\paragraph{NP shell model}
The incompressible version of the Fung model is
\eqb{l}
\tW(\tI_1,\tJ) = \ds\frac{\tc_1}{2\,c_2}\,\left\lbrace \exp \left[ c_2\,(\tI_1 - 3 )\right] - 1 \right\rbrace + \tp\,\tg~,
\label{e:fung1}\eqe
which gives
\eqb{l}
\ttau^{\alpha\beta} = \tD_1\,\ttau^{\alpha\beta}_\mathrm{NH}~,
\label{e:ttauF}\eqe
\eqb{l}
\tc^{\alpha\beta\gamma\delta} = 
\tD_1\,\left( \tc^{\alpha\beta\gamma\delta}_\mathrm{NH} + 2\,c_2\,\ttau^{\alpha\beta}_\mathrm{NH}\,\ttau^{\gamma\delta}_\mathrm{NH} \right)~, 
\label{e:tcabgdF}\eqe
with
\eqb{l}
\tD_1 := \ds\pa{\tW}{\tI_1} = \tc_1\,\exp \left[ c_2\,(\tI_1 - 3 )\right] ~.
\label{e:fungD1}\eqe

\paragraph{AP shell model}
Plugging Eq.~\eqref{e:ttauF} into Eq.~\eqref{e:ltau1}, we have
\eqb{lll}
\htau^{\alpha\beta} \is \hD_1\,\tau^{\alpha\beta}_\mathrm{NH}~,\\[2mm]
\htau^{\alpha\beta}_{,3} \is \hD_1\left(\tau^{\alpha\beta}_{\mathrm{NH},3} + c_2\left[\hI_{1,3} + \dfrac{4}{J^2}\,(H - H_0)\right]\tau^{\alpha\beta}_\mathrm{NH}\right)~,
\label{e:htauF}\eqe
where $\hI_{1,3}$ is given by Eq.~\eqref{e:I1_xi} (see Appendix~\ref{s:vxi}) and
\eqb{l}
\hD_1 := \tc_1\,\exp \left[ c_2\,\left(I_1 + \dfrac{1}{J^2} - 3 \right)\right] ~.
\label{e:funghD1}\eqe
The corresponding material tangents are $\hd^{\alpha\beta\gamma\delta} = 0$ and
\eqb{lll}
\hc^{\alpha\beta\gamma\delta} \is \hD_1\left(c^{\alpha\beta\gamma\delta}_\mathrm{NH} + 2\,c_2\,\tau^{\alpha\beta}_\mathrm{NH}\,\tau^{\gamma\delta}_\mathrm{NH}\right) ~,\\[3mm]
\hc^{\alpha\beta\gamma\delta}_{,3} \is 
\hD_1\left(\hc^{\alpha\beta\gamma\delta}_{\mathrm{NH},3} + c_2\,\left[\hI_{1,3} + \dfrac{4}{J^2}(H - H_0)\right]\,c^{\alpha\beta\gamma\delta}_\mathrm{NH}\right)
+ 2\,c_2\,\htau^{\alpha\beta}_{,3}\,\tau^{\gamma\delta}_\mathrm{NH} \\[3mm]
\plus 4\,c_2\,\hD_1\,\tau^{\alpha\beta}_\mathrm{NH} \left(B^{\gamma\delta}-\dfrac{1}{J^2}\left[2\,(H - H_0)\,a^{\gamma\delta} + b^{\gamma\delta}\right]\right) ~,\\[3mm]
\hd^{\alpha\beta\gamma\delta}_{,3} \is \hD_1\,\hd^{\alpha\beta\gamma\delta}_{\mathrm{NH},3} - 2\,c_2\,\hD_1\,\tau^{\alpha\beta}_\mathrm{NH}\,\tau^{\gamma\delta}_\mathrm{NH}~.
\label{e:hcabgdF}\eqe

\paragraph{DD shell model}
The corresponding membrane strain energy function is 
\eqb{l}
W_\mrM(\tI_1,\tJ) = \ds\frac{c_1}{2\,c_2}\,\left\lbrace \exp \left[ c_2\,(\tI_1 - 3 )\right] - 1 \right\rbrace + p\,\tg~,
\label{e:fungM1}\eqe
where $c_1 = T\,\tc_1$. Similarly, we have
\eqb{l}
\tau^{\alpha\beta} = D_1\,\tau^{\alpha\beta}_\mathrm{NH}~,
\label{e:tauF}\eqe
\eqb{l}
c^{\alpha\beta\gamma\delta} = 
D_1\,\left( c^{\alpha\beta\gamma\delta}_\mathrm{NH} + 2\,c_2\,\tau^{\alpha\beta}_\mathrm{NH}\,\tau^{\gamma\delta}_\mathrm{NH} \right) 
\eqe
and
\eqb{l}
D_1 := \ds\pa{W_\mrM}{\tI_1} = c_1\,\exp \left[ c_2\,(\tI_1 - 3 )\right] ~.
\label{e:fungMD1}\eqe

In the reference configuration, $D_1 = c_1$ and $\tau^{\alpha\beta}_{\mathrm{NH}0} = 0$, which results in 
\eqb{l}
c_0^{\alpha\beta\gamma\delta} := c_1\,c^{\alpha\beta\gamma\delta}_{\mathrm{NH}0}~.
\eqe

\subsection{Anisotropic models}\label{s:aniso}
The fibrous structure of soft tissues adds anisotropic features to their mechanical behavior. In order to capture those, different anisotropic hyperelastic models are introduced here. The various isotropic material models, introduced in Sec.~\ref{s:iso}, depend on a combination of the first three invariants of the Cauchy--Green deformation tensor, i.e.~$\tI_1$, $\tI_2$ and $\tI_3 := \tJ$. Similarly, the anisotropic constitutive laws introduced in this section depend on extra invariants, which are related to the principal direction of the fibers. The anisotropy can also be measured by components of the Green--Lagrange tensor \citep{chagnon15}, which is not discussed here. Considering that the principal direction of the $i^\mathrm{th}$ family of fibers is $\tbL_i$ in the reference configuration, the structural tensor $\tbM_i$ can be expressed according to the kinematics of Kirchhoff--Love shells as 
\eqb{l}
\tbM_i := \tbL_i \otimes \tbL_i = \abM_i + \aL^{33}_i\,\bN\otimes\bN ~.
\label{e:tM_i}\eqe
The in-plane component of structural tensor is  
\eqb{l}
\abM_i := \abL_i \otimes \abL_i = \aL_i^{\alpha\beta}\,\bG_\alpha \otimes \bG_\beta ~,
\label{e:aM_i2}\eqe
where
\eqb{lll}
\abL_i := \aL_i^\alpha\,\bG_\alpha ~, \quad
\aL^{\alpha\beta}_i := \aL_i^\alpha \, \aL_i^\beta ~, \quad
\aL_i^\alpha := \tbL_i\cdot\bG‍‍‍‍‍‍‍‍^\alpha~.
\eqe
The out-of-plane component of structural tensor is then
\eqb{l}
\aL^{33}_i := \big( \tbL_i\cdot\bN \big)^2~.
\eqe

The first invariant of the structural tensor, which is used for most anisotropic models, is\footnote{Some scholars \citep[e.g.][]{gasser06} use $\tI_4$ and $\tI_6$ for $\tI_4^{1}$ and $\tI_4^{2}$ if two family of fibers are considered. Here, we use $\tI_4$ if only one family of fibers is included and $\tI_4^{i}$ for more families of fibers.} 
\eqb{l}
\tI_4^{i} := \tr\big(\tbC\,\tbM_i\big) = \tbL_i\cdot\tbC\tbL_i = \aI_4^i + g_{33}\,\aL^{33}_i ~; 
\label{e:tI4tI5}\eqe
however, other invariants can also be used \citep{chagnon15}. Likewise, the in-plane invariant is defined as
\eqb{l}
\aI_4^{i} := \tr\big(\abC\,\abM_i\big) = \abL_i\cdot\abC\abL_i = g_{\alpha\beta}\,\aL_i^{\alpha\beta} ~.
\label{e:aI4aI5}\eqe

In the same fashion, for the membrane formulation, the corresponding quantities are defined on the shell mid-surface as
\eqb{l}
\tbM_i = \bM_i + L^{33}_i\,\bN\otimes\bN ~,
\label{e:tM_i2}\eqe
where
\eqb{l}
\bM_i := \bL_i \otimes \bL_i = {L_i}^{\alpha\beta}\,\bA_\alpha \otimes \bA_\beta ~,
\label{e:M_i}\eqe
\eqb{l}
\bL_i := L_i^\alpha\,\bA_\alpha ~,\quad
L^{\alpha\beta}_i := L_i^\alpha \, L_i^\beta 
\eqe
and $L_i^\alpha = \bL_i\cdot\bA^\alpha$. Thus, the invariants are reformulated as
\eqb{l}
\tI_4^{i} = I_4^i + \tmod{\lam^2}\,L^{33}_i~, 
\label{e:tI4tI5_2}\eqe
with
\eqb{l}
I_4^{i} := \tr\big(\bC\,\bM_i\big) = \bL_i\cdot\bC\bL_i = a_{\alpha\beta}\,L_i^{\alpha\beta}~. 
\label{e:I4I5}\eqe

\begin{rem}\label{r:fibers}
For thin membrane and shells, it is more realistic to assume that fibers are distributed layer-wise, i.e.~$\aL^{33}_i=L^{33}_i=0$. This implies that $\tI_4^i  = \aI_4^i $ on each shell layer through the thickness and $\tI_4^i  = I_4^i $ on the mid-surface. Here, for the examples shown in Sec.~\ref{s:examples}, it is assumed that $\aL^{33}_i=L^{33}_i=0$. 
\end{rem}

Anisotropic hyperelastic material models are mostly developed based on the assumption that the material is constructed from an ``isotropic'' matrix reinforced with several fibers with a given principal orientation, which induce ``anisotropy''. Hence, the strain energy function $\tW$ is composed of an isotropic part $\tW_\mrm$ and an anisotropic part $\tW_\mrf$ as
\eqb{l}
\tW = \tW_\mrm\big(\tI_1,\tI_2,\tJ\big) + \ds\sum_{i=1}^{n_\mrf} \tW_\mrf\big(\tI_1,\tI_2,\cdots,\tI_4^{i},\tI_5^{i},\cdots\big)~,
\eqe   
where $n_\mrf$ is the number of fiber families. The anisotropic part may only include the invariants of the structural tensors, like in anisotropic Mooney--Rivlin model (Sec.~\ref{s:amr}), or it may combine them with the invariants of the right Cauchy--Green deformation tensor, like in Gasser--Ogden--Holzapfel model (Sec.~\ref{s:HGO_3D}). For a detailed survey of anisotropic models for biological tissues, see \citet{chagnon15}.  

\subsubsection{Anisotropic Mooney--Rivlin (AMR)}\label{s:amr}
The anisotropic Mooney--Rivlin (AMR) material model can be obtained by generalizing the formulation of \citet{rivlin51} in terms of the invariants of the structural tensor as \citep{kaliske00}
\eqb{l}
\tW_\mrf\big(\tI_4^{i}\big) = \ds\sum_{j\geq2} \tc_j\big(\tI_4^{i}-1\big)^j~, \quad\quad (i = 1,~...,n_\mrf)~.
\label{e:AMR0}\eqe
In this study, for the isotropic part, an incompressible Mooney--Rivlin constitution is considered (see Sec.~\ref{s:incMR}). For the anisotropic part, $n_\mrf$ families of fibers with a quadratic potential are included as
\eqb{lll}
\tW \is \tW_\mrm\big(\tI_1,\tI_2,\tJ\big) + \tW_\mrf\big(\tI_4^{1},\tI_4^{2},...,\tI_4^{n_\mrf}\big) \\[3mm] 
\is \ds\frac{\tc_1}{2}(\tI_1 - 3 ) + \ds\frac{\tc_2}{2}(\tI_2 - 3 ) +  \ds\sum_{i=1}^{n_\mrf}\tc_{3i}\big(\tI_4^{i}-1\big)^2 + \tp\,\tg~.
\label{e:AMR1}\eqe

\paragraph{NP shell model}
The total Kirchhoff stress is
\eqb{l}
\ttau^{\alpha\beta} = \ttau^{\alpha\beta}_\mrm + \ttau^{\alpha\beta}_\mrf ~,
\label{e:ttauAMR}\eqe 
where 
\eqb{l}
\ttau^{\alpha\beta}_\mrf := 2\,\ds\pa{\tW_\mrf}{g_{\alpha\beta}} = 2\,\ds\sum_{i=1}^{n_\mrf}\tc_{3i}\big(\tI_4^{i}-1\big)\,\aL^{\alpha\beta}_i
\label{e:ttauAMR_f}\eqe
according to Eq.~\eqref{e:AMR1} and $\ttau^{\alpha\beta}_\mrm$ is given by Eq.~\eqref{e:ttauMR}. The total layer-wise elasticity tensor is then $\tc^{\alpha\beta\gamma\delta} = \tc^{\alpha\beta\gamma\delta}_\mrm + \tc^{\alpha\beta\gamma\delta}_\mrf$, where $\tc^{\alpha\beta\gamma\delta}_\mrm$ is given by Eq.~\eqref{e:tcabgdMR} and
\eqb{l}
\tc^{\alpha\beta\gamma\delta}_\mrf = 4\,\ds\sum_{i=1}^{n_\mrf}\tc_{3i}\,\aL^{\alpha\beta}_i\,\aL^{\gamma\delta}_i ~.
\eqe

\paragraph{AP shell model}
Likewise to Eq.~\eqref{e:ttauAMR}, the total stress is split  into the isotropic and anisotropic contributions as
\eqb{l}
\htau^{\alpha\beta} = \htau^{\alpha\beta}_\mrm + \htau^{\alpha\beta}_\mrf ~,
\label{e:htauAMR}\eqe 
where $\htau^{\alpha\beta}_\mrm$ is given by Eq.~(\ref{e:htauMR}.1) and 
\eqb{l}
\htau^{\alpha\beta}_\mrf = 2\,\ds\sum_{i=1}^{n_\mrf}\tc_{3i}\big(I_4^{i}-1\big)\,L^{\alpha\beta}_i~.
\label{e:htauAMR_f}\eqe

Similarly, the first-order approximated terms are
\eqb{l}
\htau^{\alpha\beta}_{,3} = \htau^{\alpha\beta}_{\mrm,3} + \htau^{\alpha\beta}_{\mrf,3} ~,
\label{e:htauAMR_xi}\eqe 
where $\htau^{\alpha\beta}_{\mrm,3}$ follows from Eq.~(\ref{e:htauMR}.2) and
\eqb{l}
\htau^{\alpha\beta}_{\mrf,3} = 2\,\ds\sum_{i=1}^{n_\mrf}\tc_{3i}\left[ \hI_{4,3}^{i}\,L^{\alpha\beta}_i + \big(I_4^{i}-1\big)\,\hL^{\alpha\beta}_{i,3} \right]~.
\label{e:htauAMR_f_xi}\eqe 
Here, $\hL^{\alpha\beta}_{i,3}$ and $\hI^{i}_{4,3}$ are given by Eqs.~\eqref{e:Lab_xi}~and~\eqref{e:hI4_xi}, respectively. In the same fashion, the fibers and the matrix contribute to the corresponding tangents, e.g.~$\hc^{\alpha\beta\gamma\delta} = \hc^{\alpha\beta\gamma\delta}_\mrm + \hc^{\alpha\beta\gamma\delta}_\mrf$. The isotropic tensors $\hc^{\alpha\beta\gamma\delta}_\mrm$, $\hc^{\alpha\beta\gamma\delta}_{\mrm,3}$ and $\hd^{\alpha\beta\gamma\delta}_{\mrm,3}$ are given by Eqs.~(\ref{e:hcabgd_MR}-\ref{e:hdabgd_MR}) and $\hd^{\alpha\beta\gamma\delta}_\mrm = 0$. The anisotropic tangents are then
\eqb{lllll}
\hc^{\alpha\beta\gamma\delta}_\mrf \is 4\,\ds\sum_{i=1}^{n_\mrf}\tc_{3i}\,L^{\alpha\beta}_i\,L^{\gamma\delta}_i~,\hfill \hd^{\alpha\beta\gamma\delta}_\mrf \is 0 ~, \\[5mm]
\hc^{\alpha\beta\gamma\delta}_{\mrf,3} \is 4\,\ds\sum_{i=1}^{n_\mrf}\tc_{3i}\left( L^{\alpha\beta}_i\,\hL^{\gamma\delta}_{i,3} + \hL^{\alpha\beta}_{i,3}\,L^{\gamma\delta}_i \right) ~,\quad\quad \hd^{\alpha\beta\gamma\delta}_{\mrf,3} \is -4\,\ds\sum_{i=1}^{n_\mrf}\tc_{3i}\, L^{\alpha\beta}_i\,L^{\gamma\delta}_i~.
\eqe

\paragraph{DD shell model}
The strain energy of the corresponding membrane formulation is
\eqb{lll}
W_\mrM \is W_\mrm\big(\tI_1,\tI_2,\tJ\big) + W_\mrf\big(\tI_4^{1},\tI_4^{2},...,\tI_4^{n_\mrf}\big) \\[3mm] 
\is \ds\frac{c_1}{2}(\tI_1 - 3 ) + \ds\frac{c_2}{2}(\tI_2 - 3 ) + \ds\sum_{i=1}^{n_\mrf}\,c_{3i}\big(\tI_4^{i}-1\big)^2 + p\,\tg ~.
\label{e:AMRN1}\eqe

Accordingly, the total stress is
\eqb{l}
\tau^{\alpha\beta} = \tau^{\alpha\beta}_\mrm + \tau^{\alpha\beta}_\mrf ~,
\eqe 
where $\tau^{\alpha\beta}_\mrm$ is given by Eq.~\eqref{e:tauMR} and
\eqb{l}
\tau^{\alpha\beta}_\mrf = 2\,\ds\sum_{i=1}^{n_\mrf}\, c_{3i}\big(\tI_4^{i}-1\big)\,L^{\alpha\beta}_i~.
\eqe
The corresponding tangent tensor is $c^{\alpha\beta\gamma\delta} = c^{\alpha\beta\gamma\delta}_\mrm + c^{\alpha\beta\gamma\delta}_\mrf$. The isotropic part $c^{\alpha\beta\gamma\delta}_\mrm$ is defined by Eq.~\eqref{e:cabgdMR} and the anisotropic part is   
\eqb{l}
c^{\alpha\beta\gamma\delta}_\mrf = 4\,\ds\sum_{i=1}^{n_\mrf}\,c_{3i}\,L^{\alpha\beta}_i\,L^{\gamma\delta}_i~,
\eqe
which gives
\eqb{l}
c_0^{\alpha\beta\gamma\delta} = \big(c_1+c_2\big)\,c^{\alpha\beta\gamma\delta}_{\mathrm{NH}0} + 4\,\ds\sum_{i=1}^{n_\mrf}\,c_{3i}\,L^{\alpha\beta}_i\,L^{\gamma\delta}_i ~.
\eqe

\subsubsection{Gasser--Ogden--Holzapfel (GOH)}\label{s:HGO_3D}
The Gasser--Ogden--Holzapfel (GOH) material model is an anisotropic hyperelastic material model, which is used  to model soft tissues with distributed collagen fibers, and is mainly developed for the modeling of cardiovascular arteries \citep{gasser06}. This model is constructed of an isotropic part, which represents the elastin matrix of soft tissue and is modeled by the incompressible Neo-Hookean material model, and an anisotropic part due the collagen network, which is based on the structural tensors of two families of fibers. Here, a 3D generalized structural tensor (3D~GST) is considered; however, for thin structures, 2D generalized structural tensors (2D~GST) can also be used \citep{tonge13}.

\paragraph{NP shell model}
The strain-energy density function is considered as
\eqb{l}
\tW(\tI_1, \tJ_4^{1}, \tJ_4^{2}, \tJ) =  \tW_{\mathrm{m}}(\tI_1) +  \tW_{\mathrm{f}}(\tJ_4^{1}, \tJ_4^{2})  + \tp\,\tg~,
\label{e:tWHGO1} 
\eqe
with
\eqb{rll}
 \tW_{\mathrm{m}} \is \ds\frac{\tilde{\mu}}{2}\big( \tI_1 - 3) ~, \\[3mm]
 \tW_{\mathrm{f}} \is \ds\sum_{i=1}^2 \frac{\tilde{k}_{1i}}{2\,k_{2i}}\,\big\lbrace \exp \big[ k_{2i}\,(\tJ_4^{i} -1)^2\big] -1 \big\rbrace ~,
\label{e:tWHGO2} 
\eqe
where
\eqb{l}
\tJ_4^{i} = \tilde{\bC}:\tbH_i = \kappa_i\big(\aI_1 + \tmod{g_{33}}\big) + (1-3\,\kappa_i)\,\tI_4^{i} ~,\quad \quad (i=1,2) ~,
\label{e:tJ4}\eqe
are the invariants of  the 3D generalized structural tensor $\tbH_i$. $\tbH_i$ is introduced by \citet{gasser06} to extend the structural tensor $\tbM_i$, cf.~Eq.~\eqref{e:tM_i}, by accounting for dispersion in fibers as
\eqb{l}
\tbH_i := \kappa_i\,\bone + (1 - 3\,\kappa_i)\,\tbM_i = \kappa_i\,\bone + (1 - 3\,\kappa_i)\,\tbL_i\otimes\tbL_i~, \quad \quad (i=1,2) ~,
\eqe
where $\kappa_i\in[0,~1/3]$ is the parameter determining the degree of dispersion \citep{gasser06} and $\bone$ is the usual identity tensor in $\bbR^3$. Assuming $\aL^{33}_i = 0$, the layer-wise 3D generalized structural tensor becomes
\eqb{l}
\tbH_i = \kappa_i\,\bone + (1 - 3\,\kappa_i)\,\abL_i\otimes\abL_i~, \quad \quad (i=1,2) ~.
\label{e:tHi}\eqe

Similar to the other anisotropic models, the stress tensor has two components as
\eqb{l}
\ttau^{\alpha\beta} =\ttau^{\alpha\beta}_\mrm+ \ttau^{\alpha\beta}_\mrf~,
\label{e:ttauHGO2}\eqe
where
\eqb{l}
\ttau^{\alpha\beta}_\mrm = \tilde\mu\,\ttau^{\alpha\beta}_\mathrm{NH}
\label{e:ttauHGOm}
\eqe
is the isotropic contribution. The anisotropic contribution of the fibers is then 
\eqb{lll}
\ttau^{\alpha\beta}_\mathrm{f} \is  2 \ds\sum_{i=1}^2 \tilde{E}_i\, \tilde{R}_i^{\alpha\beta}~,
\label{e:ttauHGOf}\eqe
where we have defined
\eqb{lll}
\tilde{R}_i^{\alpha\beta}:= \ds\pa{\tJ_4^i}{g_{\gamma\delta}} = \kappa_i\,\tilde\tau_\mathrm{NH}^{\alpha\beta} + (1-3\,\kappa_i)\,\aL_i^{\alpha\beta} ~,\quad \quad (i=1,2)
\label{e:tRHGO}\eqe
and
\eqb{l}
\tE_i := \ds\pa{\tW_\mrf}{\tJ_4^{i}} = \tk_{1i}\,\big(\tJ_4^{i} -1\big)\,\exp \big[ k_{2i}\,(\tJ_4^{i} -1)^2\big] ~, \quad \quad (i=1,2) ~.
\label{e:tE_i}\eqe

From Eqs.~\eqref{e:ttauHGOm} and \eqref{e:ttauHGOf}, the symmetric elasticity tensor is
\eqb{lll}
\tc^{\alpha\beta\gamma\delta} := 2\ds\pa{\tilde\tau^{\alpha\beta}}{g_{\gamma\delta}} = \Big(\tilde\mu + 2\,\ds\sum_{i=1}^2 \kappa_i\,\tilde{E}_i \Big)\,\tc^{\alpha\beta\gamma\delta}_\mathrm{NH} + 4\sum_{i=1}^2\,\tD_i\,\tilde{R}_i^{\alpha\beta}\,\tilde{R}_i^{\gamma\delta}~,
\label{e:tcabgdHGO}\eqe
with
\eqb{l}
\tD_i := \ds\pa{\tilde{E}_i}{\tJ_4^{i}} = \tilde{k}_{1i}\,\left[1 + 2\,k_{2i}\,\big(\tJ_4^{i} -1\big)^2 \right] \, \exp \left[ k_{2i}\,\big(\tJ_4^{i} -1\big)^2 \right] ~,\quad \quad (i=1,2) ~.
\label{e:tDHGO}\eqe

\paragraph{AP shell model}\label{s:HGO_AP}
On the shell mid-surface $\sS$, $\htau^{\alpha\beta} =\htau^{\alpha\beta}_\mrm+ \htau^{\alpha\beta}_\mrf$, where
\eqb{l}
\htau^{\alpha\beta}_\mrm = \tmu\,\tau^{\alpha\beta}_\mathrm{NH}~, \quad\quad
\htau^{\alpha\beta}_\mathrm{f} =  2 \ds\sum_{i=1}^2 \hE_i\, R_i^{\alpha\beta}~.
\label{e:htauHGO}\eqe
Here, we have  defined
\eqb{lll}
R_i^{\alpha\beta} := \kappa_i\,\tau_\mathrm{NH}^{\alpha\beta} + (1-3\,\kappa_i)\,L_i^{\alpha\beta}~,
\label{e:RHGO}\eqe
\eqb{l}
\hE_i = \tk_{1i}\,\big(\hJ_4^{i} -1\big)\,\exp \big[ k_{2i}\,(\hJ_4^{i} -1)^2\big]
\label{e:hE_i}\eqe
and 
\eqb{l}
\hJ_4^{i} = \kappa_i\,(I_1 + J^{-2}) + (1-3\,\kappa_i)\,I_4^{i}~, \quad \quad (i=1,2) ~,
\eqe
which depend only on the mid-surface parameters. The first-order terms are then
\eqb{l}
\htau^{\alpha\beta}_{\mathrm{m},3} =  \tmu\,\htau^{\alpha\beta}_{\mathrm{NH},3}~,\quad\quad
\htau^{\alpha\beta}_{\mathrm{f},3} =  2 \ds\sum_{i=1}^2 \left( \hE_i\, R_{i,3}^{\alpha\beta} + \hD_i\,\hJ_{4,3}^{i}\, R_i^{\alpha\beta} \right)~,
\label{e:htauHGO_xi}\eqe
where we have defined 
\eqb{l}
\hJ_{4,3}^{i} = \kappa_i\,(\hI_{1,3} + 4\,H\,J^{-2}) + (1-3\,\kappa_i)\,\hI_{4,3}^{i} ~,
\label{e:hJ4_xi}\eqe
\eqb{lll}
\hR_{i,3}^{\alpha\beta}:= \kappa_i\,\htau_\mathrm{NH,3}^{\alpha\beta} + (1-3\,\kappa_i)\,\hL_{i,3}^{\alpha\beta}
\label{e:hRHGO_xi}\eqe
and
\eqb{l}
\hD_i := \ds\pa{\hE_i}{\hJ_4^{i}} = \tilde{k}_{1i}\,\left[1 + 2\,k_{2i}\,\big(\hJ_4^{i} -1\big)^2 \right] \, \exp \left[ k_{2i}\,\big(\hJ_4^{i} -1\big)^2 \right] ~,\quad \quad (i=1,2)~, 
\label{e:hDHGO}\eqe
on $\sS$. Further, $\hI_{1,3}$, $\hL^{\alpha\beta}_{i,3}$ and $\hI^{i}_{4,3}$ are given by Eqs.~\eqref{e:I1_xi},~\eqref{e:Lab_xi}~and~\eqref{e:hI4_xi}, respectively (See Appendices~\ref{s:vxi}~and~\ref{s:switch_lin}). 

The components of the anisotropic tangent tensors are $\hd^{\alpha\beta\gamma\delta}_\mrf = 0 $ and
\eqb{lll}
\hc^{\alpha\beta\gamma\delta}_\mrf \is 2\ds\sum_{i=1}^2 \,\kappa_i\,\hE_i\,c^{\alpha\beta\gamma\delta}_\mathrm{NH} + 4\sum_{i=1}^2\,\hD_i\,R_i^{\alpha\beta}\,R_i^{\gamma\delta}~, \\[3mm]
\hc^{\alpha\beta\gamma\delta}_{\mrf,3} \is 4 \ds\sum_{i=1}^2\left(  
\hD_i\, \hR_{i,3}^{\alpha\beta}\,R_i^{\gamma\delta} + 
\hF_i\,\hJ_{4,3}^{i}\, R_i^{\alpha\beta}\,R_i^{\gamma\delta} + 
\hD_i\, R_i^{\alpha\beta}\,J^{\gamma\delta}_{ai}\right) \\
\plus 2 \ds\sum_{i=1}^2\,\kappa_i\left(
\hE_i\,\hc^{\alpha\beta\gamma\delta}_{\mathrm{NH},3} + 
\hD_i\,\hJ_{4,3}^{i}\,c^{\alpha\beta\gamma\delta}_\mathrm{NH}\right)~,\\[3mm]
\hd^{\alpha\beta\gamma\delta}_{\mrf,3} \is 2 \ds\sum_{i=1}^2\left(
\hD_i\, R_i^{\alpha\beta}\,J^{\gamma\delta}_{bi} +
\kappa_i\,\hE_i\,\hd^{\alpha\beta\gamma\delta}_{\mathrm{NH},3}\right)~.
\label{e:hcabgdHGO_f}\eqe
Here, we have defined
\eqb{l}
\hF_i := \ds\pa{\hD_i}{\hJ_4^{i}} =  2\,\tilde{k}_{1i}\,k_{2i}\,\big(\hJ_4^{i} -1\big)\,\left[3 + 2\,k_{2i}\,\big(\hJ_4^{i} -1\big)^2 \right] \, \exp \left[ k_{2i}\,\big(\hJ_4^{i} -1\big)^2 \right] 
\label{e:hFHGO}\eqe
and
\eqb{lll}
J^{\gamma\delta}_{ai} \dis \ds\pa{\hJ_{4,3}^{i}}{\augd} = 2\,\kappa_i\left[\Bab - \ds\frac{1}{J^2}\left(\bab + 2\,H\,\aab\right) \right] + \left(1-3\,\kappa_i\right)\,\hat{L}^{\alpha\beta}_{i,3} ~,\\[4mm]
J^{\gamma\delta}_{bi} \dis \ds\pa{\hJ_{4,3}^{i}}{\bugd} = -2\,\kappa_i\,\htau^{\alpha\beta}_\mathrm{NH} - 2\,(1-3\,\kappa_i)\,L^{\alpha\beta}_i ~.
\eqe

\paragraph{DD shell model}
For the membrane part, the projected membrane formulation of \citet{biomembrane} is adopted. In this setup, the generalized structural tensor $\tbH_i $, is defined on the shell mid-surface as
\eqb{l}
\tbH_i := \kappa_i\,\bone+ \big(1 - 3\,\kappa_i\big)\,\bL_i\otimes\bL_i ~, \quad \quad (i=1,2) ~,
\label{e:Hi}\eqe
where $\bL_i = L_i^\alpha\,\bA_\alpha$ ans $ L_i^\alpha = \bL_i\cdot\bA_\alpha$. Thus, the first invariant of the generalized structural tensor is 
\eqb{l}
\tJ_4^{i} = \tilde{\bC}:\tbH_i = \kappa_i\,(I_1 + \lam^2) + (1-3\,\kappa_i)\,I_4^{i} ~,\quad \quad (i=1,2) ~.
\eqe

Similar to the other shell models, the in-plane Kirchhoff stress is split as $\tau^{\alpha\beta} =\tau^{\alpha\beta}_\mathrm{m}+ \tau^{\alpha\beta}_\mathrm{f}$, where $\tau^{\alpha\beta}_\mrm = \mu\,\tau^{\alpha\beta}_\mathrm{NH}$ and
\eqb{lll}
\tau^{\alpha\beta}_\mathrm{f} \is  2 \ds\sum_{i=1}^2 E_i\, R_i^{\alpha\beta}~.
\label{e:tauHGOf}\eqe
Here, $R_i^{\alpha\beta}$ is given by Eq.~\eqref{e:RHGO} and
\eqb{l}
E_i := \ds\pa{W_\mrM}{\tJ_4^{i}} = k_{1i}\,\big(\tJ_4^{i} -1\big)\,\exp \big[ k_{2i}\,(\tJ_4^{i} -1)^2\big] ~, \quad \quad (i=1,2) ~.
\label{e:E_i}\eqe

The total elasticity tensor is then
\eqb{lll}
c^{\alpha\beta\gamma\delta} := 2\ds\pa{\tau^{\alpha\beta}}{a_{\gamma\delta}} = \Big(\mu + 2\,\ds\sum_{i=1}^2 \kappa_i\,E_i \Big)\,c^{\alpha\beta\gamma\delta}_\mathrm{NH} + 4\ds\sum_{i=1}^2D_i\,R_i^{\alpha\beta}\,R_i^{\gamma\delta}~,
\label{e:cabgdHGO}\eqe
where we have defined
\eqb{lll}
D_i:= k_{1i}\,\left[1 + 2\,k_{2i}\,(\tJ_4^{i}-1)^2\right]\,\exp\big[k_{2i}\,(\tJ_4^{i}-1)^2\big]~.
\label{e:DHGO}\eqe 

In the reference configuration, $I_4^{i} = 1 $, which gives $D_i = k_{1i}$, and $R^{\alpha\beta} = (1-3\,\kappa_i)\,L_i^{\alpha\beta}$. Further, $\tJ_4^{i} = 1$, which results in $E_i = 0 $. Hence, considering Eqs.~\eqref{e:MabKinc} and \eqref{e:fabgdKinc}, we get  
\eqb{l}
c_0^{\alpha\beta\gamma\delta} := \mu\,c^{\alpha\beta\gamma\delta}_{\mathrm{NH}0} + 4\,\ds\sum_{i=1}^2\,k_{1i}\,(1-3\,\kappa_i)^2\,L_i^{\alpha\beta}\,L_i^{\gamma\delta} ~.
\label{e:c0HGO}\eqe

\subsubsection{GOH model with compression/tension switch}\label{s:switch}
The anisotropic strain energy density function of Eq.~(\ref{e:tWHGO2}.2) is polyconvex if $\tI_4^{i} > 1$ \citep{balzani06,prot07}. This issue is also addressed by \citet{gasser06} who argue that their model will predict non-physical behavior if $\tI_4^{i} < 1$. This is due to the fact that the fibers bear no compressive force; therefore, they are active only if being extended. Thus, such models can be equipped with a compression/tension switch to exclude the compressed fibers. For instance, the GOH model should be modified as \citep[cf.][]{melnik15}
\eqb{l}
\tW_{\mathrm{f}} = \ds\sum_{i=1}^2 \aH_i\,\frac{\tilde{k}_{1i}}{2\,k_{2i}}\,\big\lbrace \exp \big[ k_{2i}\,(\tmod{\tJ}_4^{i} -1)^2\big] -1 \big\rbrace ~,
\label{e:tWHGO_SW} 
\eqe
where on the shell layer $\asS$, the compression/tension switch $\aH_i$ is formulated by the Heaviside step function
\eqb{l} 
H(x) :=
\left\{
\begin{array}{ll}
1~, \quad  & x > 0~, \\
0~, \hfill & x \leq 0
\end{array} \right.
\label{e:tHGO_SW}\eqe
as
\eqb{l} 
\aH_i = \aH_i(\guab) = H\big(\aI_4^{i} - 1\big)~.
\eqe

In the same fashion, for the membrane constitution, we have
\eqb{l}
W_{\mathrm{f}} = \ds\sum_{i=1}^2 H_i\,\frac{k_{1i}}{2\,k_{2i}}\,\big\lbrace \exp \big[ k_{2i}\,(\tmod{\tJ}_4^{i} -1)^2\big] -1 \big\rbrace ~,
\label{e:WHGO_SW}\eqe
in which, the compression/tension switch $H_i$ is defined on the shell mid-surface as 
\eqb{l} 
H_i = H_i(\auab) = H\big(I_4^{i} - 1\big)~.
\eqe

\begin{rem}
As discussed by \citet{holzapfel15}, the compression/tension switch has also been implemented in terms of $\tI_4^i$ instead of $\aI_4^i$ or $I_4^i$. But this may give erroneous results.
\end{rem}

In the following, the application of the compression/tension switch to the GOH material model is discussed for the three introduced shell models. 

\paragraph{NP shell model}
In this approach, one can directly plug the switch definition \eqref{e:tHGO_SW} into the strain energy density function \eqref{e:tWHGO_SW}. Similarly, the in-plane stress and elasticity tensors should be augmented with the compression/tension switch as
\eqb{lll}
\ttau^{\alpha\beta}_\mathrm{f} \is  2 \ds\sum_{i=1}^2 \aH_i\,\tilde{E}_i\, \tilde{R}_i^{\alpha\beta}
\label{e:ttauHGOf_SW}\eqe
and
\eqb{lll}
\tc^{\alpha\beta\gamma\delta} := 2\ds\pa{\tilde\tau^{\alpha\beta}}{g_{\gamma\delta}} = \Big(\tilde\mu + 2\,\ds\sum_{i=1}^2 \,\kappa_i\,\aH_i\,\tilde{E}_i \Big)\,\tc^{\alpha\beta\gamma\delta}_\mathrm{NH} + 4\,\sum_{i=1}^2\,\aH_i\,\tD_i\,\tilde{R}_i^{\alpha\beta}\,\tilde{R}_i^{\gamma\delta}~.
\label{e:tcabgdHGO_SW}\eqe

However, the numerical integration should be performed more carefully to assure that there are enough number of Gaussian quadrature points in the locations the switch is active.

\paragraph{AP shell model}
If the compression/tension switch is considered, the anisotropic stresses might be non-zero only within a portion of the shell thickness; however, the isotropic stresses are non-zero across the whole thickness. Thus, for the isotropic part, one can use the formulation introduced in Sec.~\ref{s:fullsh} with the corresponding stress and tangent tensors given in Sec.~\ref{s:HGO_AP}. For the anisotropic part, a partially-stressed shell formulation (see Sec.~\ref{s:parsh}) is required. Hence, the stress and moment tensors due to the fibers are
\eqb{rcl}
\tau^{\alpha\beta}_\mrf \is \ds\sum_{i=1}^2\,\left\lbrace \Big(T_2^{i} -T_1^{i}\Big)\,\hat{\tau}^{\alpha\beta}_{\mrf i} + \ds\frac{1}{2}\Big[\left(T_2^{i}\right)^2 -\left(T_1^{i}\right)^2\Big]\,\hat{\tau}^{\alpha\beta}_{\mrf i,3}\right\rbrace ~, \\[5mm]
M_{\tmod{\mrf}0}^{\alpha\beta} \is \ds\sum_{i=1}^2\,\left\lbrace \ds\frac{1}{2}\Big[\left(T_2^{i}\right)^2 -\left(T_1^{i}\right)^2\Big]\,\hat{\tau}^{\alpha\beta}_{\mrf i} + \ds\frac{1}{3}\Big[\left(T_2^{i}\right)^3 -\left(T_1^{i}\right)^3\Big]\,\hat{\tau}^{\alpha\beta}_{\mrf i,3}\right\rbrace~,
\eqe
where we have defined 
\eqb{l}
\hat{\tau}^{\alpha\beta}_{\mrf i} := 2\,\hE_i\, R_i^{\alpha\beta}~, \quad\quad
\hat{\tau}^{\alpha\beta}_{\mrf i,3} := 2\,\left( \hE_i\, R_{i,3}^{\alpha\beta} + \hD_i\,\hJ_{4,3}^{i}\, R_i^{\alpha\beta} \right)~, \quad\quad i=1,2~,
\eqe
according to Eqs.~(\ref{e:htauHGO}.2)~and~(\ref{e:htauHGO_xi}.2).  As the anisotropic part is partially-stressed, one needs to find the thickness interval $[T^{i}_1,T^{i}_2]\in[-T/2,T/2]$, where $\tI^{i}_4 > 1$. The algorithm to find $T^{i}_1$ and $T^{i}_2$ is given in Appendix~\ref{s:switch_lin}. Likewise to the stress and moment tensors, the material tangents are also derived following the formulation of Sec.~\ref{s:parsh}. For instance,
\eqb{lll}
c^{\alpha\beta\gamma\delta}_\mrf \is 
\ds\sum_{i=1}^2\,\left\lbrace
\Big(T_2^{i} -T_1^{i}\Big)\,\hat{c}^{\alpha\beta\gamma\delta}_{\mrf i} + \ds\frac{1}{2}\Big[\left(T_2^{i}\right)^2 -\left(T_1^{i}\right)^2\Big]\,\hat{c}^{\alpha\beta\gamma\delta}_{\mrf i,3}\right\rbrace \\[3mm]
\plus 2\,\ds\sum_{i=1}^2\,\left( \ttau^{\alpha\beta}_{2i}\,U_{2i}^{\gamma\delta} - \ttau^{\alpha\beta}_{1i}\,U_{1i}^{\gamma\delta} \right)~,
\eqe
where $\hat{c}^{\alpha\beta\gamma\delta}_{\mrf i}$ and $\hat{c}^{\alpha\beta\gamma\delta}_{\mrf i,3}$ are given by Eq.~\eqref{e:hcabgdHGO_f}; $\ttau^{\alpha\beta}_{1i}$ and $\ttau^{\alpha\beta}_{2i}$ are defined according to Eq.~\eqref{e:ttau_12} and $U^{\alpha\beta}_{1i}$ and $U^{\alpha\beta}_{2i}$ can be found in Appendix~\ref{s:switch_lin}. The other material tangent tensors can be derived similarly.

\paragraph{DD shell model}
The compression/tension switch is not fully consistent with the DD shell model since the assumptions made to derive Eqs.~\eqref{e:WL}~and~\eqref{e:Wb} are not necessarily valid if the switch is applied. In fact, as the material model is no longer symmetric w.r.t.~the shell mid-surface, due to the unsymmetric structure of the switch, the expressions with mixed term, i.e.~$E_{\alpha\beta}\,K_{\gamma\delta} $, do not vanish, e.g.~in Eq.~\eqref{e:WL}. Hence, the membrane and bending strains cannot be fully decoupled. Nonetheless, if the directly-decoupled approach is followed, in the reference configuration, $I_4^{i} = 1 $, which implies that $H_i =0$. Thus, the anisotropic part do not contribute to the bending energy and this reduces Eqs.~\eqref{e:c0HGO} to a purely isotropic formulation, i.e.~$c_0^{\alpha\beta\gamma\delta} = \mu\,c^{\alpha\beta\gamma\delta}_{\mathrm{NH}0}$.
%
%
Put differently, the directly-decoupled approach cannot capture the effects of the compression/tension switch if the bending moments are dominant or the anisotropic forces are much stronger than the isotropic ones; however, it is accurate if only the membrane forces are influential \citep{biomembrane}.  

%
%
\section{Numerical examples}\label{s:examples}
In this section, for each of the introduced material models, different numerical examples are considered to study the performance of the three presented shell models, i.e.~the numerically-projected (NP), analytically-projected (AP) and directly-decoupled (DD) shell models. First, a uniaxial tension test is performed to compare the membrane response of different shell models. Second, the pure bending of a cantilever subjected to a given rotation on its free end is considered, which shows how the models behave if the bending forces are dominant. Third, a square plate under pressure is studied to examine the coupled membrane and bending modes. Then, the formulation is tested for two specific applications: Large indentation of a strip under a rigid spherical indenter and an angioplasty example, which involves contact between two deformable bodies.

For the NP shell model, the through-the-thickness integration is evaluated by two Gaussian quadrature points unless specified otherwise. Furthermore, for all the examples, the material constants are set according to the Tab.~\ref{t:const}. For the anisotropic material models (GOH and AMR), two families of fibers are considered, i.e.~$n_\mrf=2$. For the GOH material model, $\kappa_i \in \{0.0,~0.226,~1/3\}$ following \citet{gasser06}.
\begin{table}[ht!]
\tabulinesep = 2pt
\centering
\begin{tabu} to 0.75\textwidth { X[-2,r,m] | X[l,m] | X[l,m] | X[l,m] }
\toprule
NH   & $\tc_1 = 10~[\mathrm{kPa}]$ & - & - \\
MR   & $\tc_1 = 10~[\mathrm{kPa}]$ & $\tc_2 = 2\,\tc_1~[\mathrm{kPa}]$ & - \\  
Fung & $\tc_1 = 10~[\mathrm{kPa}]$ & $c_2 = 10$ & - \\ 
AMR & $\tc_1 = 10~[\mathrm{kPa}]$ & $\tc_2 = 2\,\tc_1~[\mathrm{kPa}]$ & $\tc_3 = 100\,\tc_1~[\mathrm{kPa}]$ \\   
GOH & $\tmu = 10~[\mathrm{kPa}]$ & $\tk_{1i} = 100\,\tc_1~[\mathrm{kPa}]$ & $k_{2i} = 500$ \\   
\bottomrule  
\end{tabu}
\caption{Material constants of the considered material models}
\label{t:const}
\end{table}

\subsection{Uniaxial tension test}\label{s:uni}
\begin{figure}[H]
\begin{center} \unitlength1cm
\begin{picture}(15.0,5.0)
\put(0.0,0.0){\includegraphics[height=50mm]{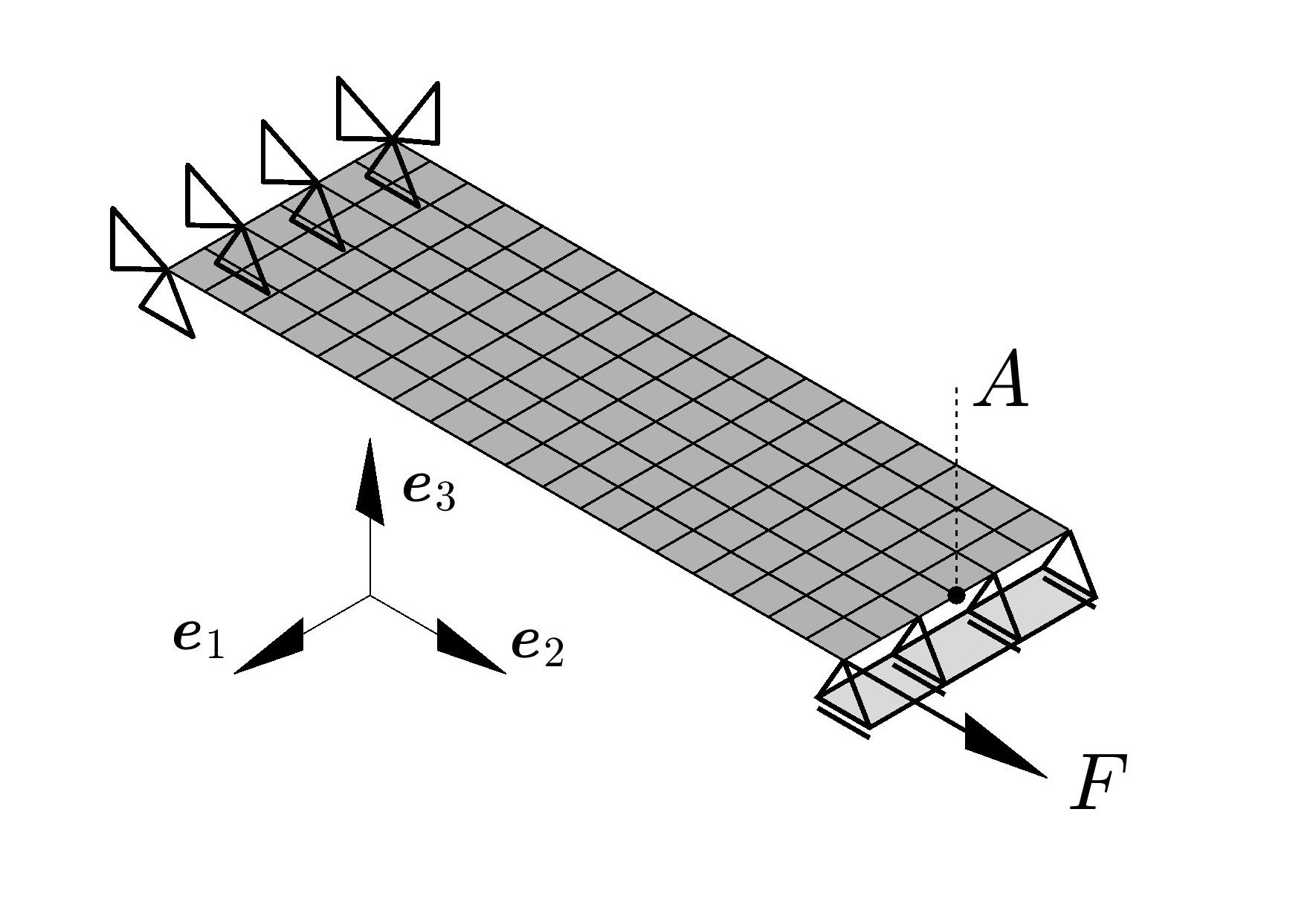}}
\put(3.0,0.0){(a)}
\put(7.0,0.5){\includegraphics[width=80mm,trim={0 100px 0 150px},clip]{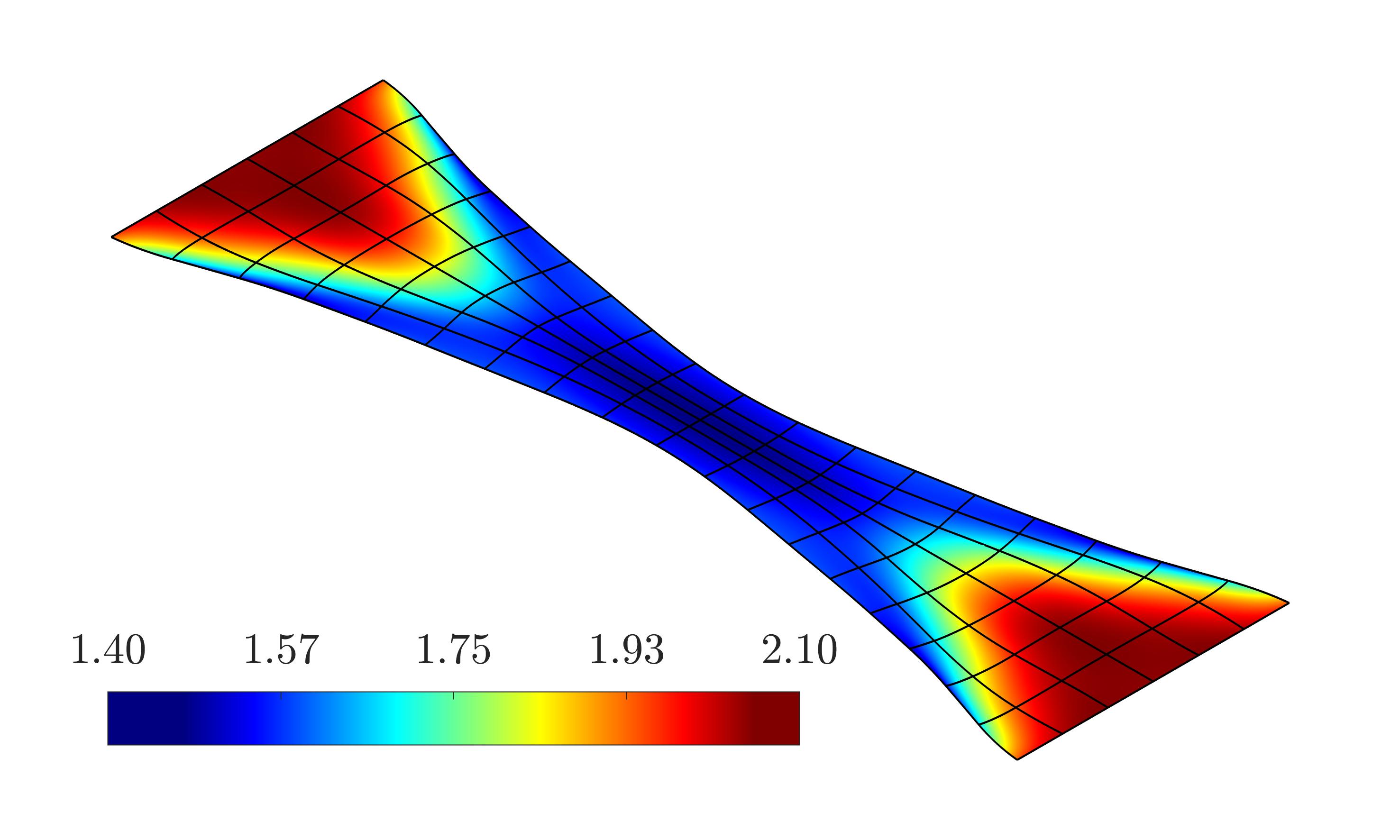}}
\put(8.5,0.0){(b)}
\end{picture}
\caption{Uniaxial tension test: (a) Reference configuration with boundary conditions and (b) deformed configuration for the GOH model (with $\kappa_i = 0$) colored by $I_1 := \tr \bC$.}
\label{f:tensile_undeformed}
\end{center}
\end{figure}

To examine the presented shell models for the case that membrane forces are dominating, a rectangular strip of $T \times W \times L = 0.3 \times 3 \times 9~[\mathrm{mm}^3]$ is pulled as shown in Fig.~\ref{f:tensile_undeformed}.a. On the pulled edged, the displacements in $\be_2$ direction are enforced to be equal. The pulling force $F$ is applied at the corner of the same edge. The strip is meshed by $6 \times 18$ quadratic NURBS elements (see Fig.~\ref{f:tensile_undeformed}.a). 

For the anisotropic materials, the principal directions of fibers are defined as 
\eqb{l}
\tbL_i = \sin \theta_i\,\be_1 + \cos \theta_i\,\be_2,\quad(i=1,2)~,
\label{e:tbL_i}\eqe
where $\be_1$ and $\be_2$ are the unit vectors of the Cartesian coordinate system (shown in Fig.~\ref{f:tensile_undeformed}.a). For this example, $\theta_1,\theta_2 = \pm 45^\circ$ \tmod{and $\pm 30^\circ$ for the AMR and GOH models, respectively}. As already mentioned, here it is assumed that $\aL^{33}_i=L^{33}_i=0$. Henceforth, the displacements in $\be_1$, $\be_2$ and $\be_3$ directions are denoted by $u$, $v$ and $w$, respectively.  

\begin{figure}[H]
\begin{center} \unitlength1cm
%
%
\begin{picture}(15.0,8.0)
\put(0.0,3.8){\includegraphics[height=42mm]{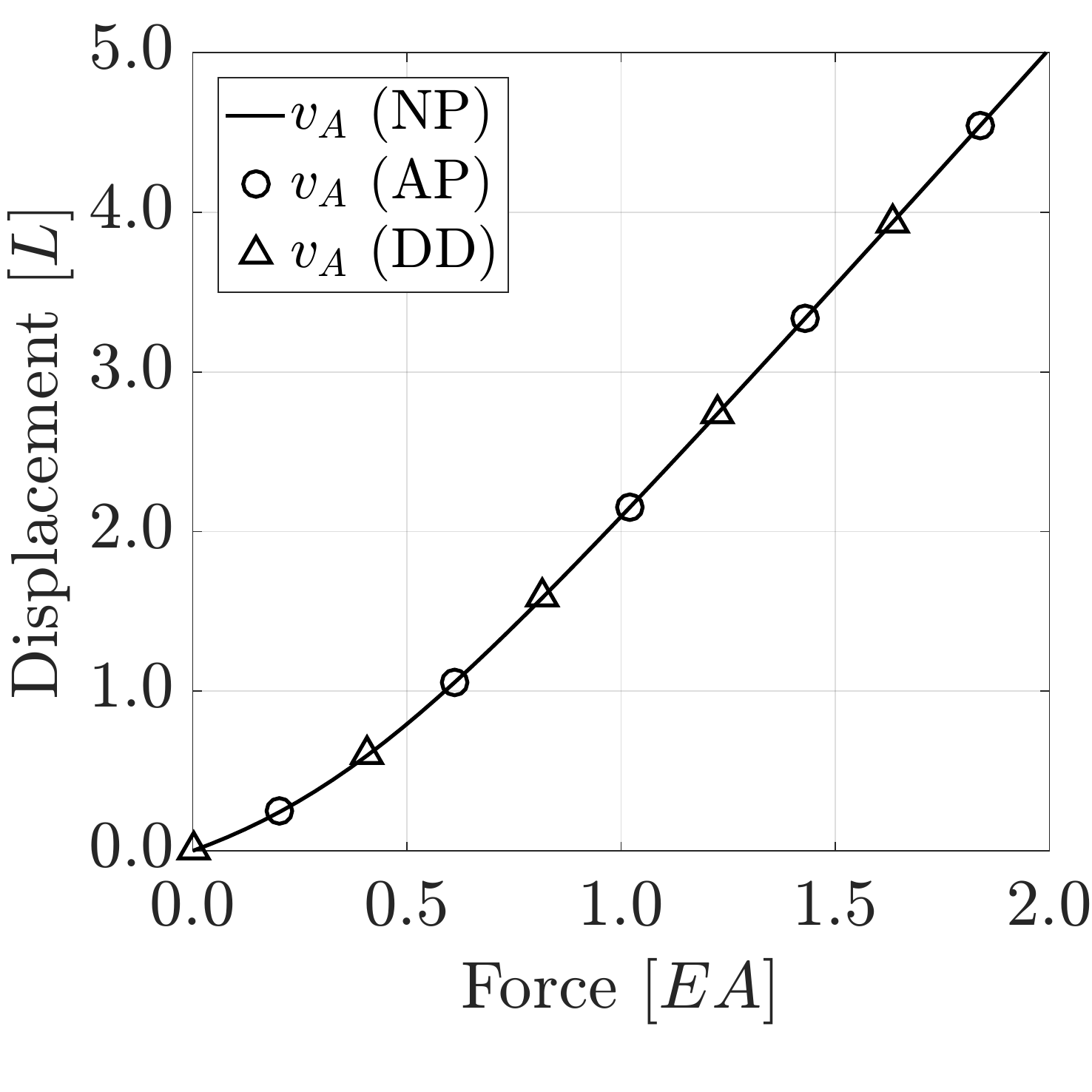}}
\put(-0.3,4.3){(a)}
\put(5.3,3.8){\includegraphics[height=42mm]{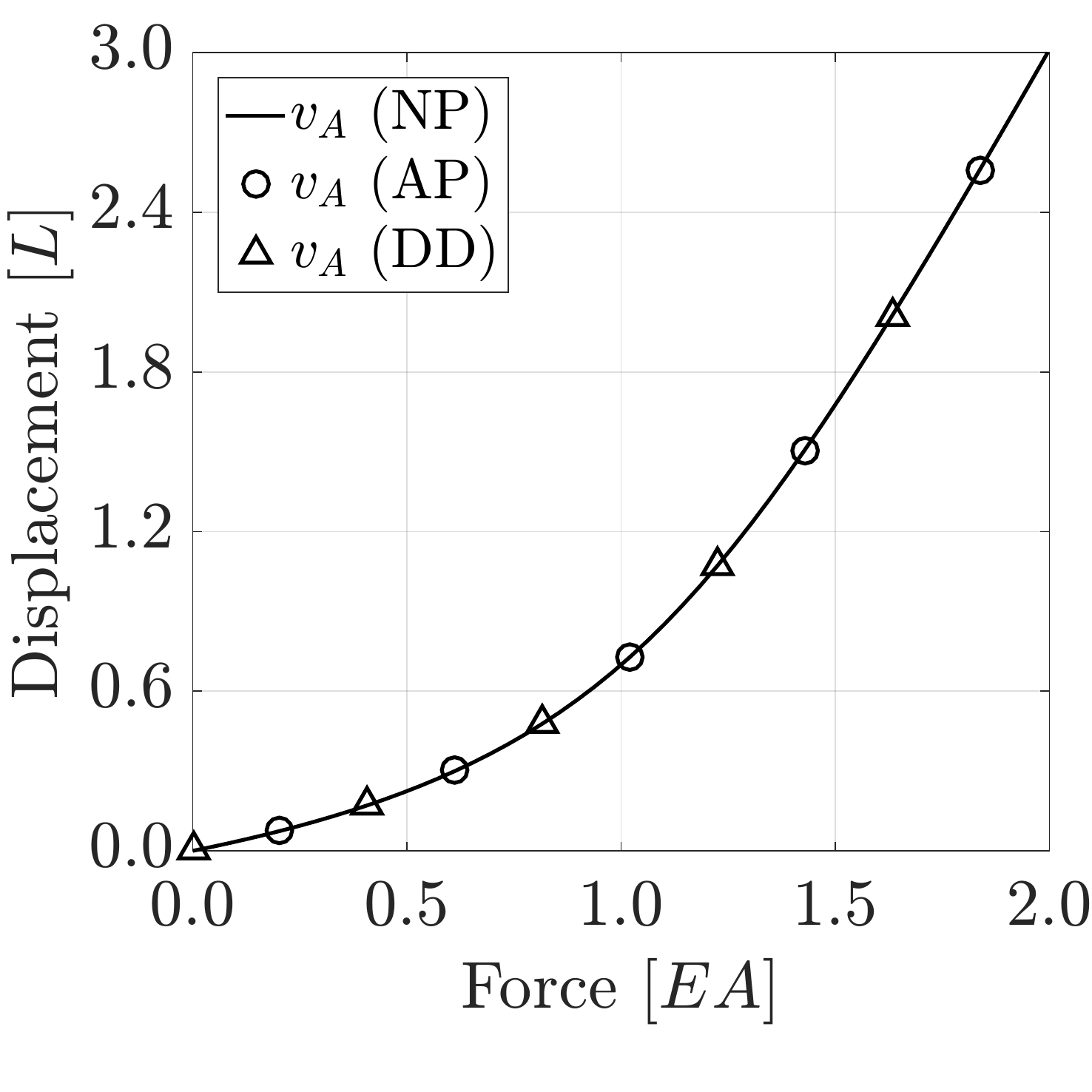}}
\put(5.0,4.3){(b)}
\put(10.5,3.8){\includegraphics[height=42mm]{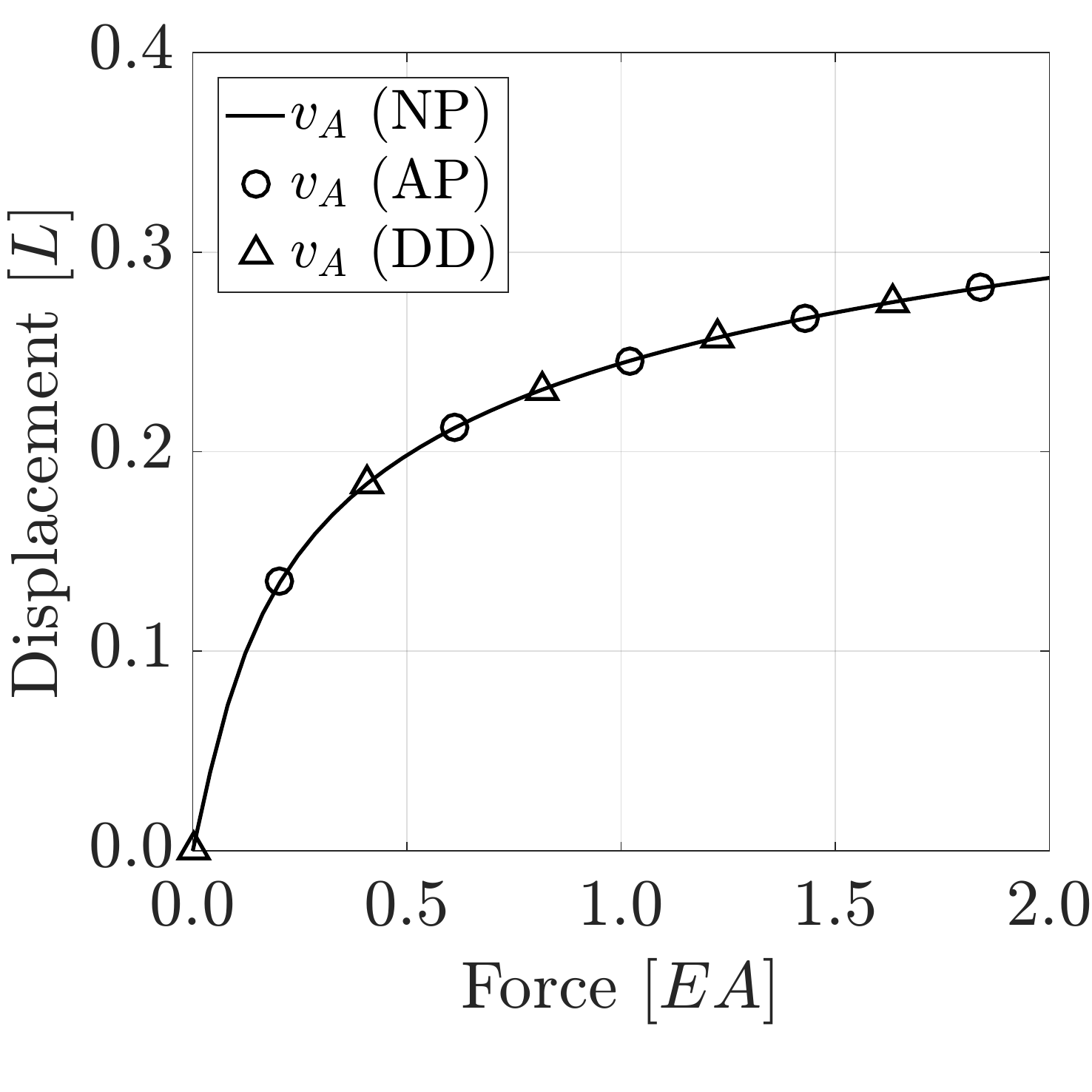}}
\put(10.2,4.3){(c)}
\put(2.0,-0.2){\includegraphics[height=42mm]{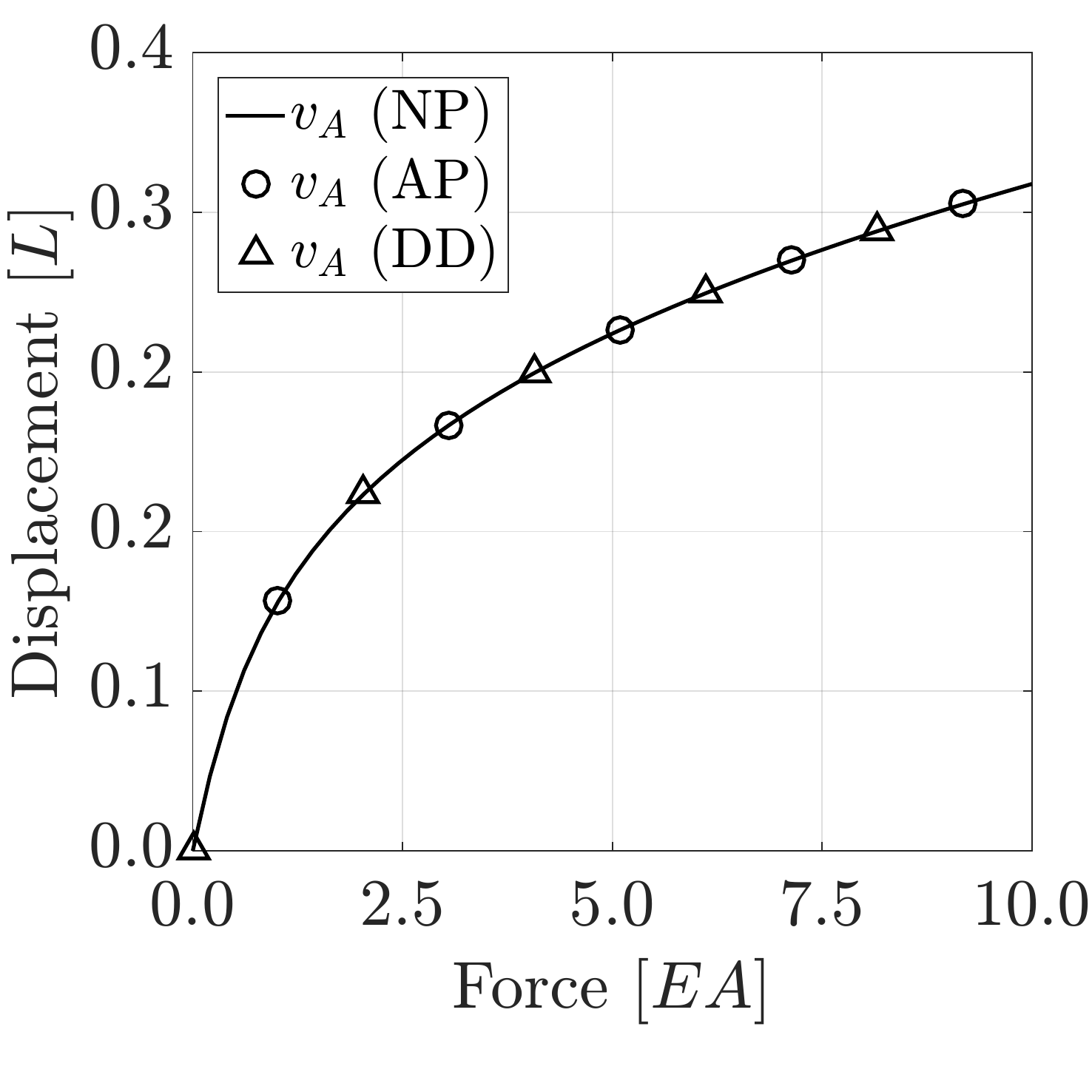}}
\put(1.7, 0.3){(d)}
\put(7.5,-0.2){\includegraphics[height=42mm]{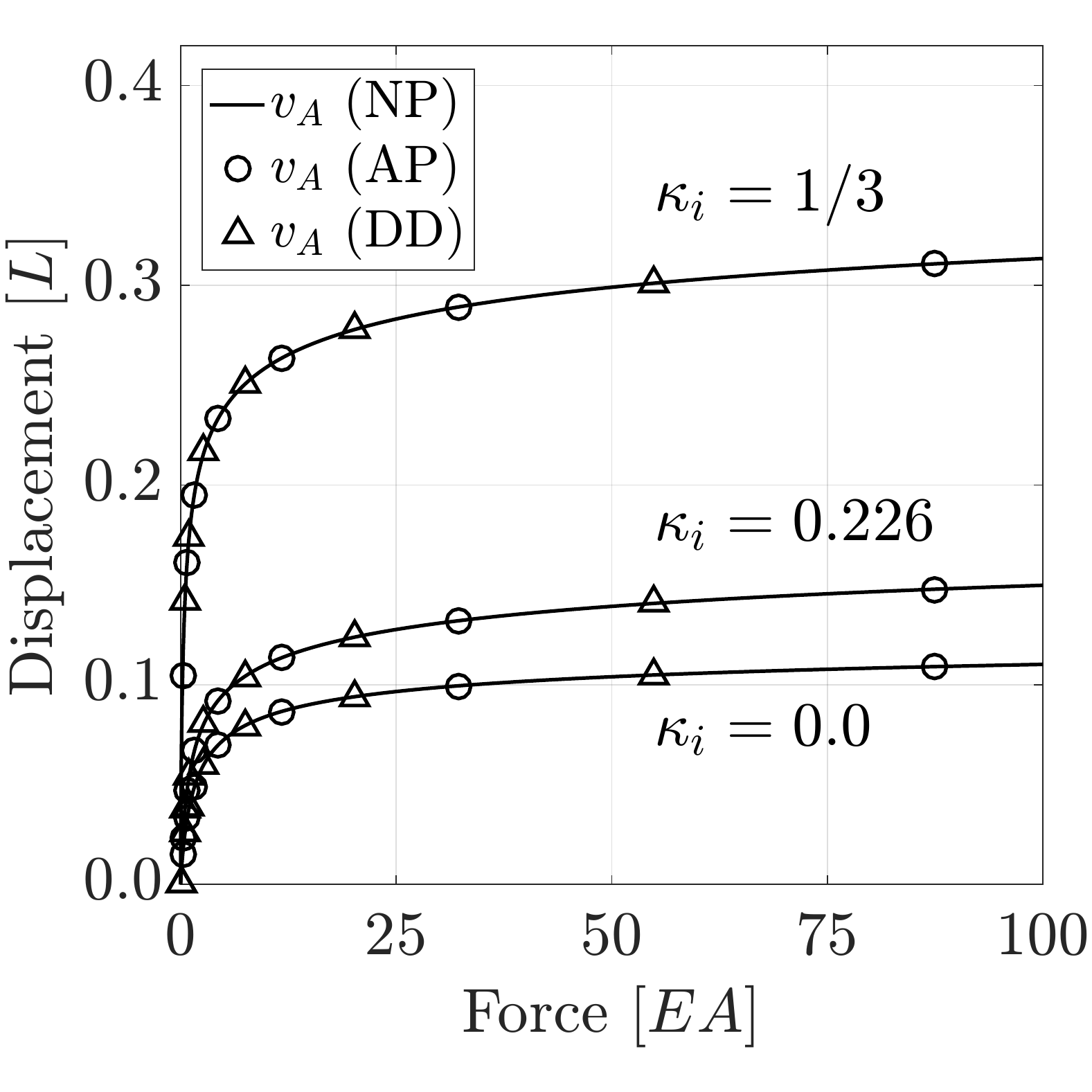}}
\put(7.2, 0.3){(e)}
\end{picture}
\caption{Uniaxial tension test -- the displacement of the tip vs.~the applied force: (a) NH, (b) MR, (c) Fung, (d) AMR and (e) GOH material model for the three constitutive approaches (NP, AP and DD) presented in Sec.~\ref{s:shells}.}
\label{f:tensile_curve}
\end{center}
\end{figure}
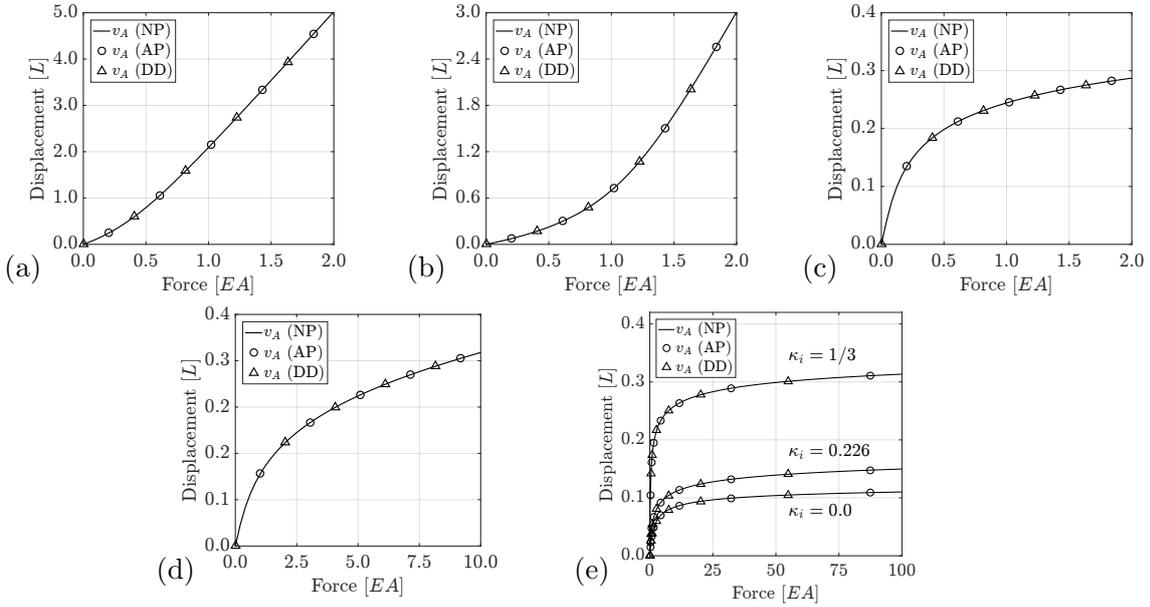

Fig.~\ref{f:tensile_curve} shows the displacement of point A (shown in Fig.~\ref{f:tensile_undeformed}.a) versus the applied total force. The applied force is normalized by $EA$, where $E=3\,\tc_1$ corresponds to an infinitesimal Young's modulus and $A = W \, T$ is the cross section area. As expected, for all the isotropic and anisotropic materials, the AP and DD shell models give exactly the same results as the NP shell model.

\subsection{Cantilever bending}\label{s:canti}
\begin{figure}[H]
\begin{center} \unitlength1cm
\begin{picture}(15.0,5.0)
\put(0.0,0.0){\includegraphics[height=50mm]{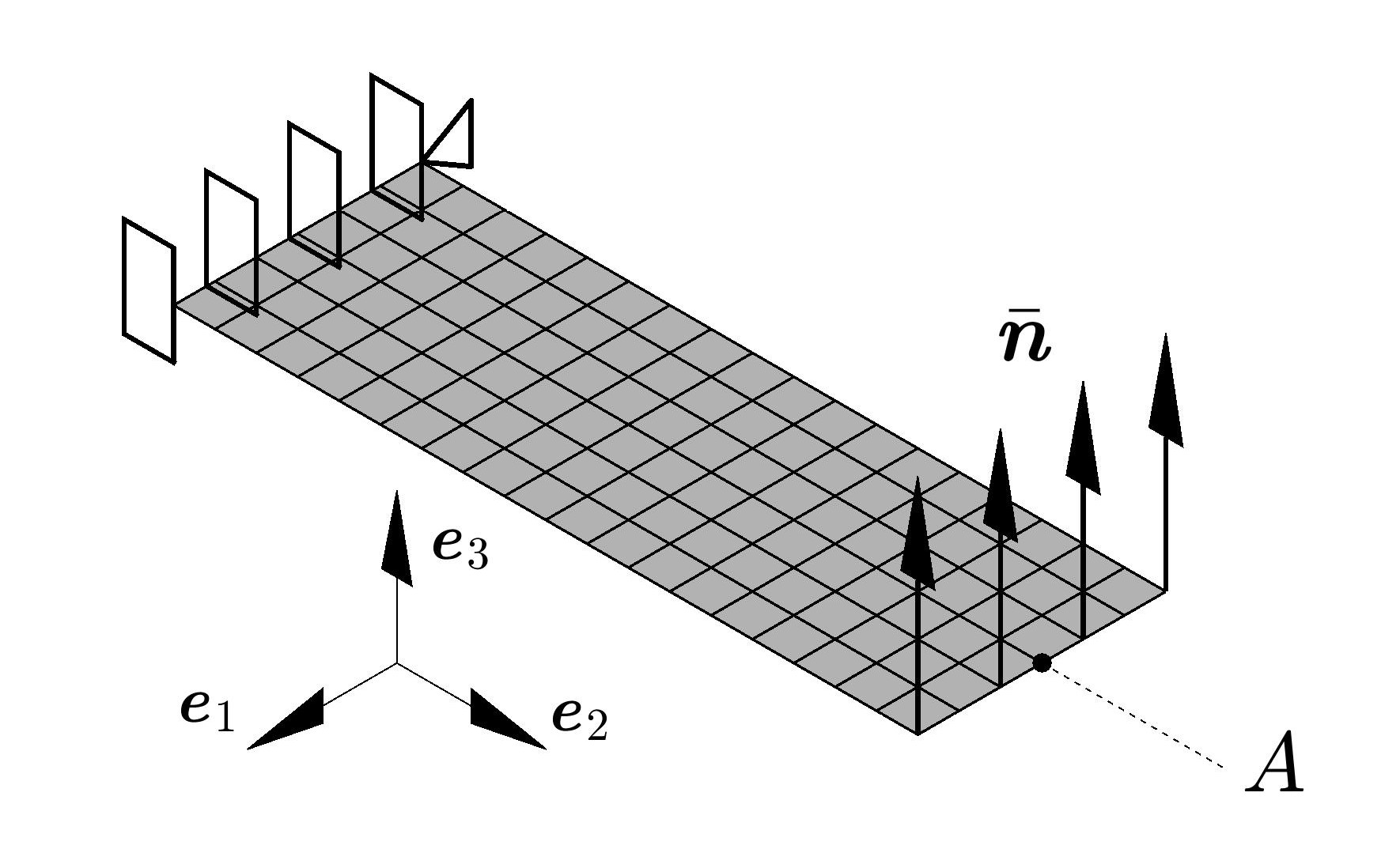}}
\put(3.0,0.0){(a)}
\put(9.0,0.0){\includegraphics[height=50mm]{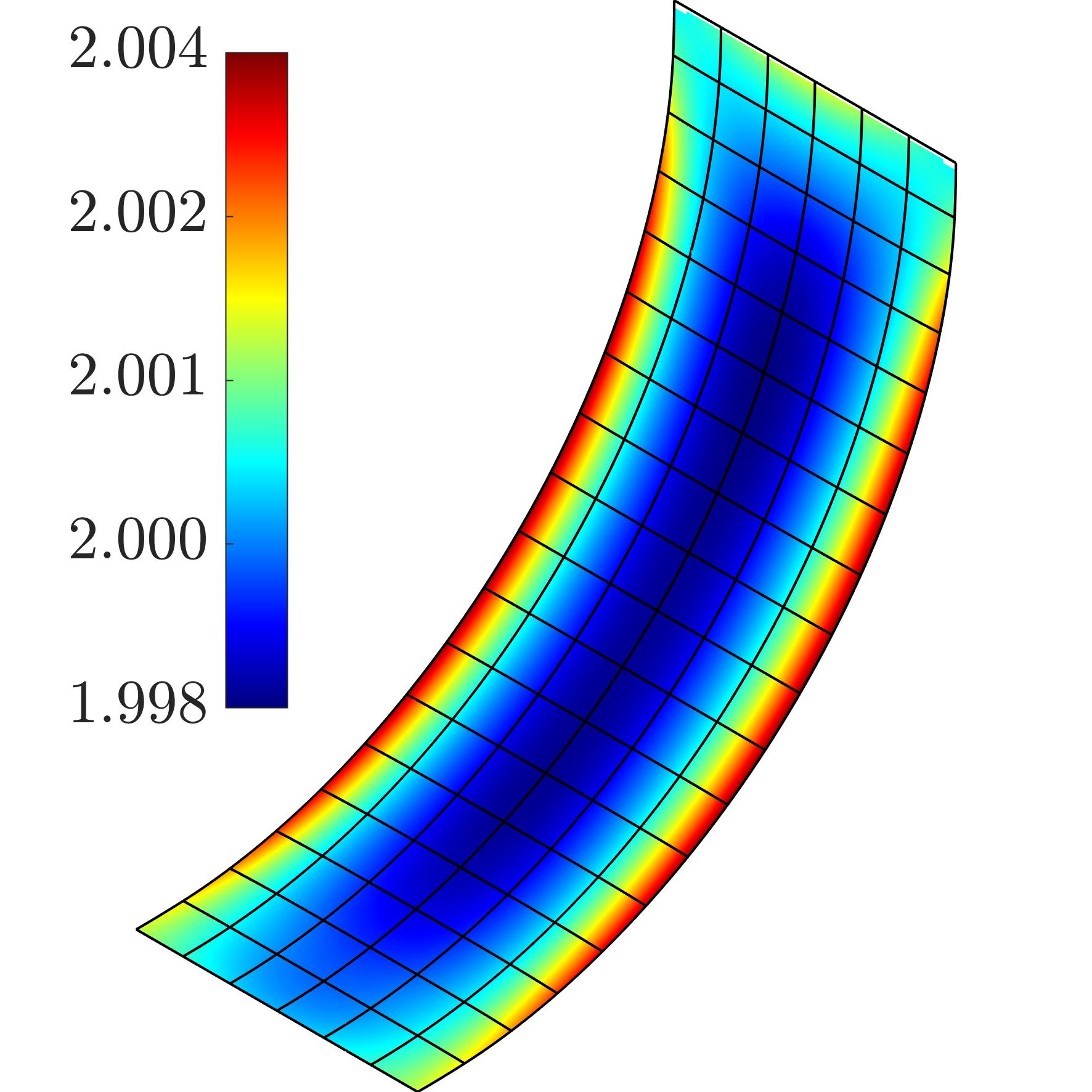}}
\put(8.5,0.0){(b)}
\end{picture}
\caption{Cantilever bending test: (a) Reference configuration with boundary conditions and (b) deformed configuration for the GOH model (with $\kappa_i = 0.226$ and the compression/tension switch) colored by $I_1 := \tr \bC$.}
\label{f:beam_undeformed}
\end{center}
\end{figure}

The cantilever has the same geometry and mesh properties as the strip of Sec.~\ref{s:uni} although here $T=W/20$. On the clamped edge (see Fig.~\ref{f:beam_undeformed}.a), the rotations are restricted following a penalty formulation. On the free end, the surface normal $\bn$ is constrained to be equal to the given normal $\bar\bn$ using the constraint of \citet{solidshell}. Here, $\bar\bn = \cos\alpha\,\be_3 - \sin\alpha\,\be_2$, where $\alpha$ is the angle of rotation around $\be_1$. In the reference configuration, $\bar\bn = \bN$ and $\alpha = 0$ (see Fig.~\ref{f:beam_undeformed}.a). Here, the maximum rotation is set to $\alpha = 90^\circ$ (see Fig.~\ref{f:beam_undeformed}.b).

The total bending moment corresponding to this rotation is determined following the constraint formulation. The corresponding bending moment is normalized by $E\,I/L$, where $I = W\,T^3/12$ is the second moment of area of the cross section. The orientation of the fibers is defined based on Eq.~\eqref{e:tbL_i}. Here, $\theta_1,\theta_2 = \pm 45^\circ $ for the AMR model and $\theta_1,\theta_2 = \pm 30^\circ $ for the GOH model. 

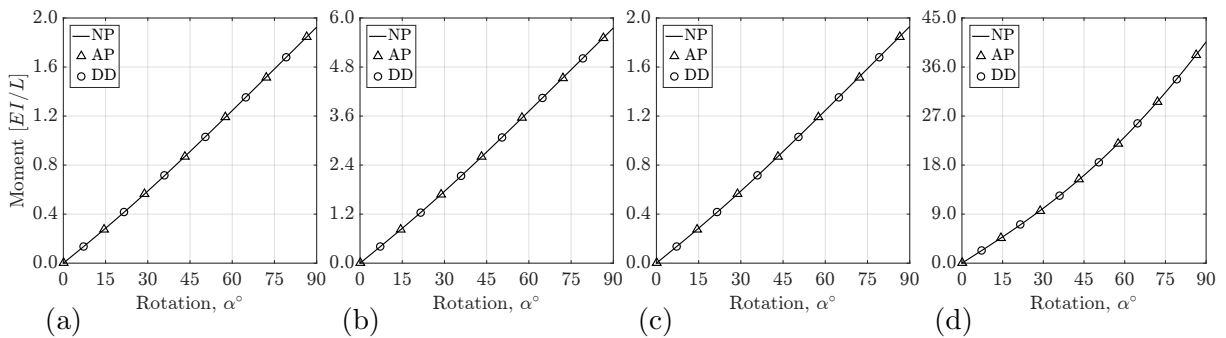
\begin{figure}[H]
\begin{center} \unitlength1cm
%
%
%
\begin{picture}(15.0,4.2)
\put(-0.5,0.0){\includegraphics[height=42mm]{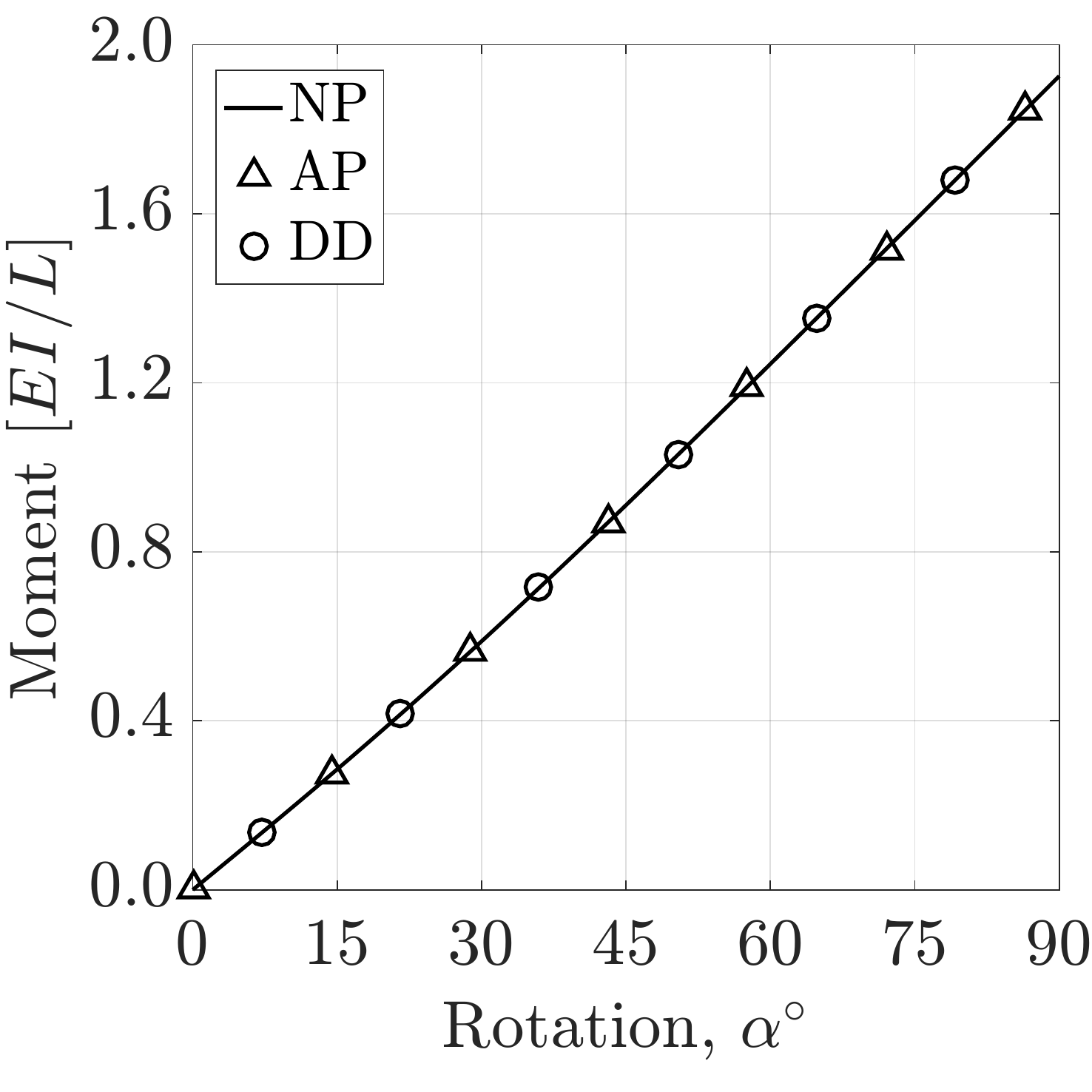}}
\put(0.0,-0.1){(a)}
\put(3.7,0.0){\includegraphics[height=42mm,trim={30px 0px 0px 0px},clip]{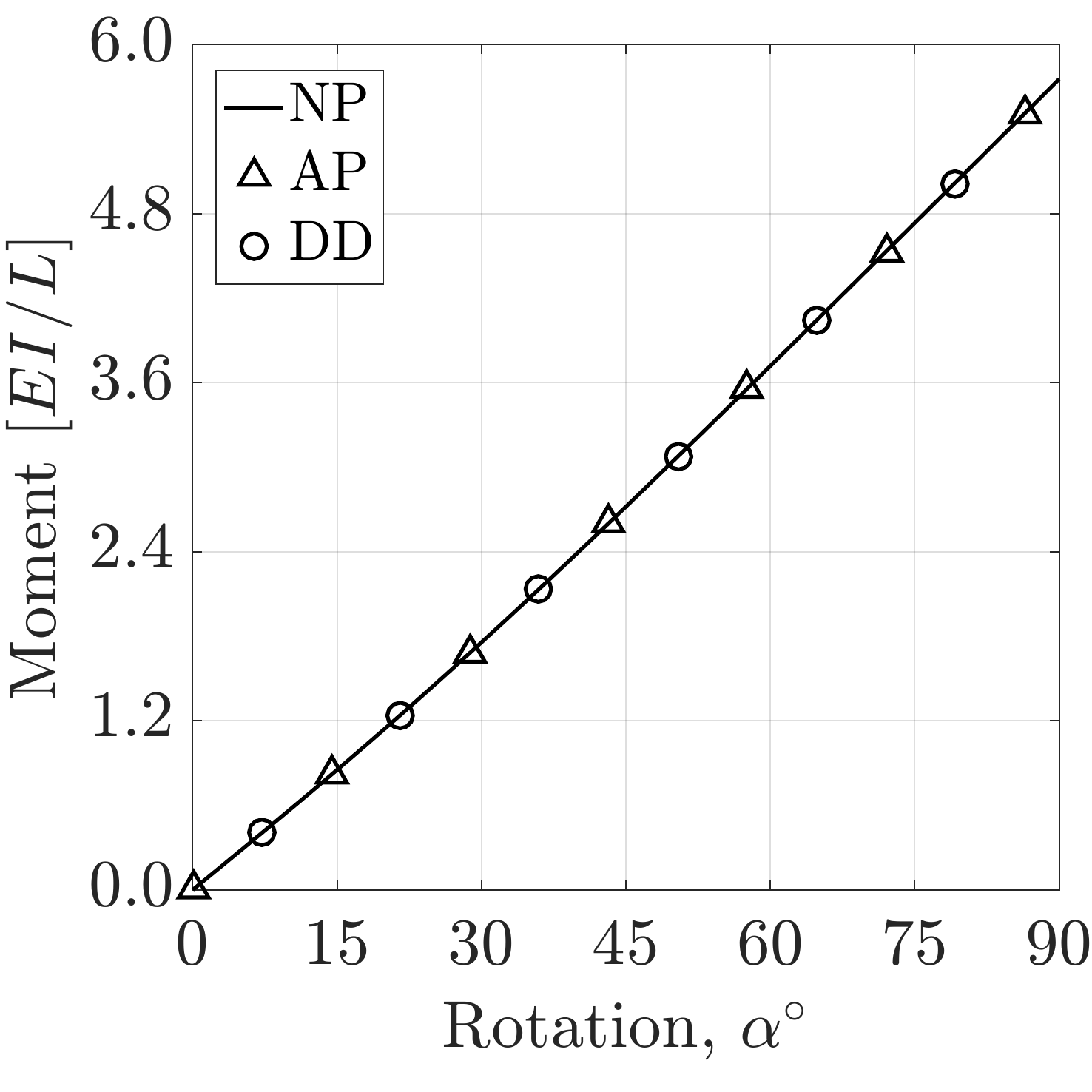}}
\put(3.9,-0.1){(b)}
\put(7.6,0.0){\includegraphics[height=42mm,trim={30px 0px 0px 0px},clip]{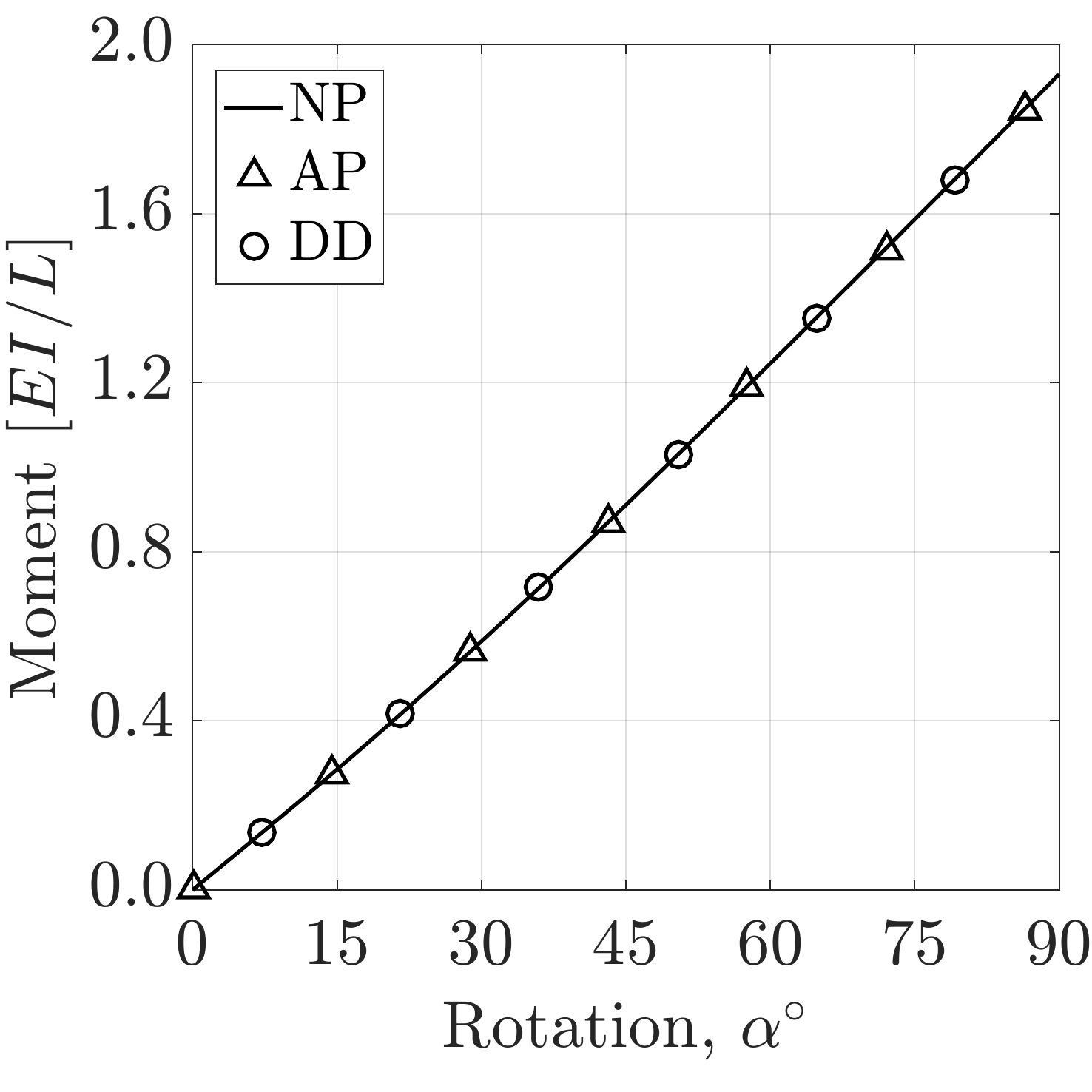}}
\put(7.8,-0.1){(c)}
\put(11.5,0.0){\includegraphics[height=42mm,trim={30px 0px 0px 0px},clip]{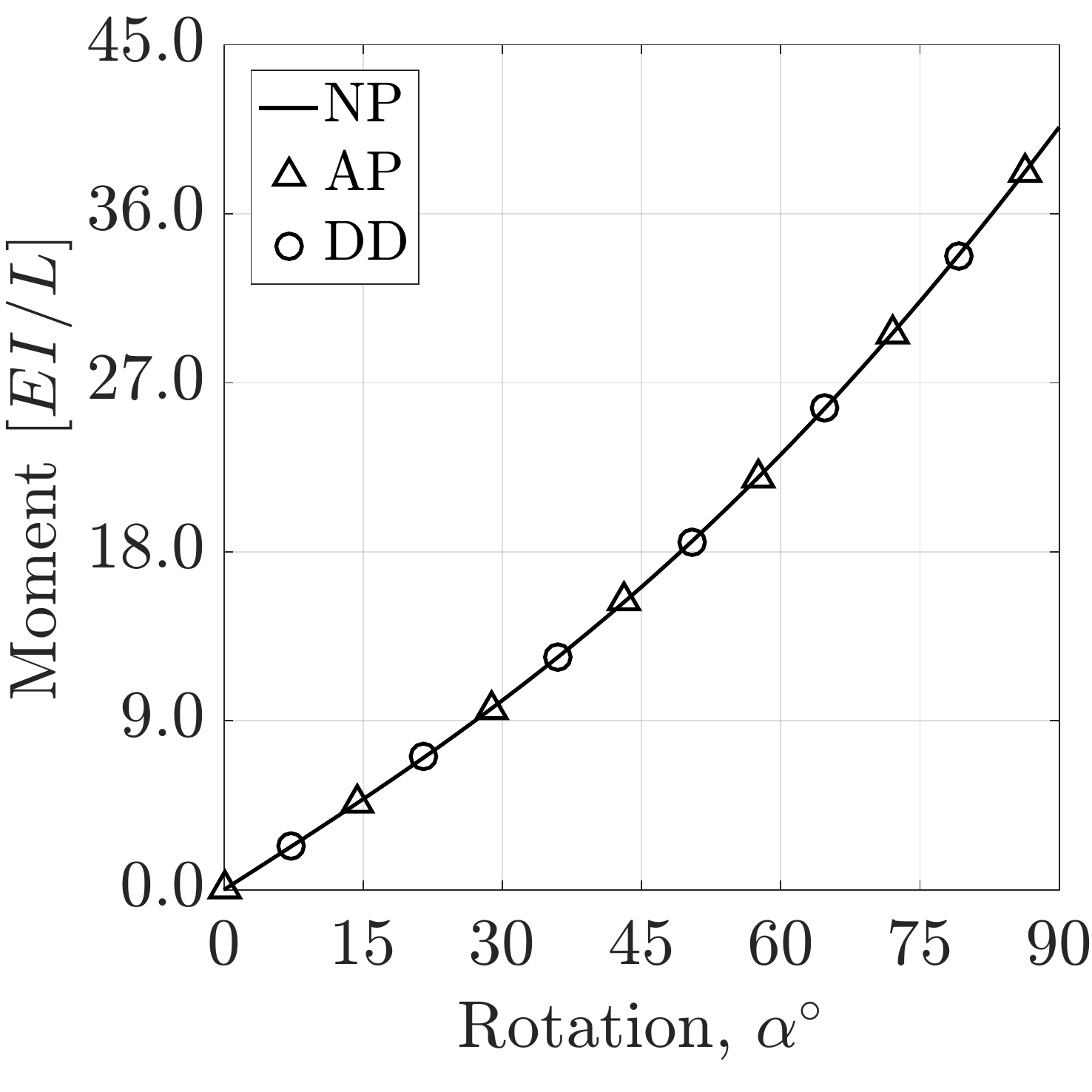}}
\put(11.7,-0.1){(d)}
\end{picture}
\caption{Cantilever bending test -- the corresponding bending moment vs.~the applied rotation: (a) NH, (b) MR, (c) Fung and (d) AMR material model for the three constitutive approaches (NP, AP and DD) presented in Sec.~\ref{s:shells}.}
\label{f:beam_curve1}
\end{center}
\end{figure}

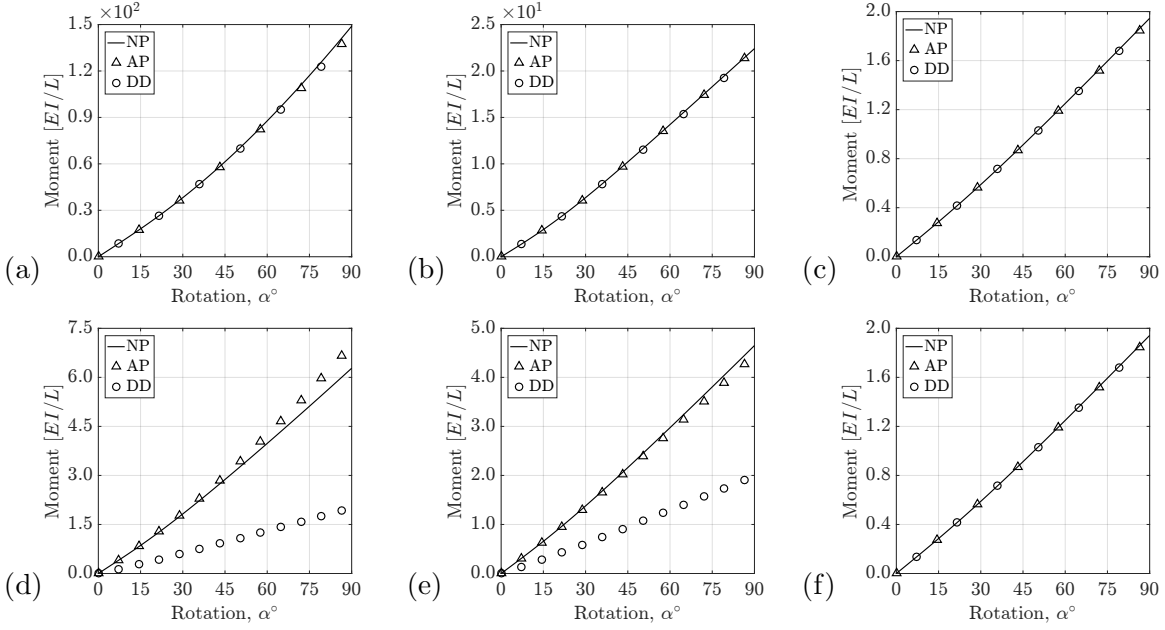
\begin{figure}[ht]
\begin{center} \unitlength1cm
%
%
\begin{picture}(15.0,8.4)
\put(0.0,4.2){\includegraphics[height=42mm]{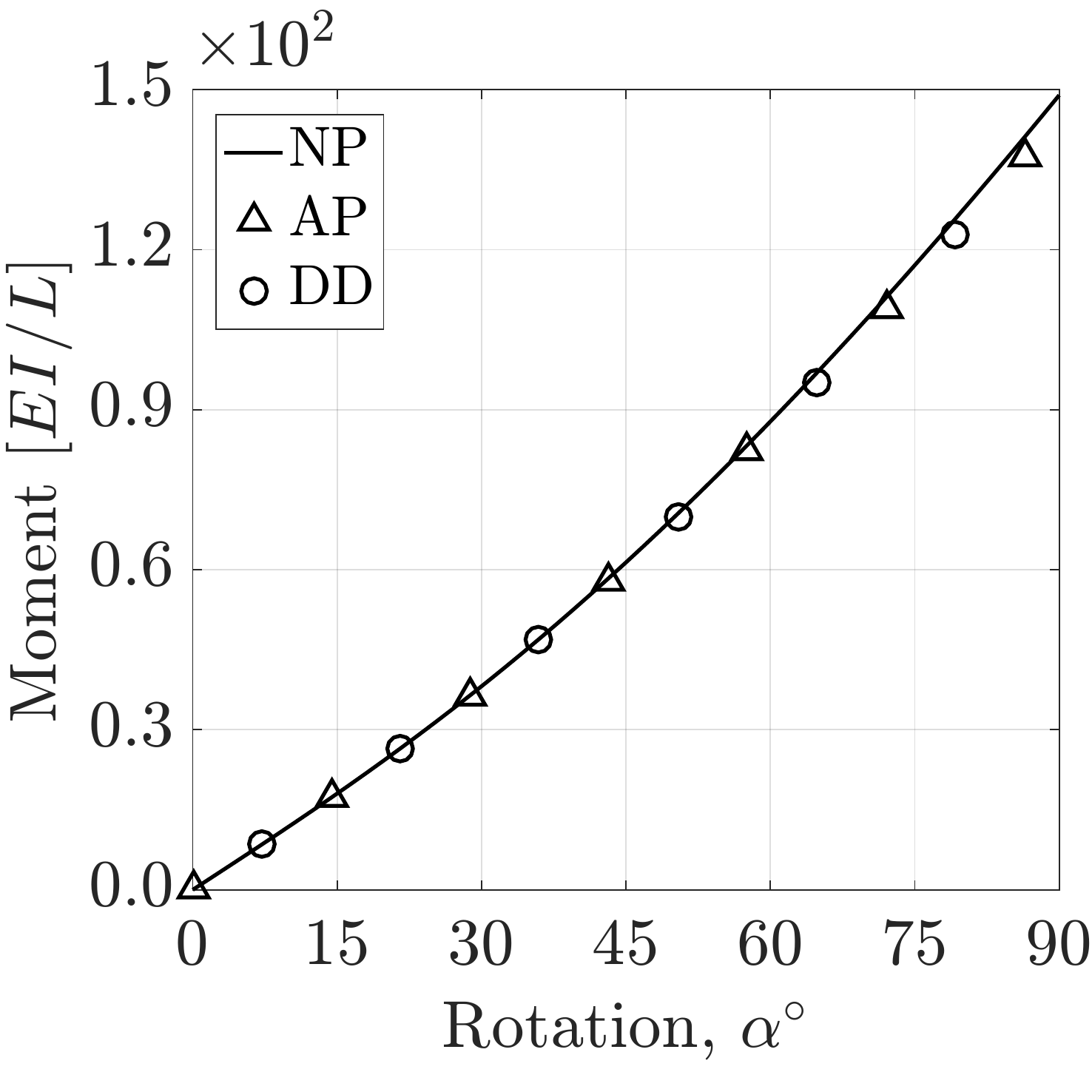}}
\put(-0.5,4.7){(a)}
\put(5.3,4.2){\includegraphics[height=42mm]{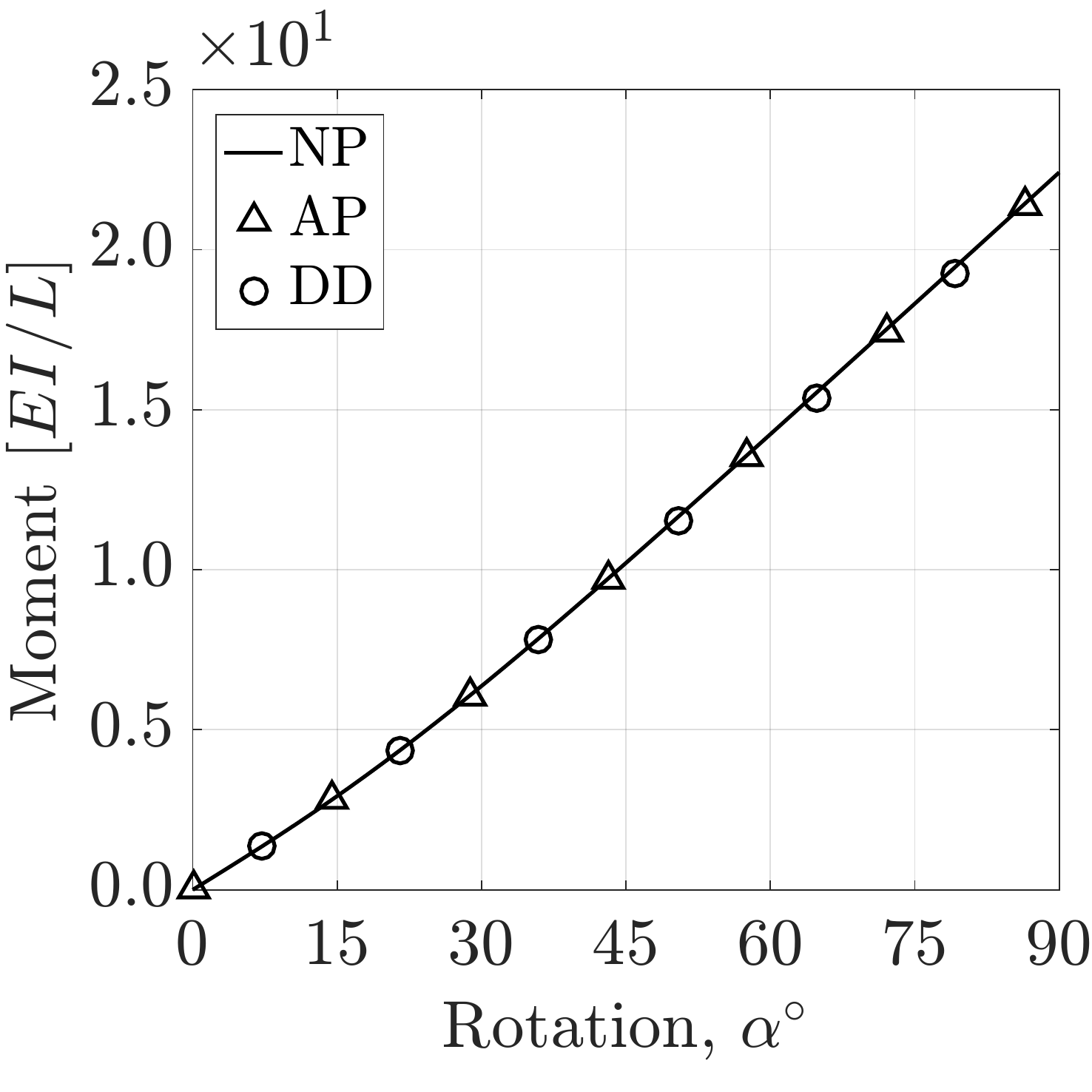}}
\put(4.8,4.7){(b)}
\put(10.5,4.2){\includegraphics[height=42mm]{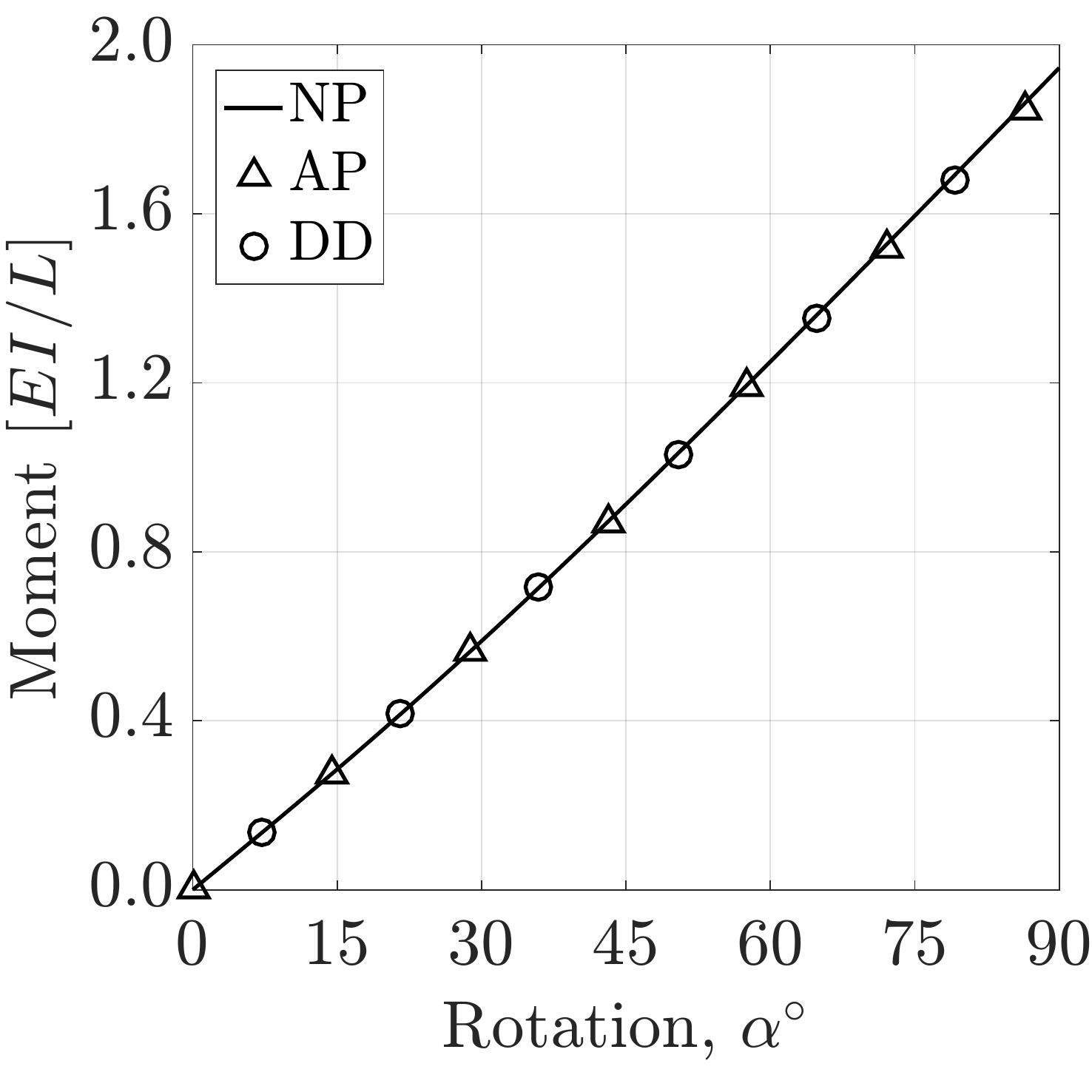}}
\put(10.0,4.7){(c)}
\put(0.0,0.0){\includegraphics[height=42mm]{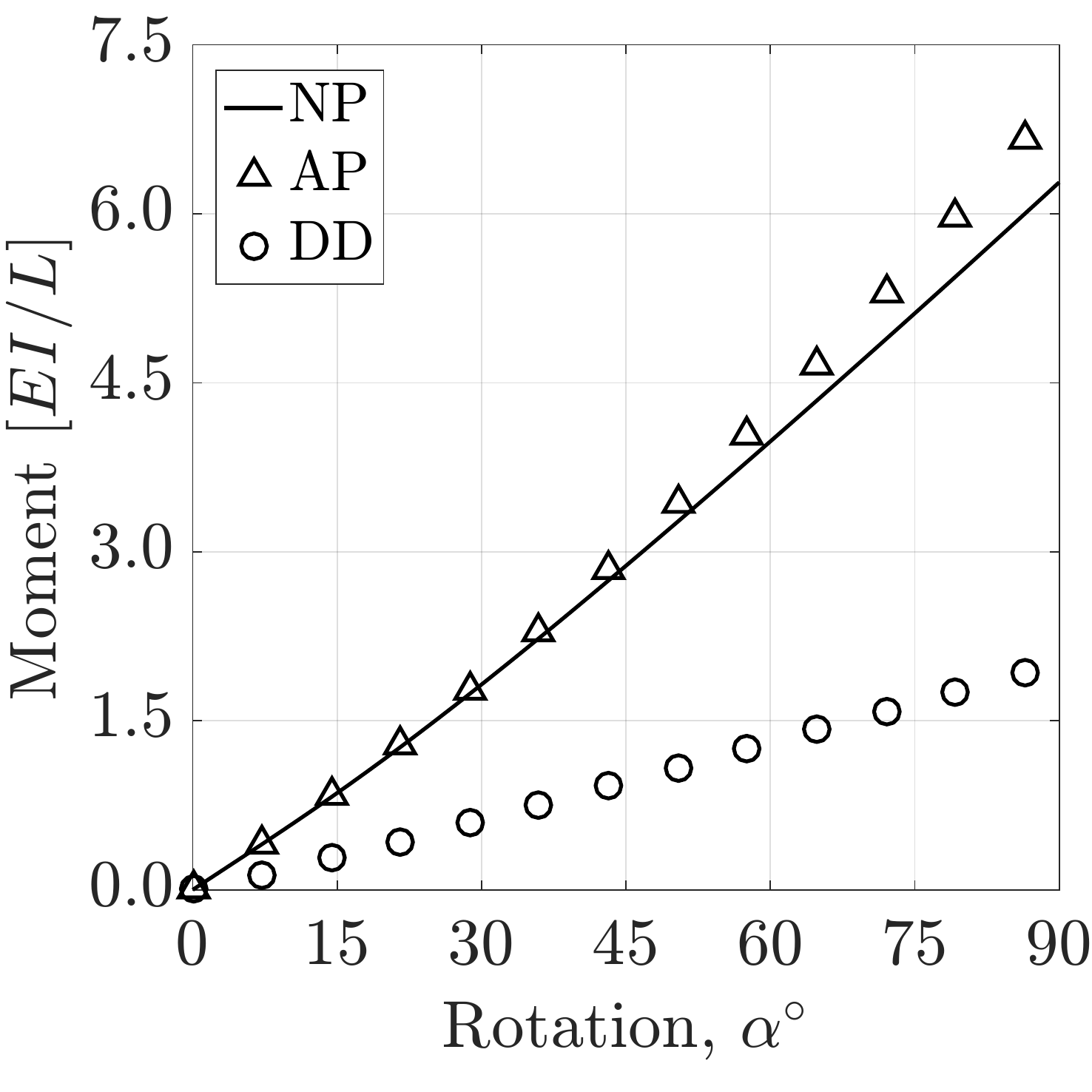}}
\put(-0.5,0.5){(d)}
\put(5.3,0.0){\includegraphics[height=42mm]{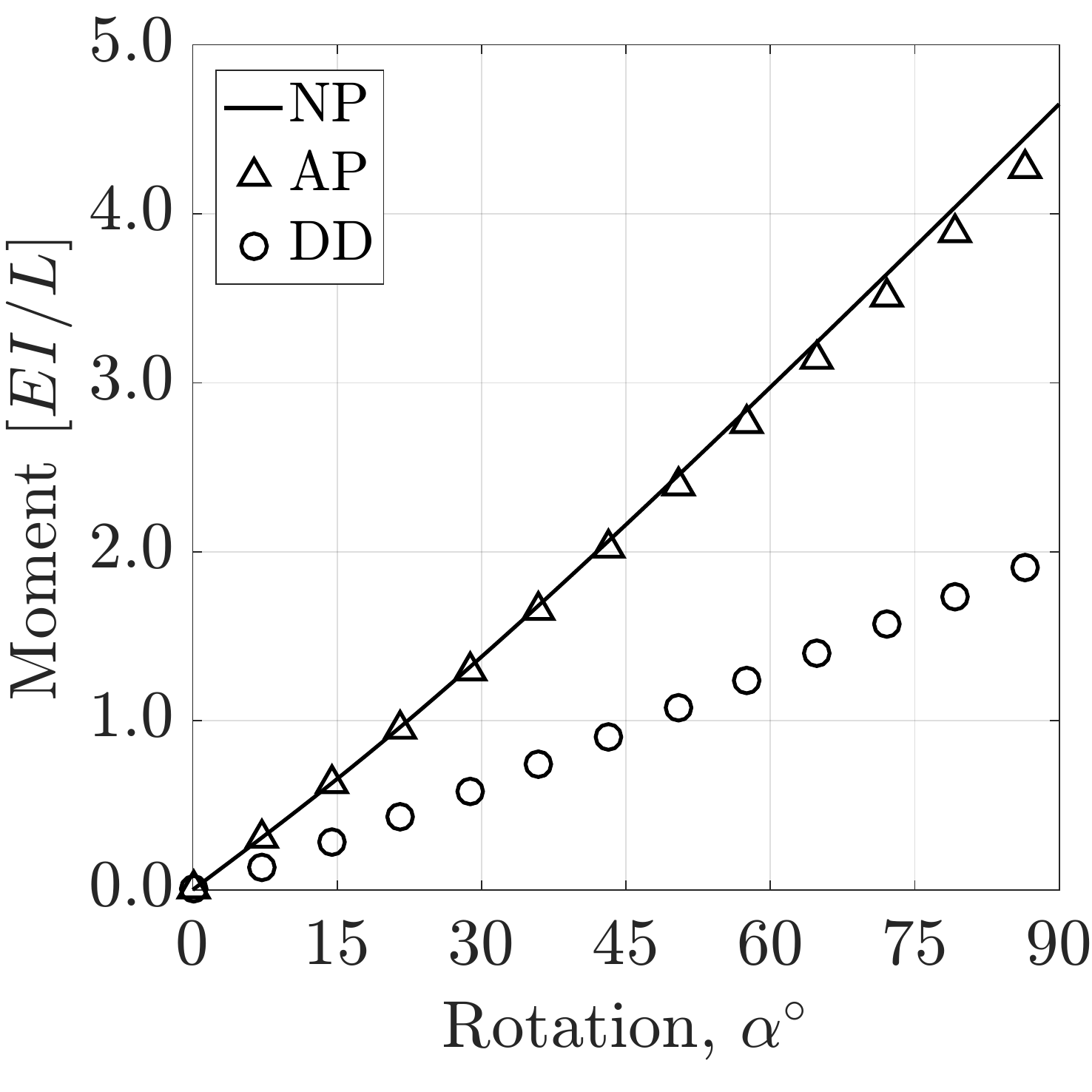}}
\put(4.8,0.5){(e)}
\put(10.5,0.0){\includegraphics[height=42mm]{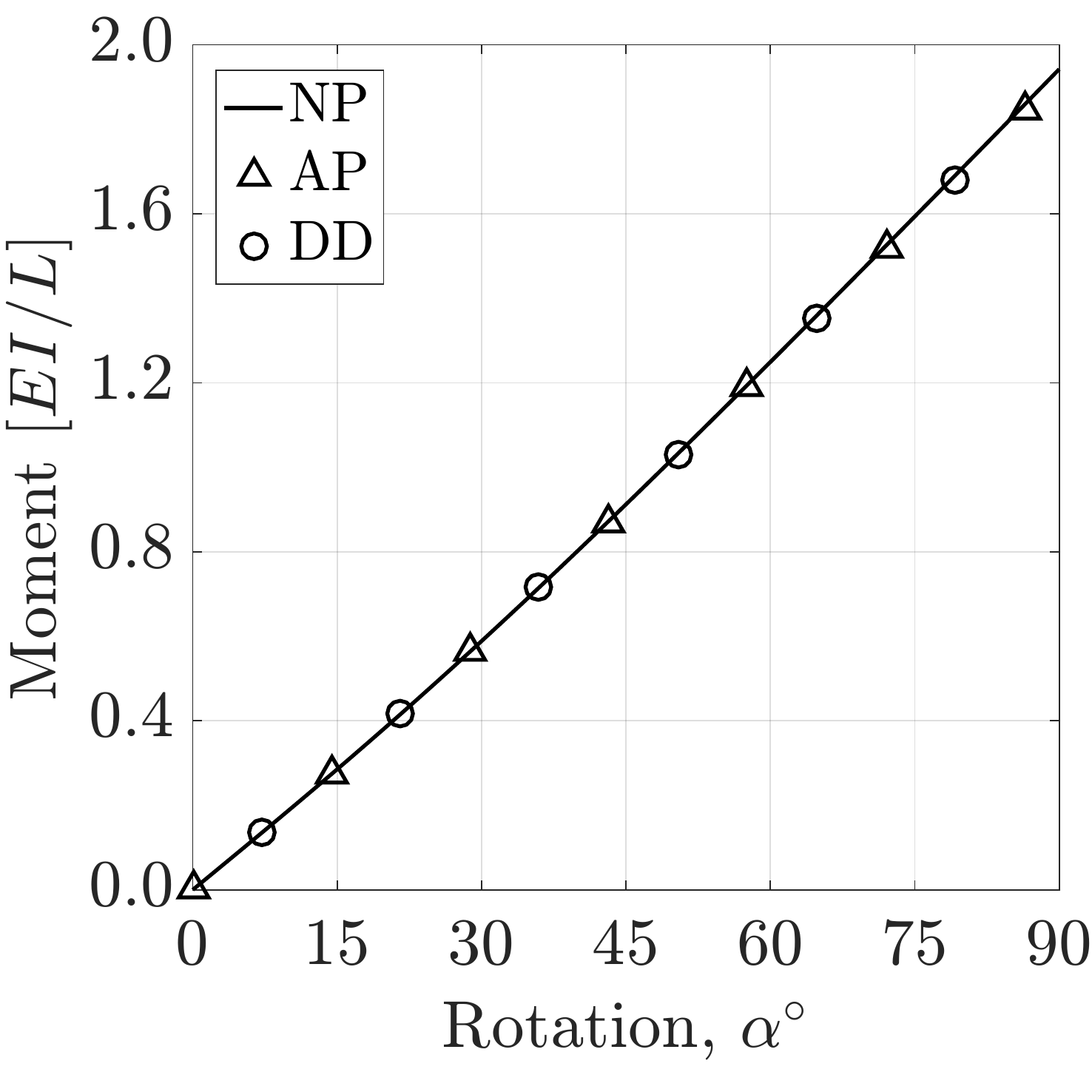}}
\put(10.0,0.5){(f)}
\end{picture}
\caption{Cantilever bending test for the GOH material model -- the corresponding bending moment vs.~the applied rotation for the three constitutive approaches NP, AP and DD: (a-c) without and (d-f) with the compression/tension switch. Further, for (a) and (d) $\kappa_i = 0.0$, for (b) and (e) $\kappa_i = 0.226$ and for (c) and (f) $\kappa_i = 1/3$.}
\label{f:beam_curve2}
\end{center}
\end{figure}

In Figs.~\ref{f:beam_curve1}~and~\ref{f:beam_curve2}, the corresponding bending moment is plotted against the applied rotation. Similar to the previous example, for the Neo--Hookean, Mooney--Rivlin and Fung material models, which are isotropic, as well as for the anisotropic Mooney--Rivlin material model, the AP and DD shell models are as accurate as the NP shell model (see Fig.~\ref{f:beam_curve1}). 

Fig.~\ref{f:beam_curve2} shows the results for the Gasser--Ogden--Holzapfel material model. If the compression/tension switch is excluded (see Figs.~\ref{f:beam_curve2}.a-\ref{f:beam_curve2}.c), all the three introduced shell models behave very similarly. In this case, for the NP shell model, 2 Gaussian quadrature points are sufficient to evaluate the integration through the shell thickness. However, if the compression/tension switch is included (see Figs.~\ref{f:beam_curve2}.d-\ref{f:beam_curve2}.f), the DD shell model cannot capture the switch effect since the material model is no longer symmetric w.r.t.~the shell mid-surface. Although if the material model is completely isotropic (i.e.~setting $\kappa_i = 1/3$), the fibers are excluded and trivially the switch has no effect on the constitutive equations (see Fig.~\ref{f:beam_curve2}.f). By increasing the anisotropy (i.e.~$\kappa_i \to 0$), the AP and NP shell models behave very similarly although the DD shell model deviates from the correct solution (see Figs.~\ref{f:beam_curve2}.d-\ref{f:beam_curve2}.e). 

\begin{figure}[H]
\begin{center} \unitlength1cm
%
%
\begin{picture}(15.0,3.8)
\put(2.6,-0.2){\includegraphics[height=42mm]{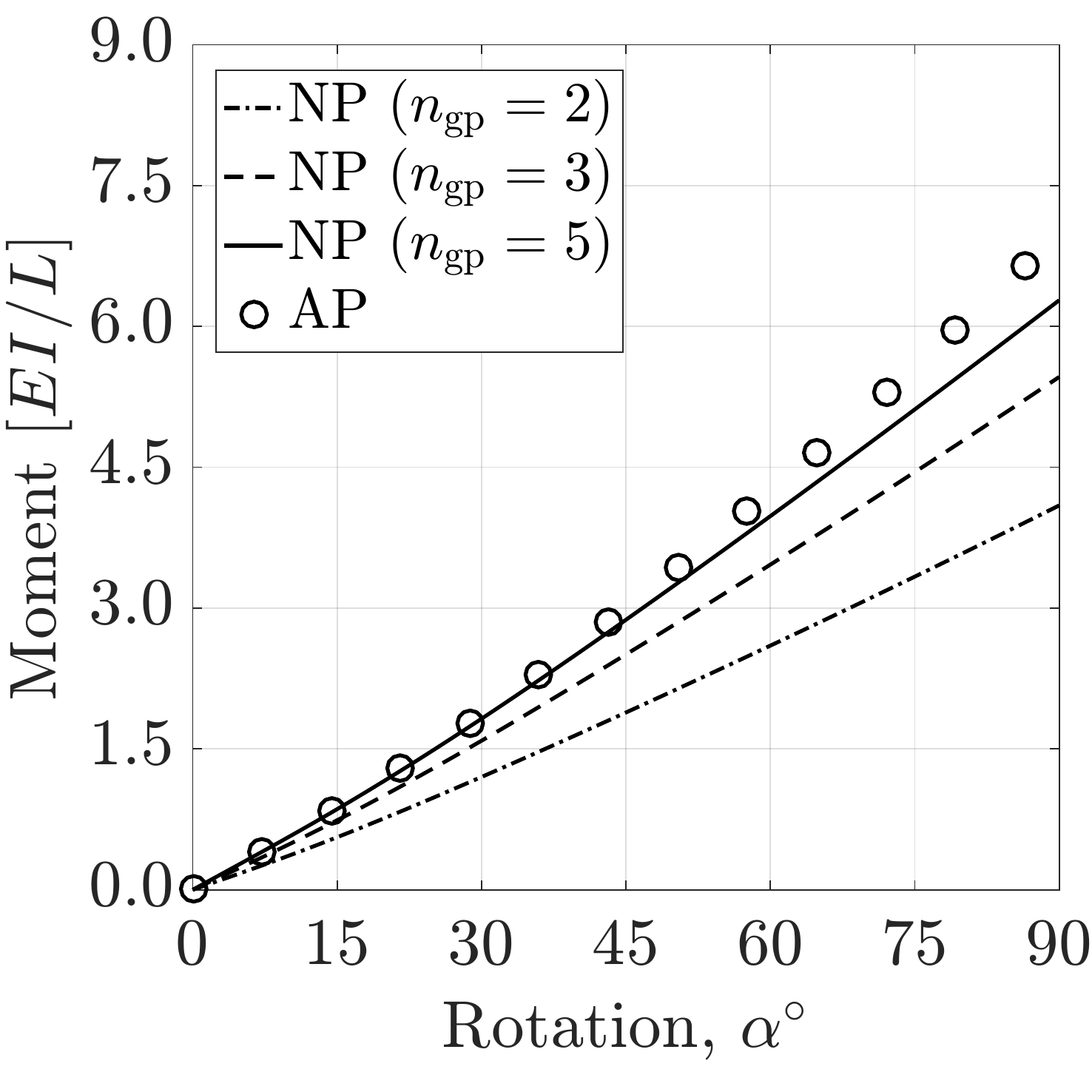}}
\put(2.1,0.3){(a)}
\put(8.0,-0.2){\includegraphics[height=42mm]{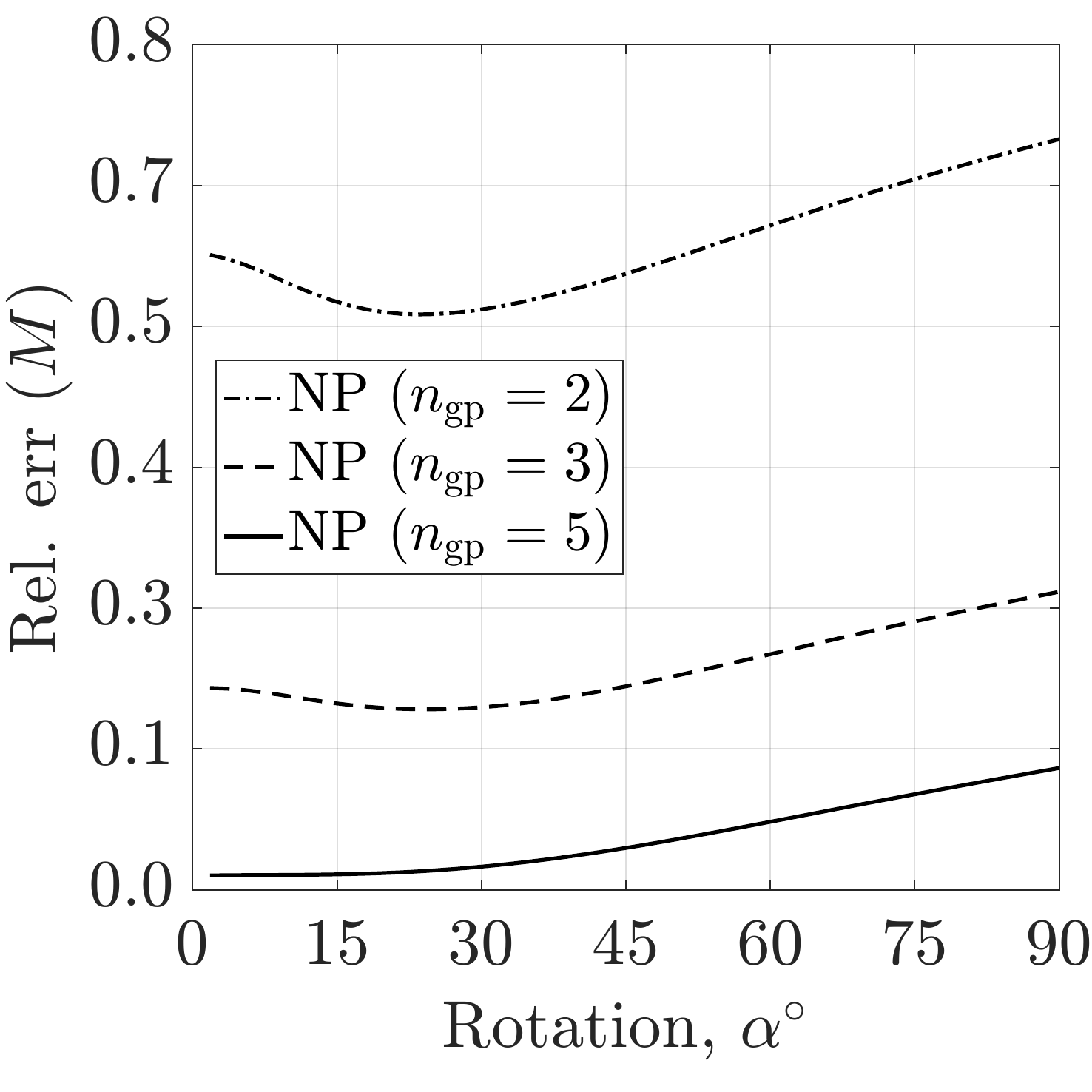}}
\put(7.5,0.3){(b)}
\end{picture}
\caption{Cantilever bending test for the GOH material model (with $\kappa_i = 0.0$ and the compression/tension switch): The corresponding bending moment (a) and its error (b) vs.~the applied rotation for the NP and AP shell model with different number of Gaussian quadrature points $n_\mathrm{gp}$ considered across the thickness.}
\label{f:beam_hgo_k000}
\end{center}
\end{figure}

If the compression/tension switch is included, more Gaussian quadrature points are needed to capture the discontinuity of switch through the shell thickness. For the results shown in Figs.~\ref{f:beam_curve2}.d-\ref{f:beam_curve2}.f, 5 quadrature points are used, which is computationally more expensive compared to the cases that no switch is considered. This issue is further investigated in Fig.~\ref{f:beam_hgo_k000}, which shows how the NP shell model approaches the AP shell model by increasing the number of Gaussian quadrature points. 

\begin{figure}[ht]
\begin{center} \unitlength1cm
%
%
\begin{picture}(15.0,4.0)
\put(-0.5,0.0){\includegraphics[width=0.25\textwidth]{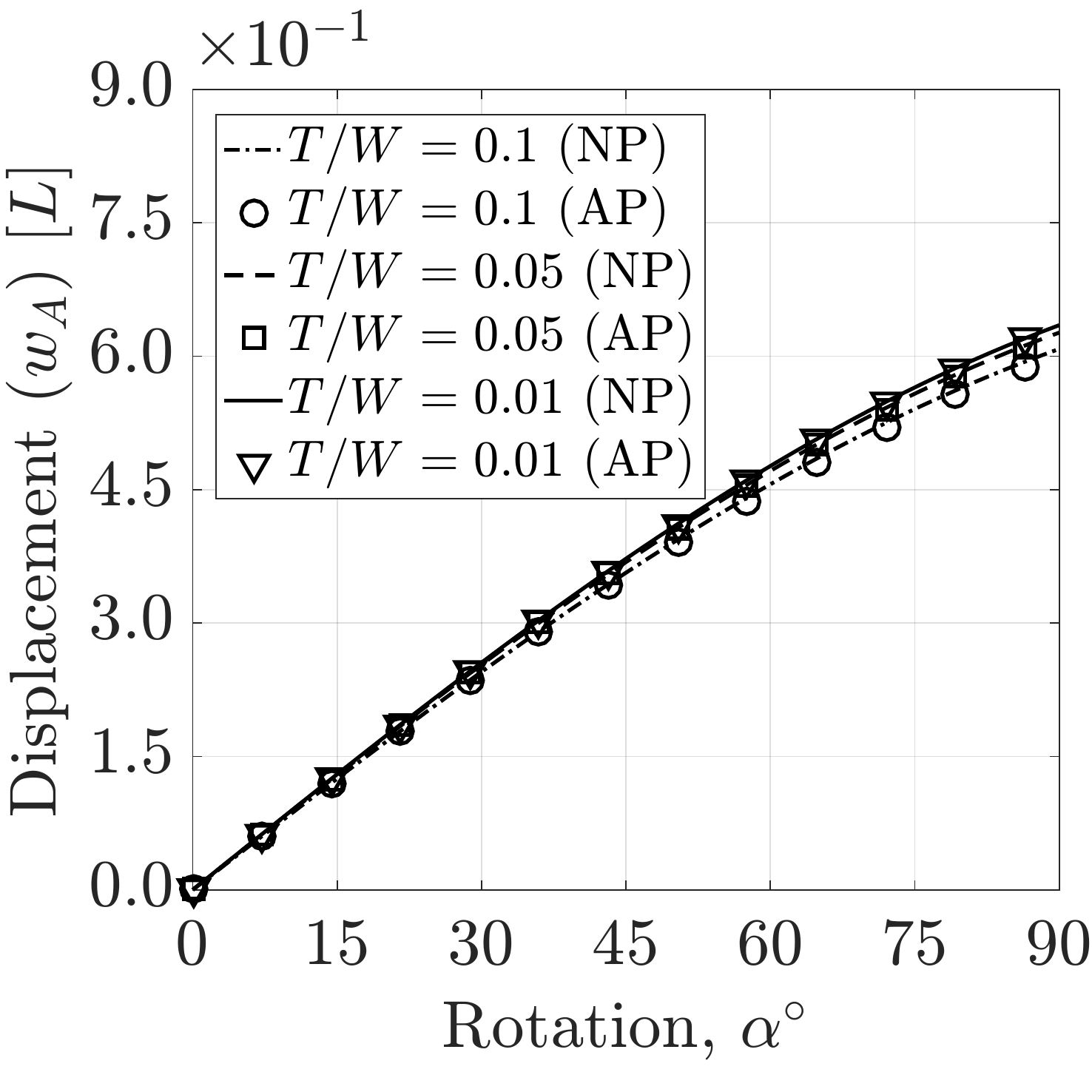}}
\put(-0.1,-0.1){(a)}
\put(3.5,0.0){\includegraphics[width=0.25\textwidth]{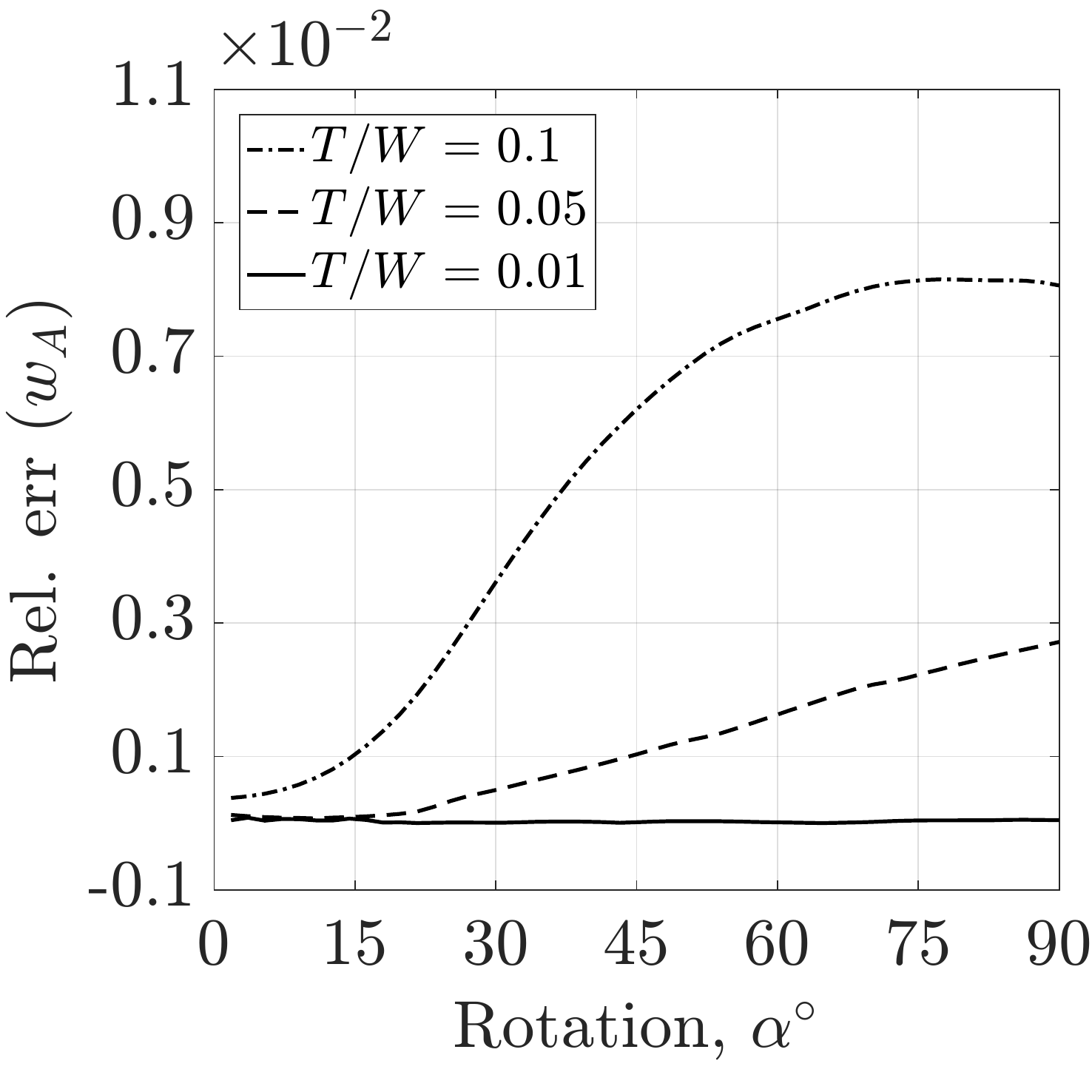}}
\put(3.9,-0.1){(b)}
\put(7.5,0.0){\includegraphics[width=0.25\textwidth]{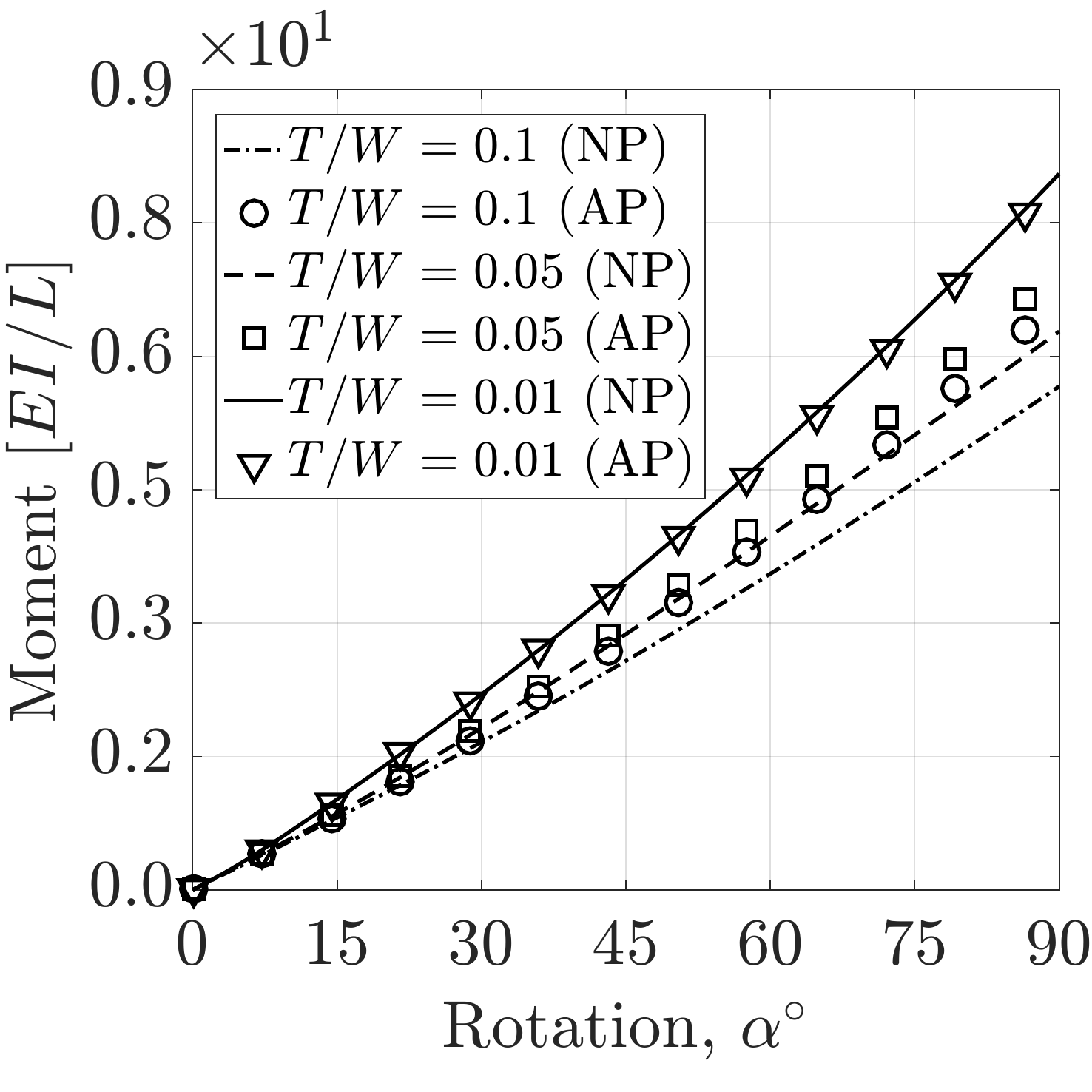}}
\put(7.9,-0.1){(c)}
\put(11.5,0.0){\includegraphics[width=0.25\textwidth]{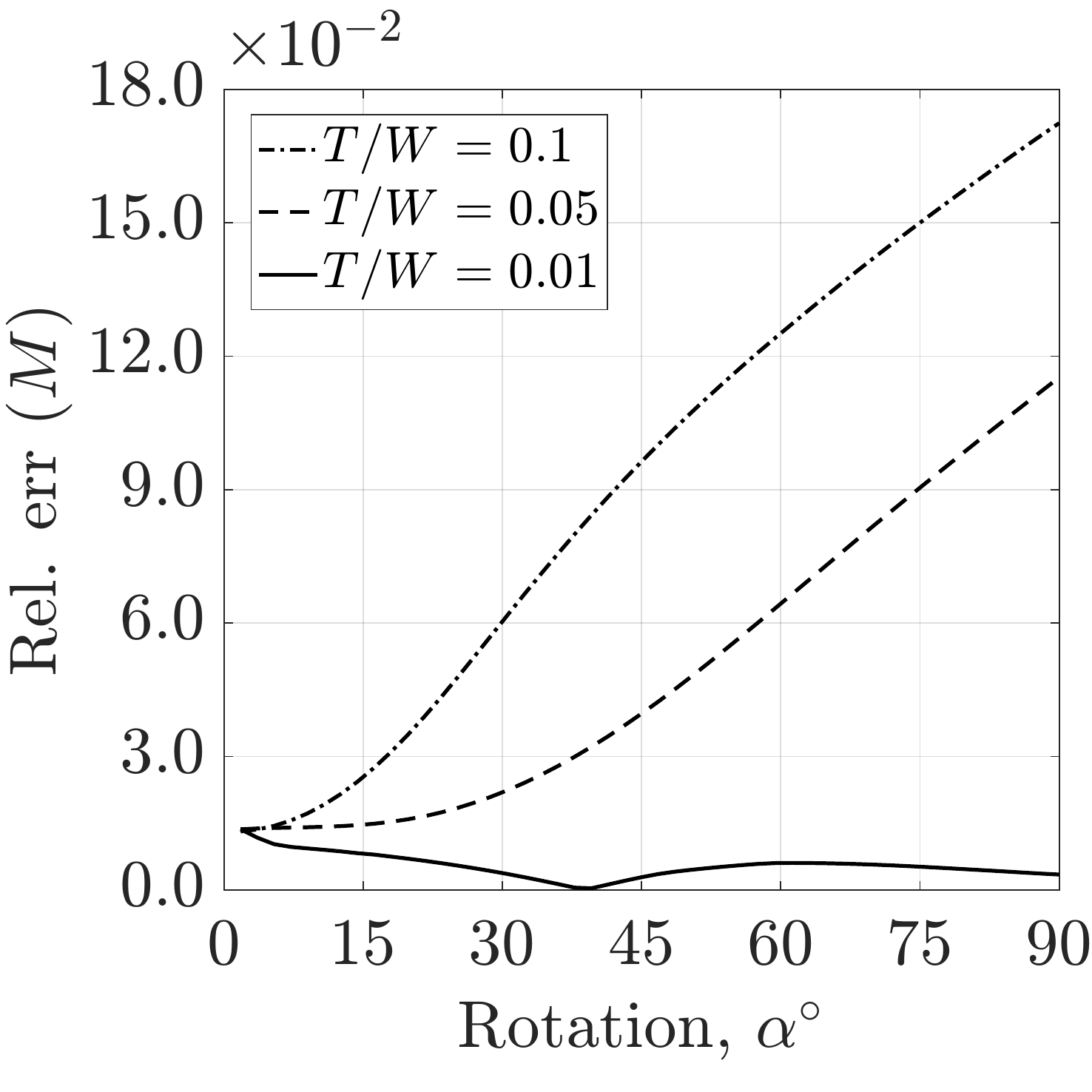}}
\put(11.9,-0.1){(d)}
\end{picture}
\caption{Cantilever bending test for the GOH material model (with $\kappa_i = 0.0$ and the compression/tension switch): Comparison of the NP and AP shell models with different thickness to width ratios ($T/W$).}
\label{f:beam_hgo_wpt}
\end{center}
\end{figure}
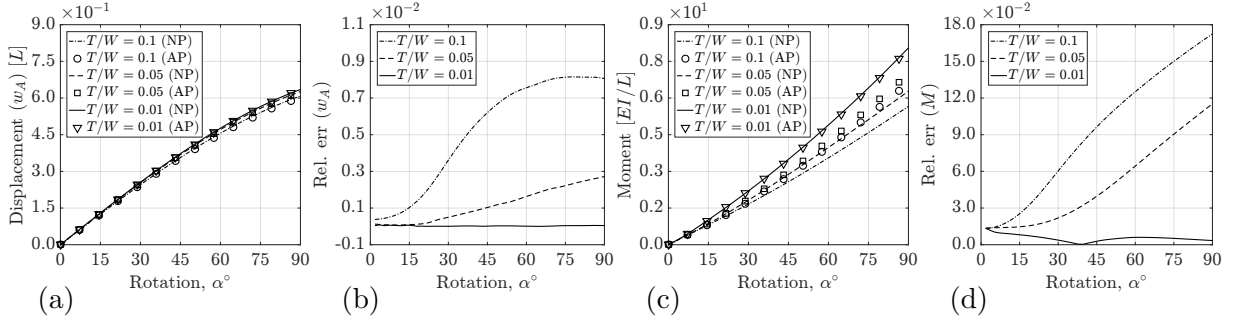

Furthermore, as the shell thickness decreases, the AP shell model becomes more accurate. Fig.~\ref{f:beam_hgo_wpt}.a shows the displacement of the tip versus the applied rotation for different thickness-to-width ($T/W$) ratios. Fig.\ref{f:beam_hgo_wpt}.c shows how the corresponding bending moments change. Here, for all the cases modeled by the NP shell model, 5 Gaussian quadrature points are considered through the shell thickness. Fig.~\ref{f:beam_hgo_wpt}.b~and~\ref{f:beam_hgo_wpt}.d show the corresponding relative errors evaluated w.r.t.~the solution of the NP shell model. As can be observed, the AP shell model becomes inaccurate for thick shells; however, such shells are not covered by the Kirchhoff-Love hypothesis.

\subsection{A clamped plate under pressure}\label{s:plate}
\begin{figure}[H]
\begin{center} \unitlength1cm
\begin{picture}(15.0,5.0)
\put(0.5,0.0){\includegraphics[height=50mm,trim={350px 300px 0 300px},clip]{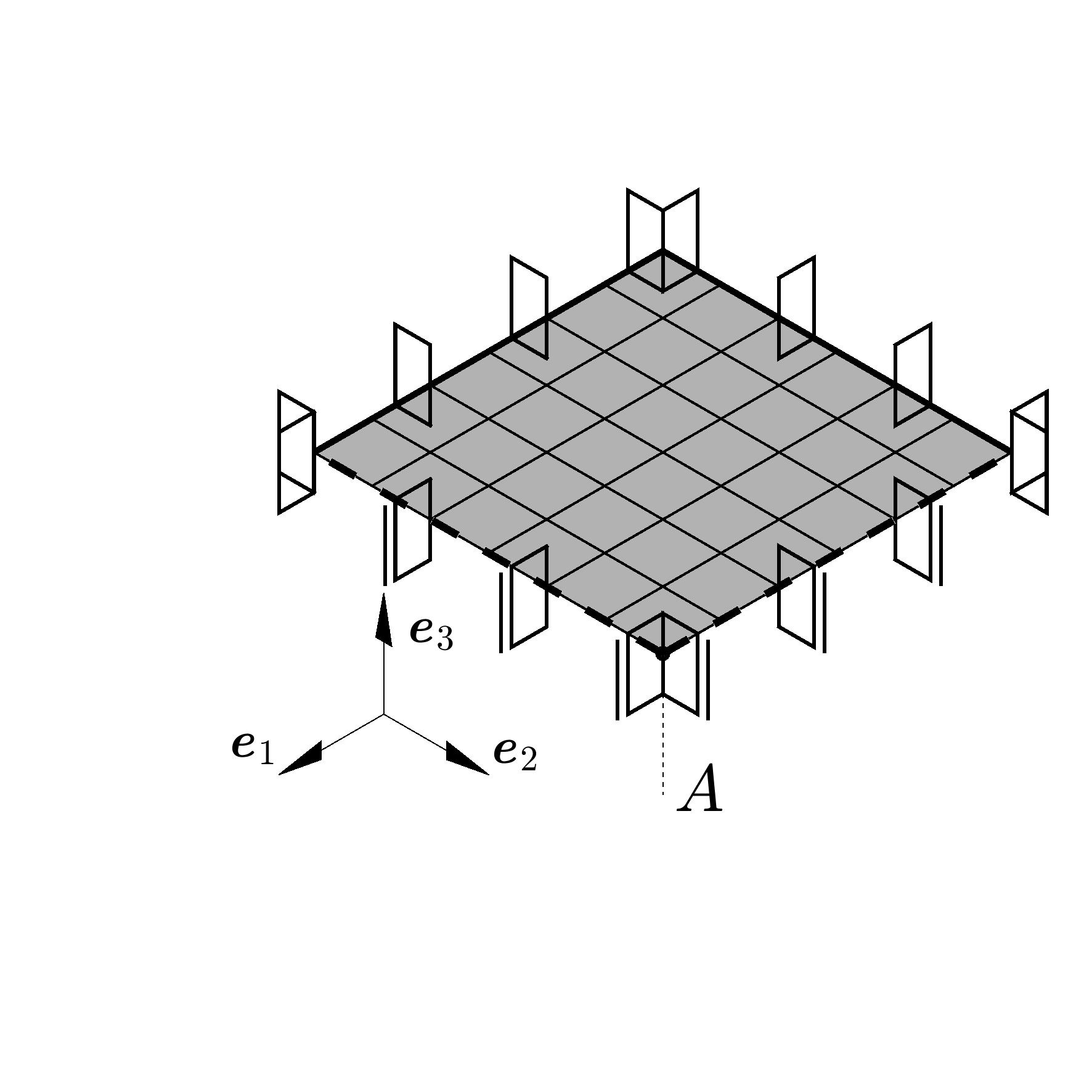}}
\put(0.0,0.5){(a)}
\put(7.5,0.0){\includegraphics[height=50mm,trim={0 250px 0 300px},clip]{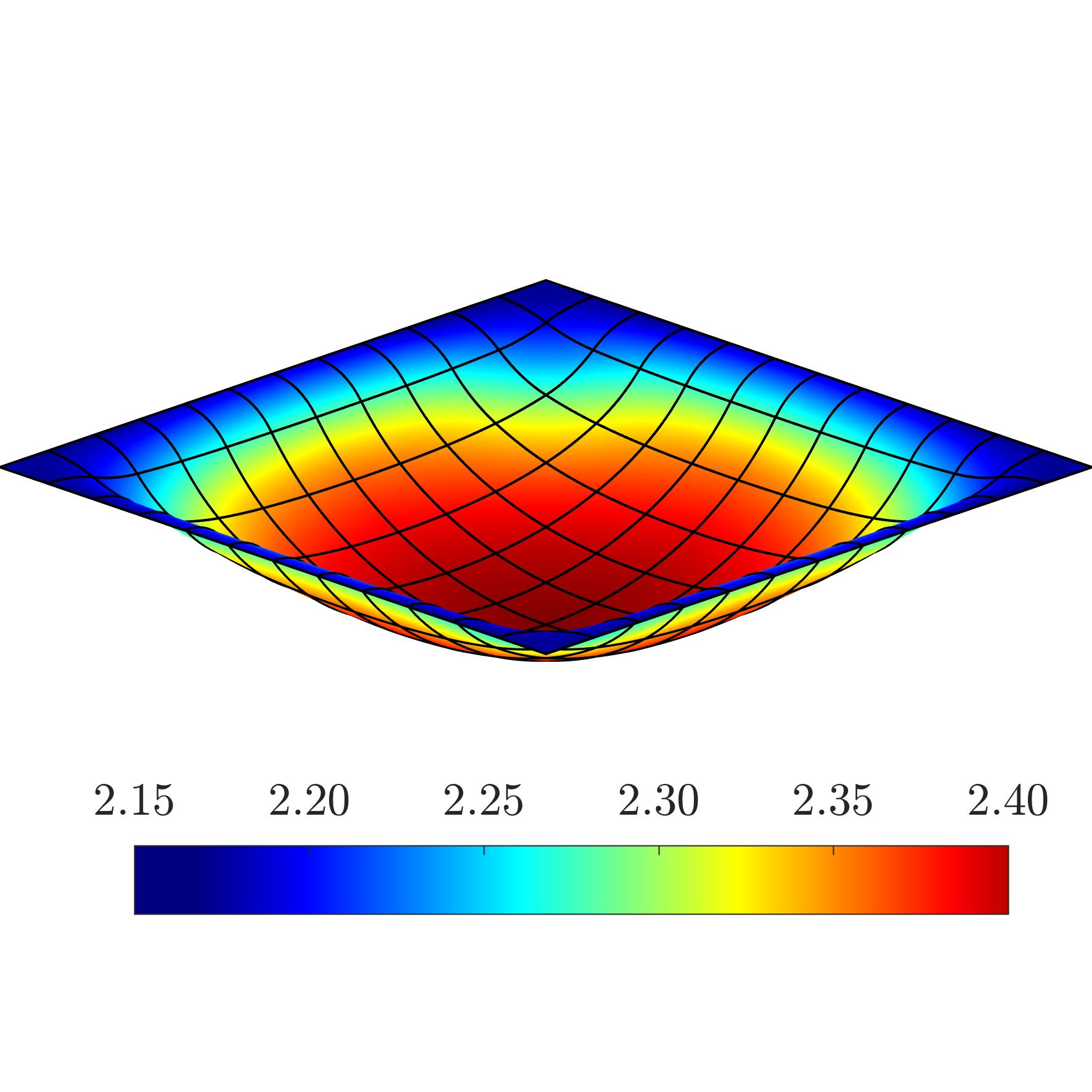}}
\put(7.0,0.5){(b)}
\end{picture}
\caption{Clamped plate under pressure: (a) Reference configuration (1/4 system) with boundary conditions and (b) deformed configuration (full system) for the GOH model (with $\kappa = 1/3$ and the compression/tension switch) colored by $I_1 := \tr \bC$.}
\label{f:plate_undeformed}
\end{center}
\end{figure}

Large deformation of a clamped plate under live pressure is a challenging example. Such an example is used here to compare the capabilities of the three introduced shell models to capture the membrane and bending forces together. As shown in Fig.~\ref{f:plate_undeformed}.a, a square plate, with $T \times L \times L = 0.25 \times 10  \times 10~[\mathrm{mm}^3]$, is clamped with appropriate boundary conditions. As the problem is symmetric, only $1/4$ of the whole system is modeled and symmetry constraints are applied along the corresponding boundaries. On the clamped and symmetry edges, the rotations are fixed following the constraint formulation of \citet{solidshell}. The plate quarter is meshed by $6 \times 6$ quadratic NURBS-based elements. Furthermore, for both the anisotropic material models, the fibers are oriented according to Eq.~\eqref{e:tbL_i} with $\theta_1,\theta_2 = \pm 45^\circ $. 

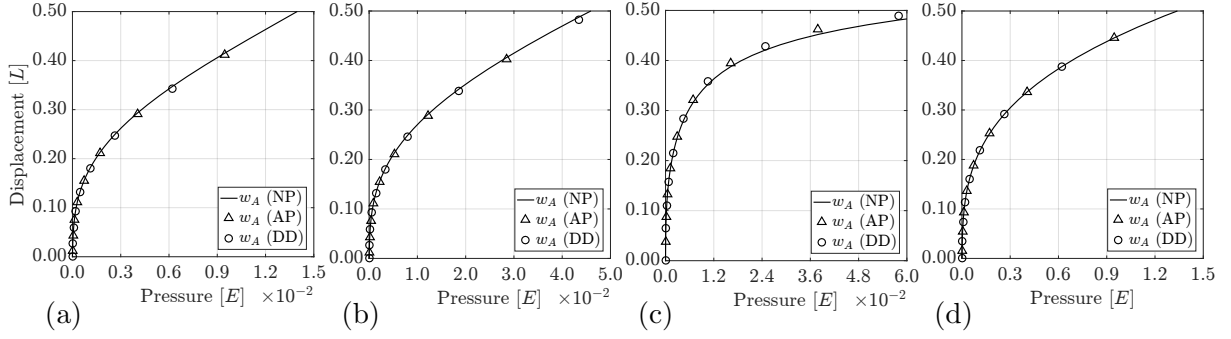
\begin{figure}[ht]
\begin{center} \unitlength1cm
%
%
\begin{picture}(15.0,4.2)
\put(-0.5,0.0){\includegraphics[height=42mm]{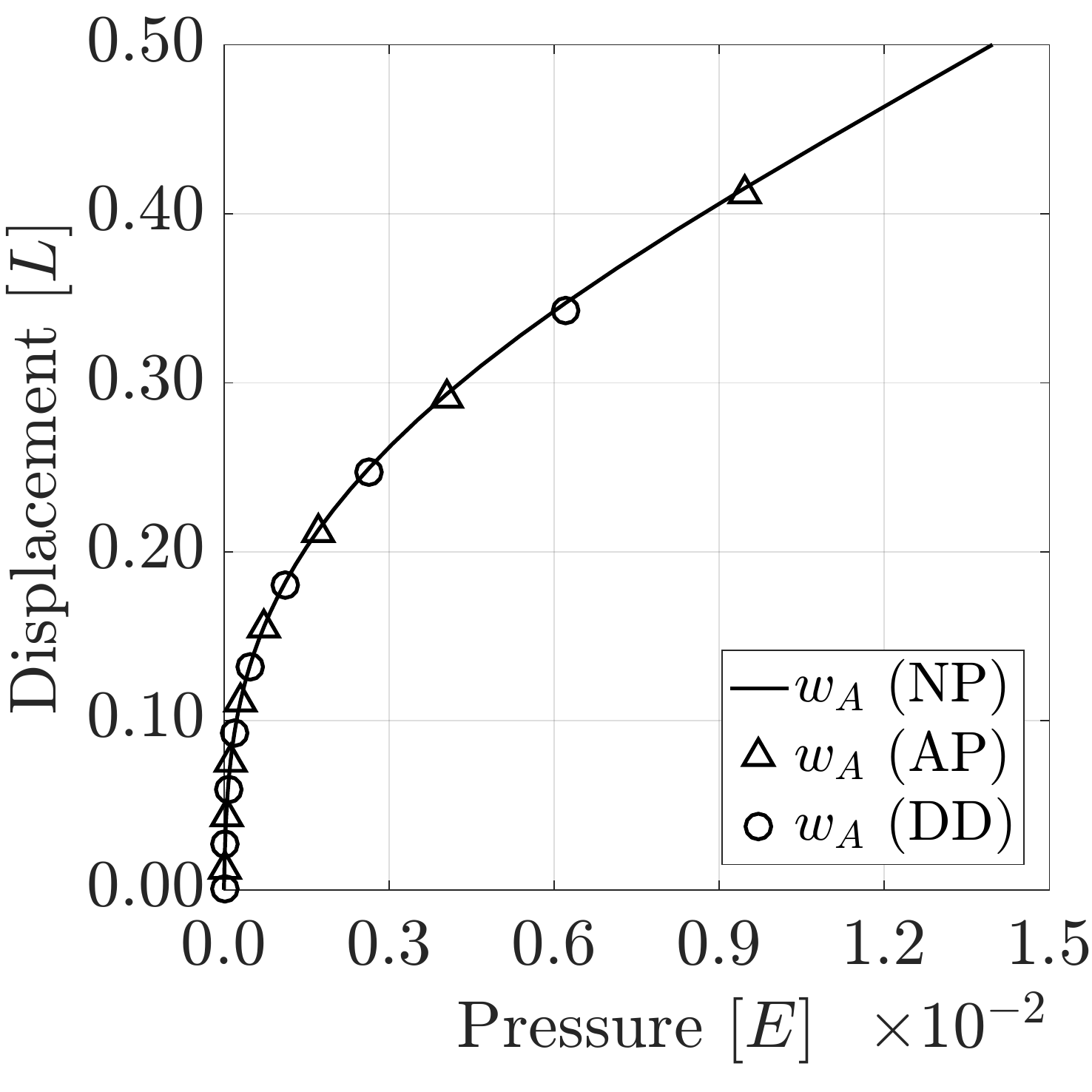}}
\put(0.0,-0.1){(a)}
\put(3.7,0.0){\includegraphics[height=42mm,trim={30px 0px 0px 0px},clip]{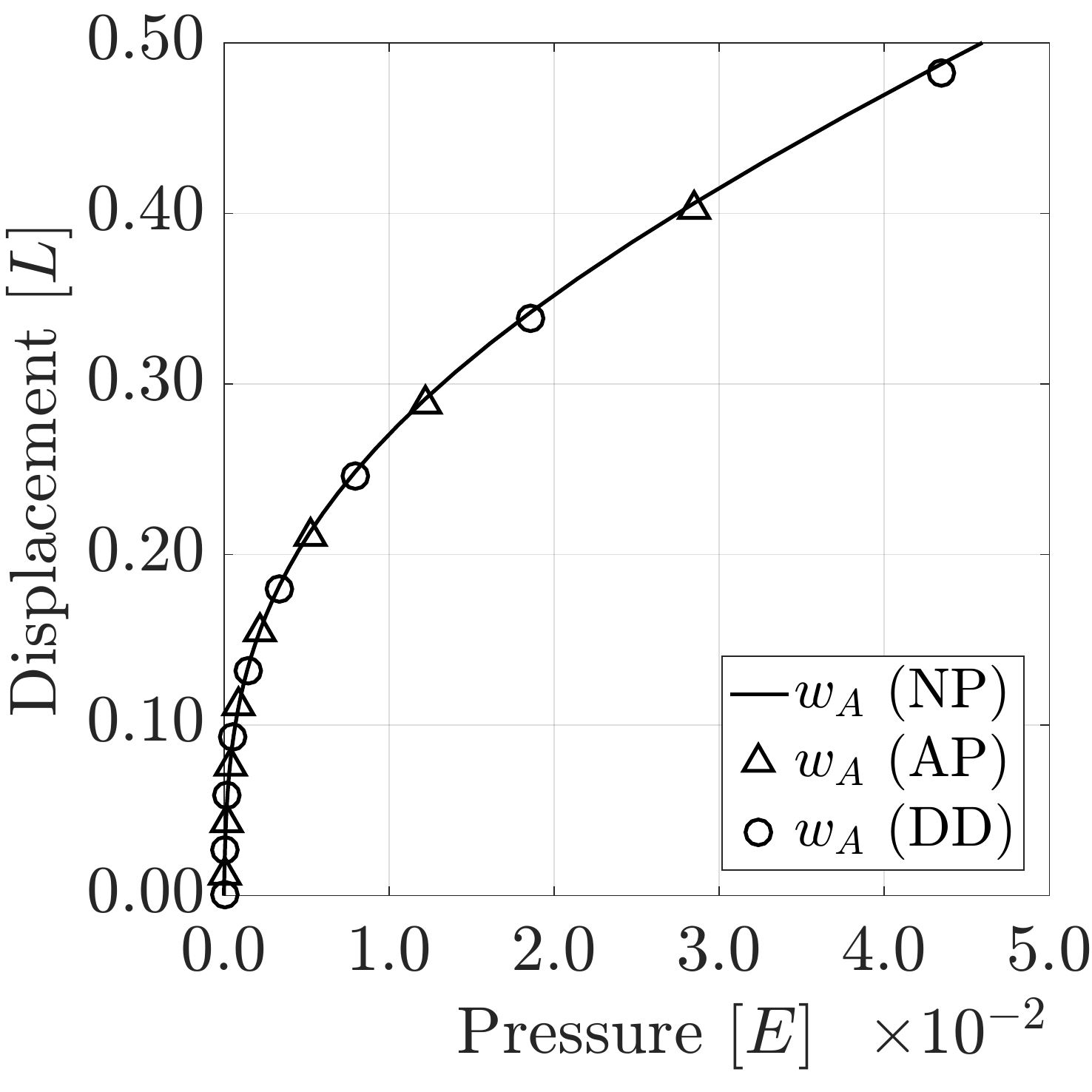}}
\put(3.9,-0.1){(b)}
\put(7.6,0.0){\includegraphics[height=42mm,trim={30px 0px 0px 0px},clip]{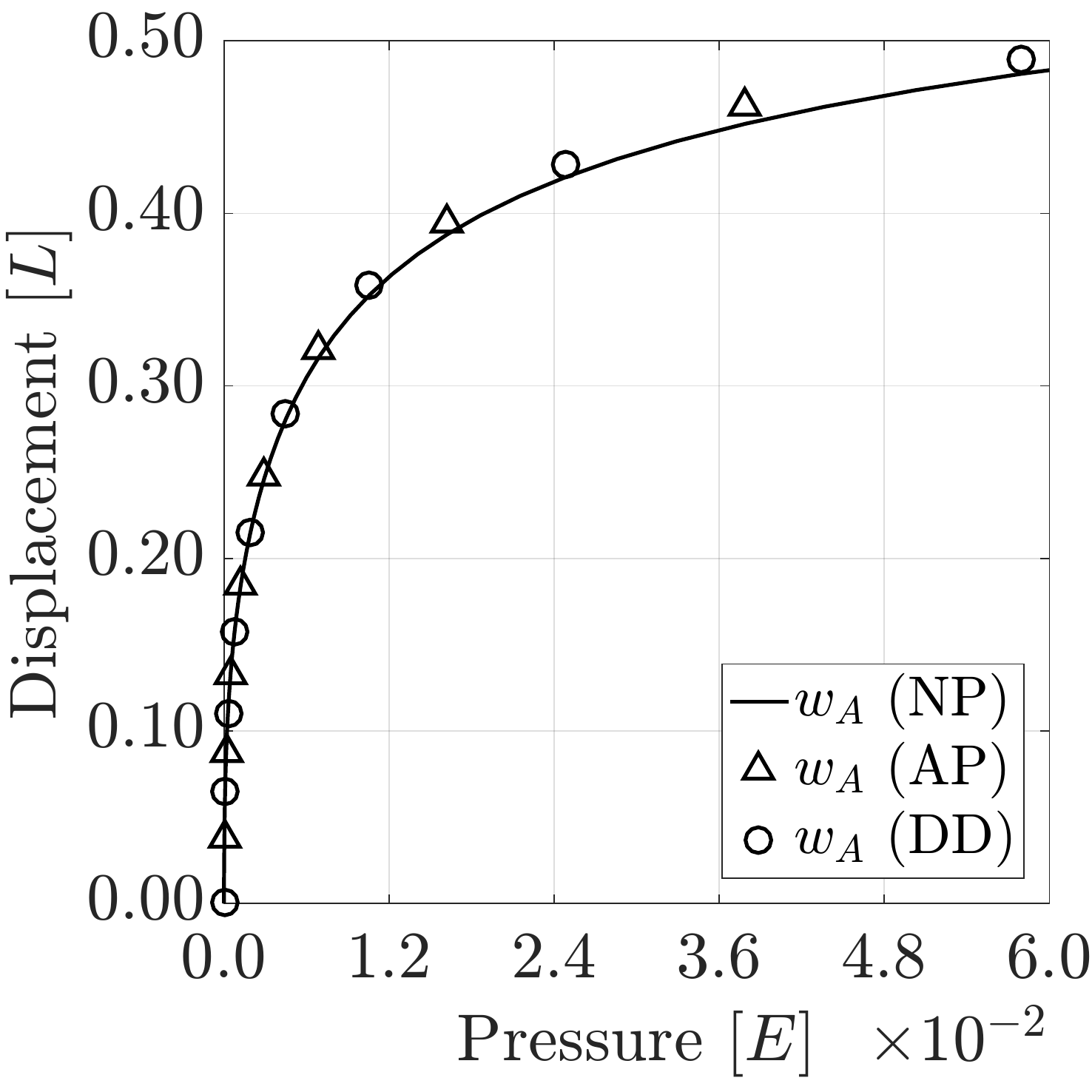}}
\put(7.8,-0.1){(c)}
\put(11.5,0.0){\includegraphics[height=42mm,trim={30px 0px 0px 0px},clip]{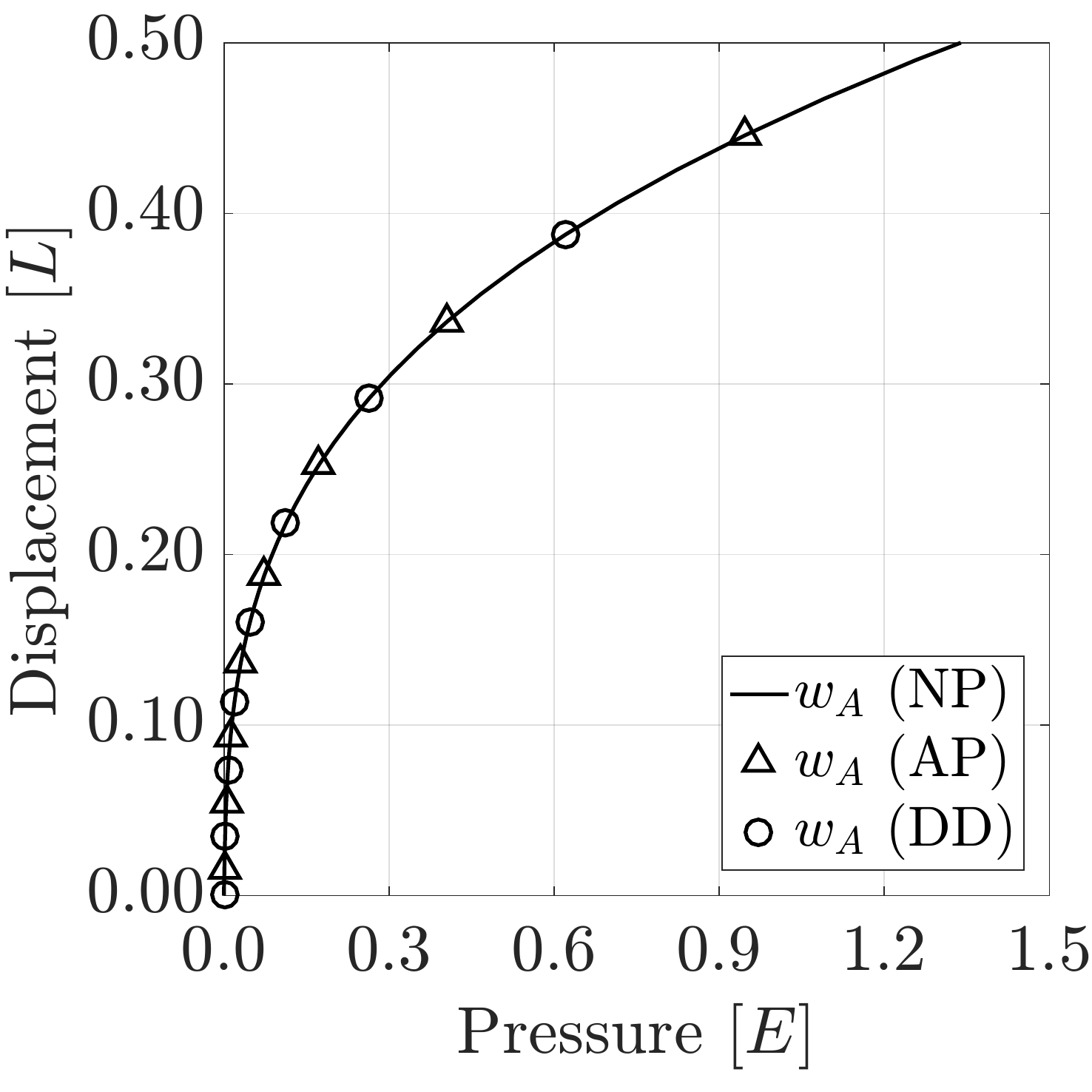}}
\put(11.7,-0.1){(d)}
\end{picture}
\caption{Clamped plate under pressure -- the displacement of the middle point vs.~the applied pressure:  (a) NH, (b) MR, (c) Fung and (d) AMR material model for the three constitutive approaches (NP, AP and DD) presented in Sec.~\ref{s:shells}.}
\label{f:plate_curve_1}
\end{center}
\end{figure}

\begin{figure}[H]
\begin{center} \unitlength1cm
%
%
\begin{picture}(15.0,8.4)
\put(0.0,4.2){\includegraphics[height=42mm]{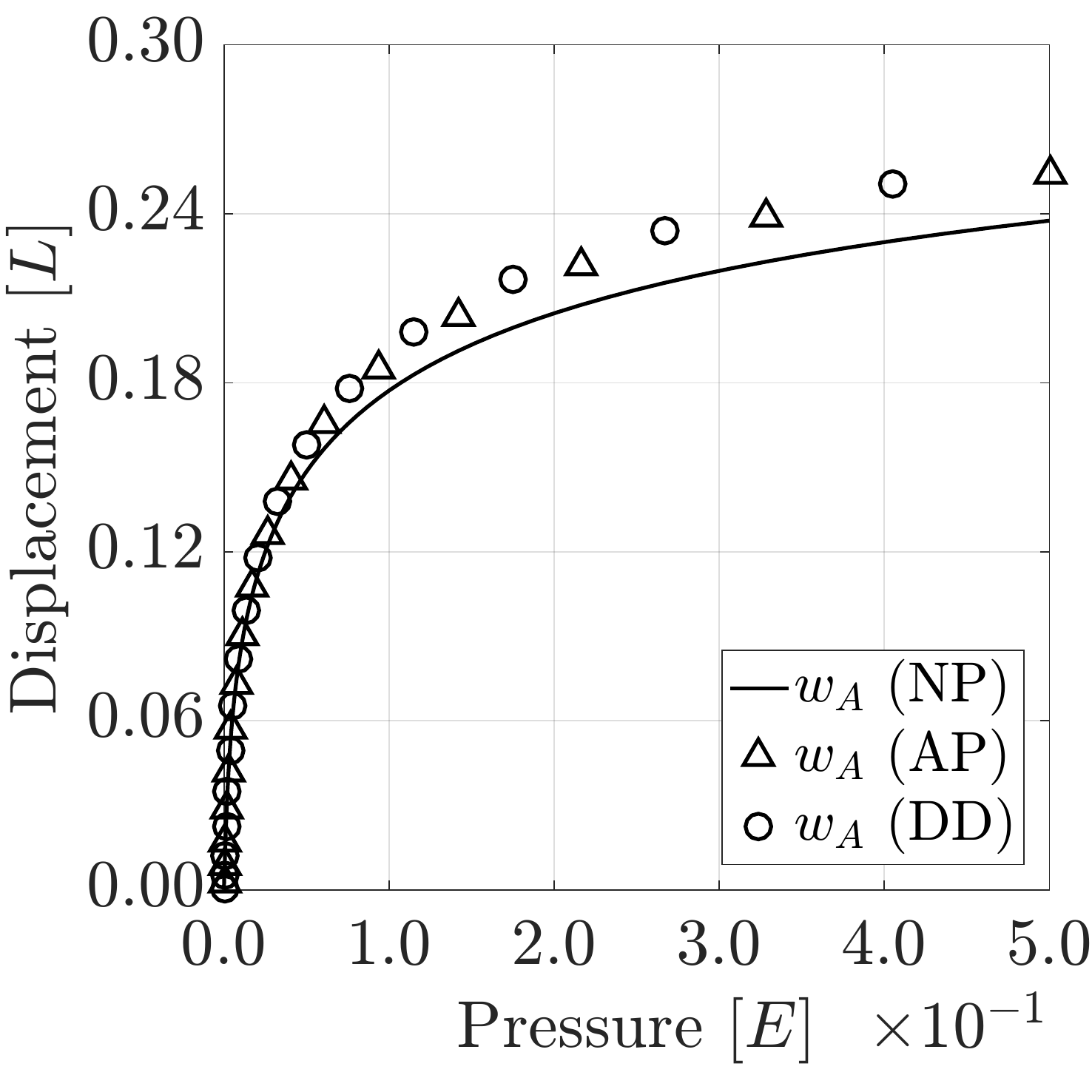}}
\put(-0.5,4.7){(a)}
\put(5.3,4.2){\includegraphics[height=42mm]{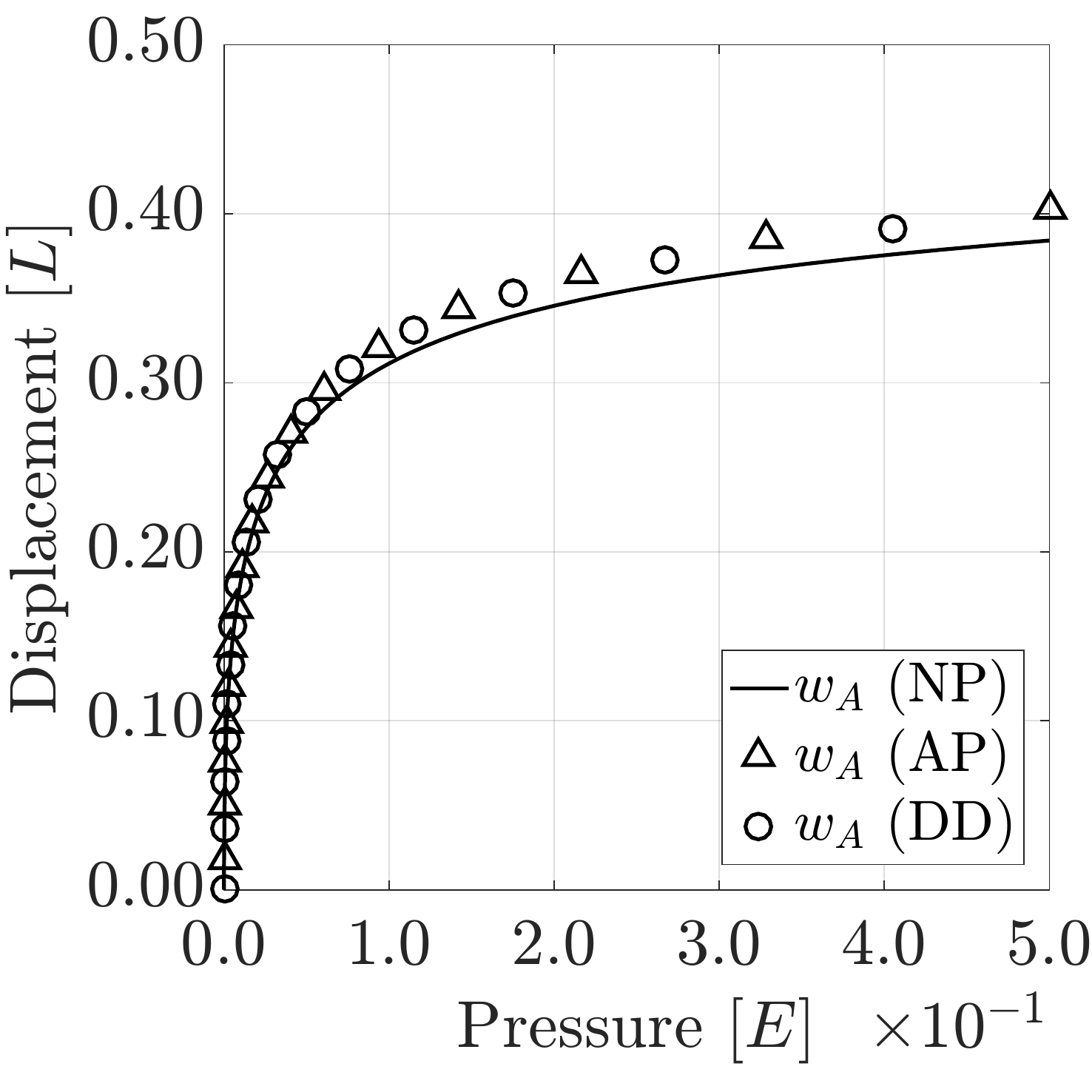}}
\put(4.8,4.7){(b)}
\put(10.5,4.2){\includegraphics[height=42mm]{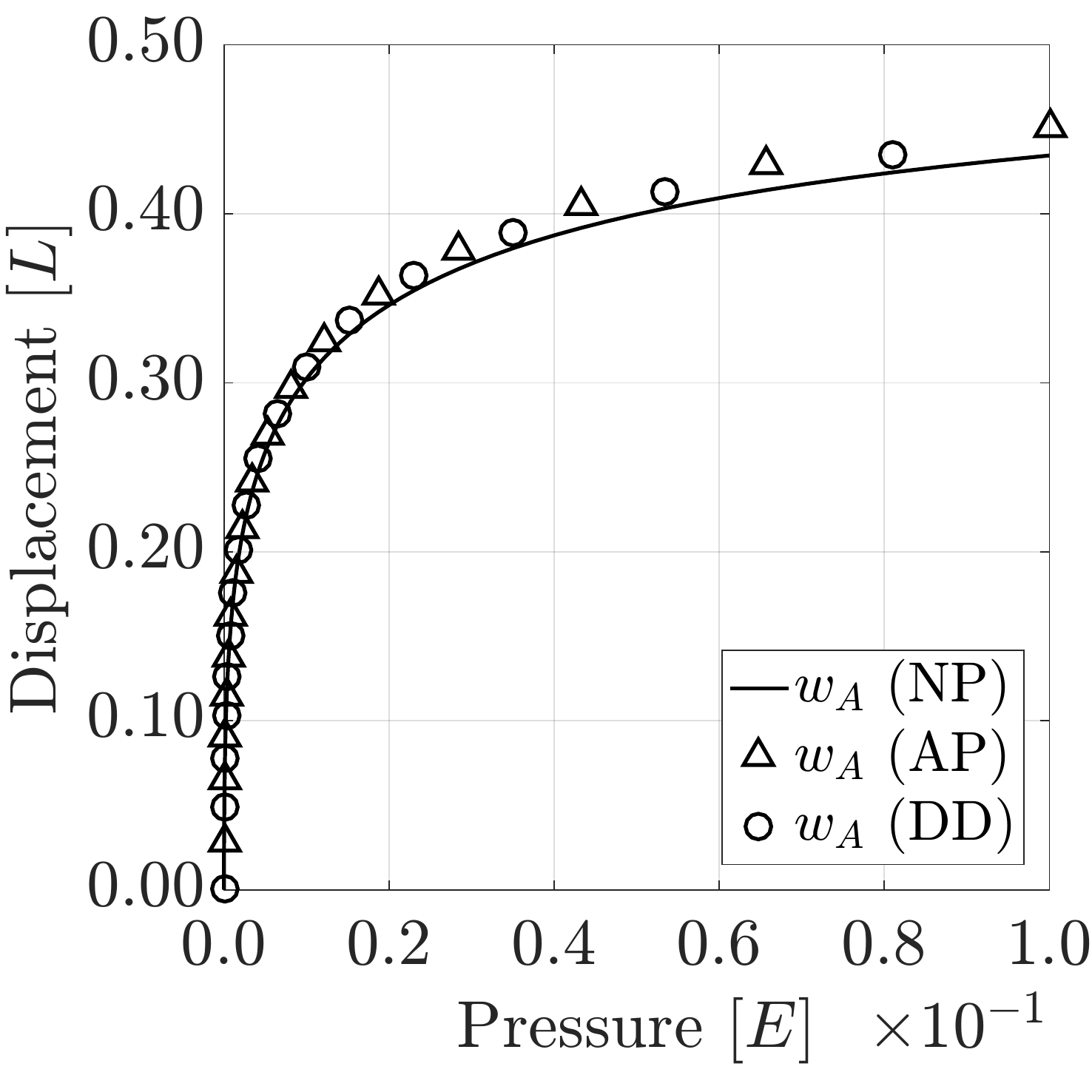}}
\put(10.0,4.7){(c)}
\put(0.0,0.0){\includegraphics[height=42mm]{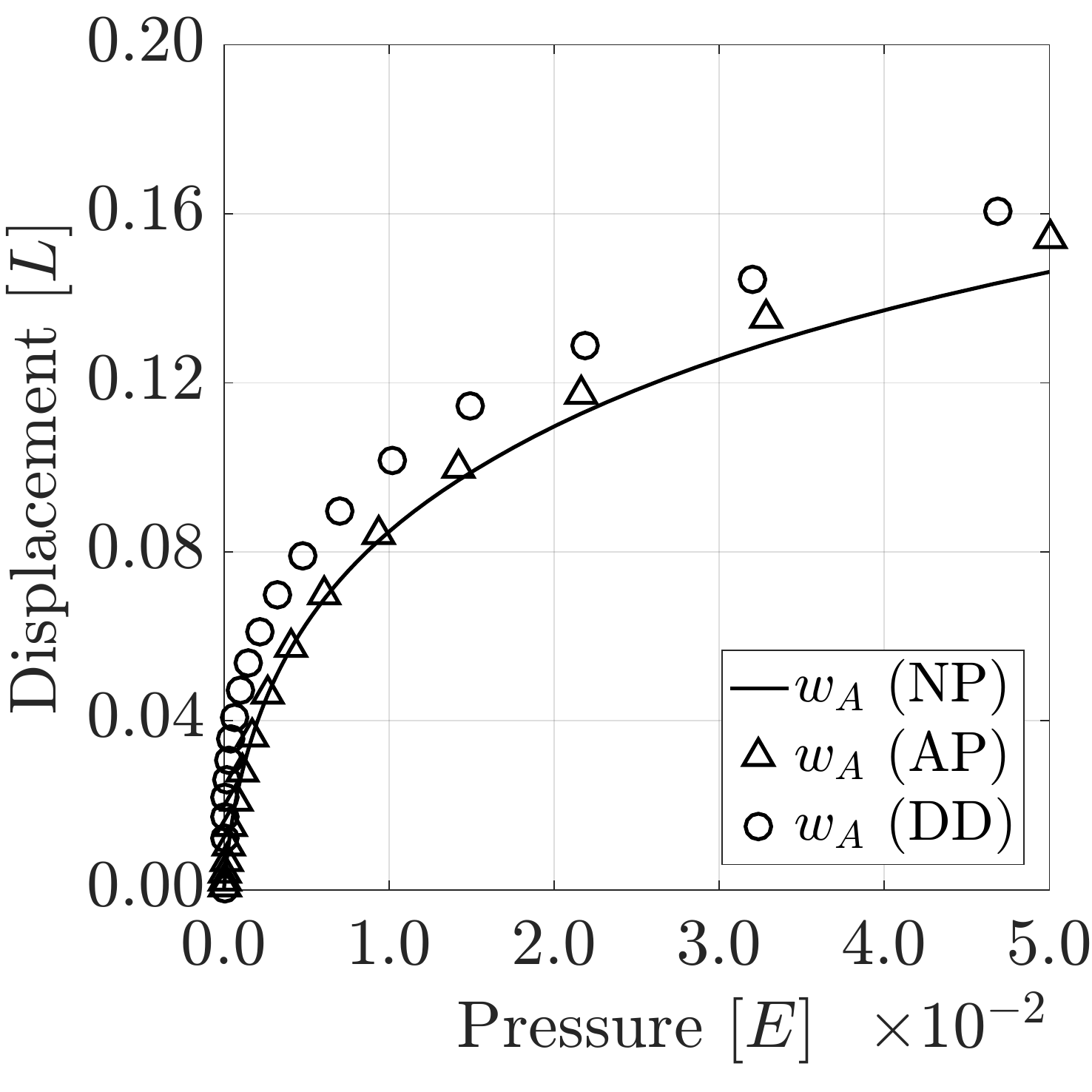}}
\put(-0.5,0.5){(d)}
\put(5.3,0.0){\includegraphics[height=42mm]{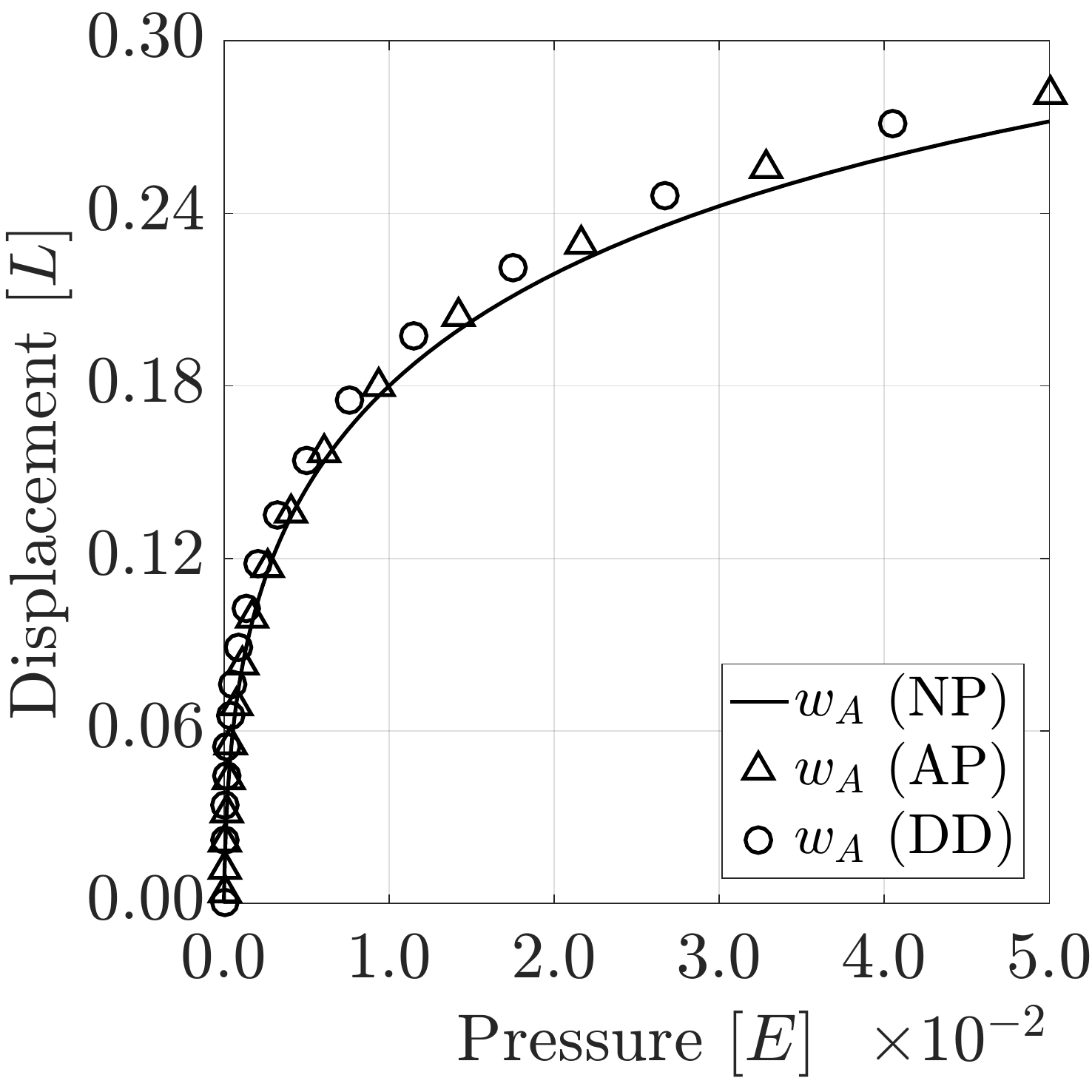}}
\put(4.8,0.5){(e)}
\put(10.5,0.0){\includegraphics[height=42mm]{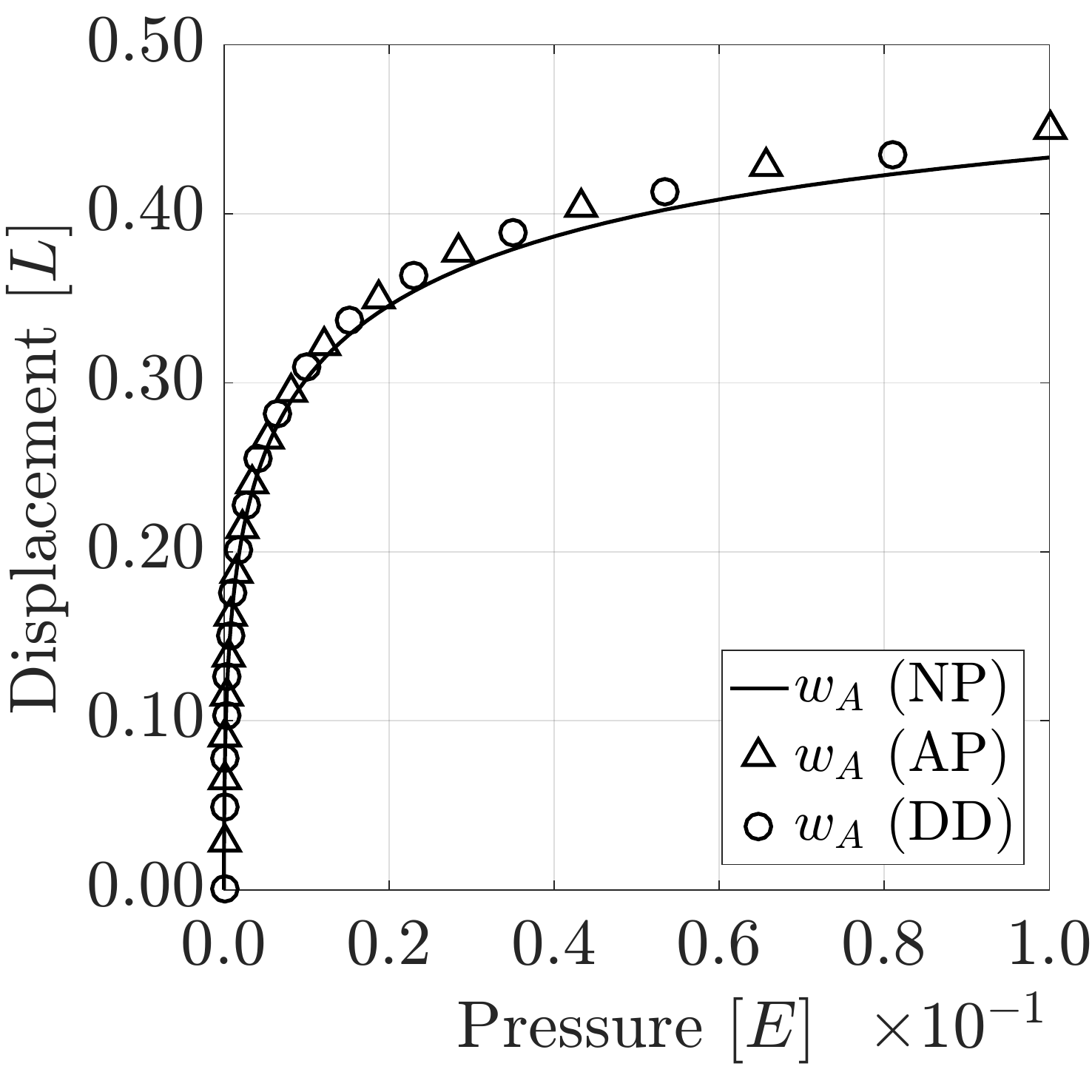}}
\put(10.0,0.5){(f)}
\end{picture}
\caption{Clamped plate under pressure for the GOH material model -- the displacement of the middle point vs.~the applied pressure for the three constitutive approaches NP, AP and DD: (a-c) without and (d-f) with the compression/tension switch. Further, for (a) and (d) $\kappa_i = 0.0$, for (b) and (e) $\kappa_i = 0.226$ and for (c) and (f) $\kappa_i = 1/3$.}
\label{f:plate_curve_2}
\end{center}
\end{figure}
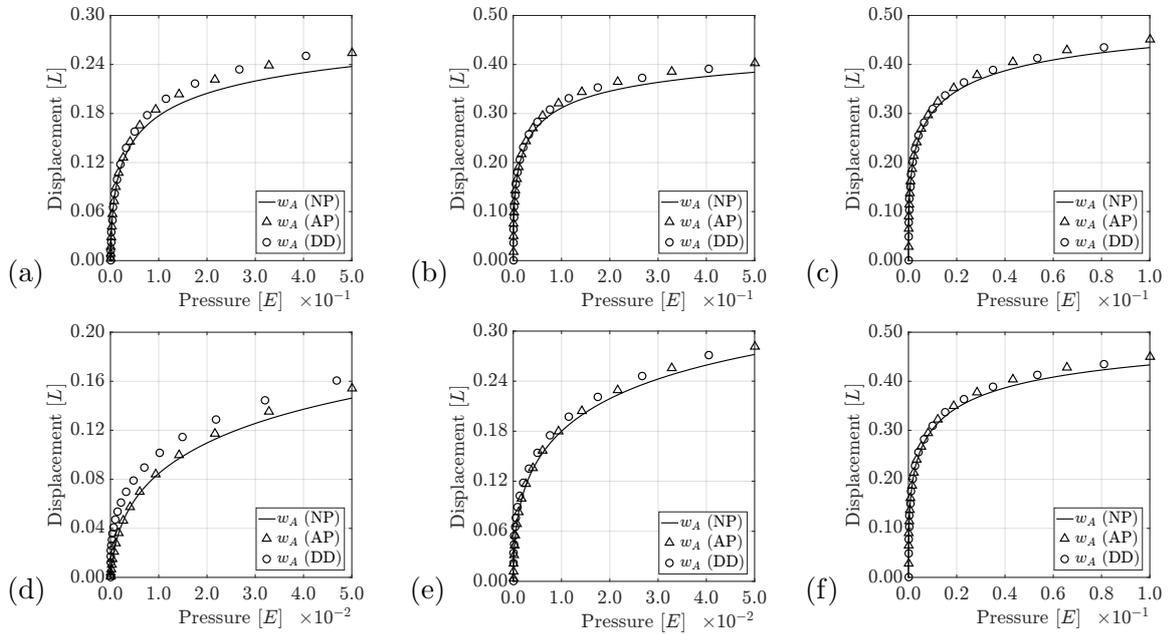

Figs.~\ref{f:plate_curve_1}~and~\ref{f:plate_curve_2} represent the deflection of the mid point A (shown in Fig.~\ref{f:plate_undeformed}.a) under the applied live pressure. As expected, all the three presented shell models predict similar displacements. Further, the results for the Gasser--Ogden--Holzapfel material model with and without the compression/tension switch are shown in Figs.~\ref{f:plate_curve_2}.a-\ref{f:plate_curve_2}.c and Figs.~\ref{f:plate_curve_2}.d-\ref{f:plate_curve_2}.f, respectively. Here, for the cases modeled by the NP shell model, 3 Gaussian quadrature points are considered across the thickness if the compression/tension switch is excluded and 5 Gaussian quadrature points are used if the switch is included.

\subsection{Indentation of a sheet}\label{s:indent}
In vitro and in silico indentation tests are widely used to empirically and numerically determine the mechanical characteristics of soft tissues \citep{zhang97,liu04indentation,choi05estimation,mckee11,lu12}. For instance, puncture testing has been applied frequently for the mechanical characterization of the human fetal membrane tissue \citep{burzle14puncture}.

\begin{figure}[ht]
\begin{center} \unitlength1cm
\begin{picture}(15,9.5)
\put(0.0,4.5){\includegraphics[height=50mm,trim={350px 300px 0 300px},clip]{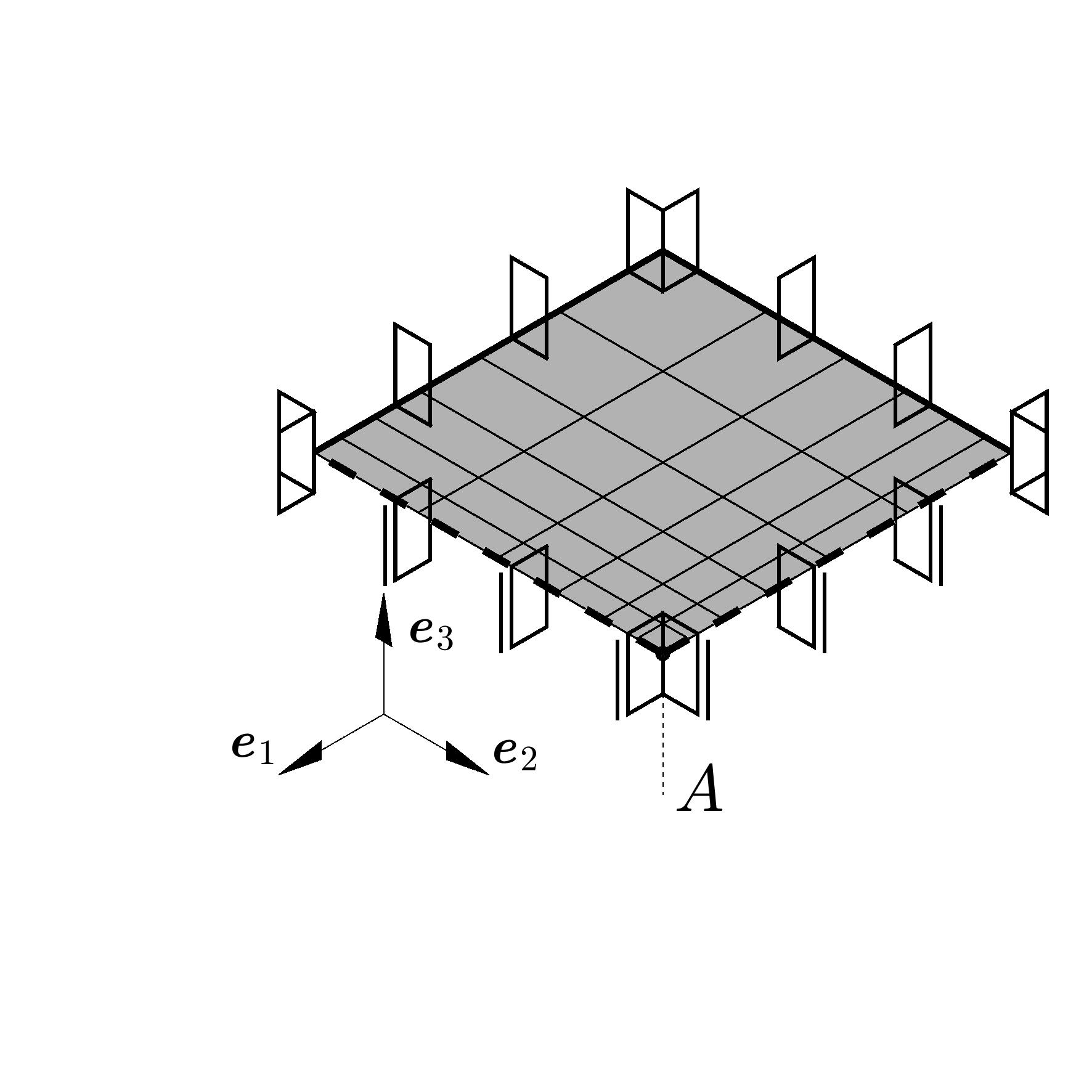}}
\put(6.2,5.5){\includegraphics[width=90mm,trim={0 500px 0 550px},clip]{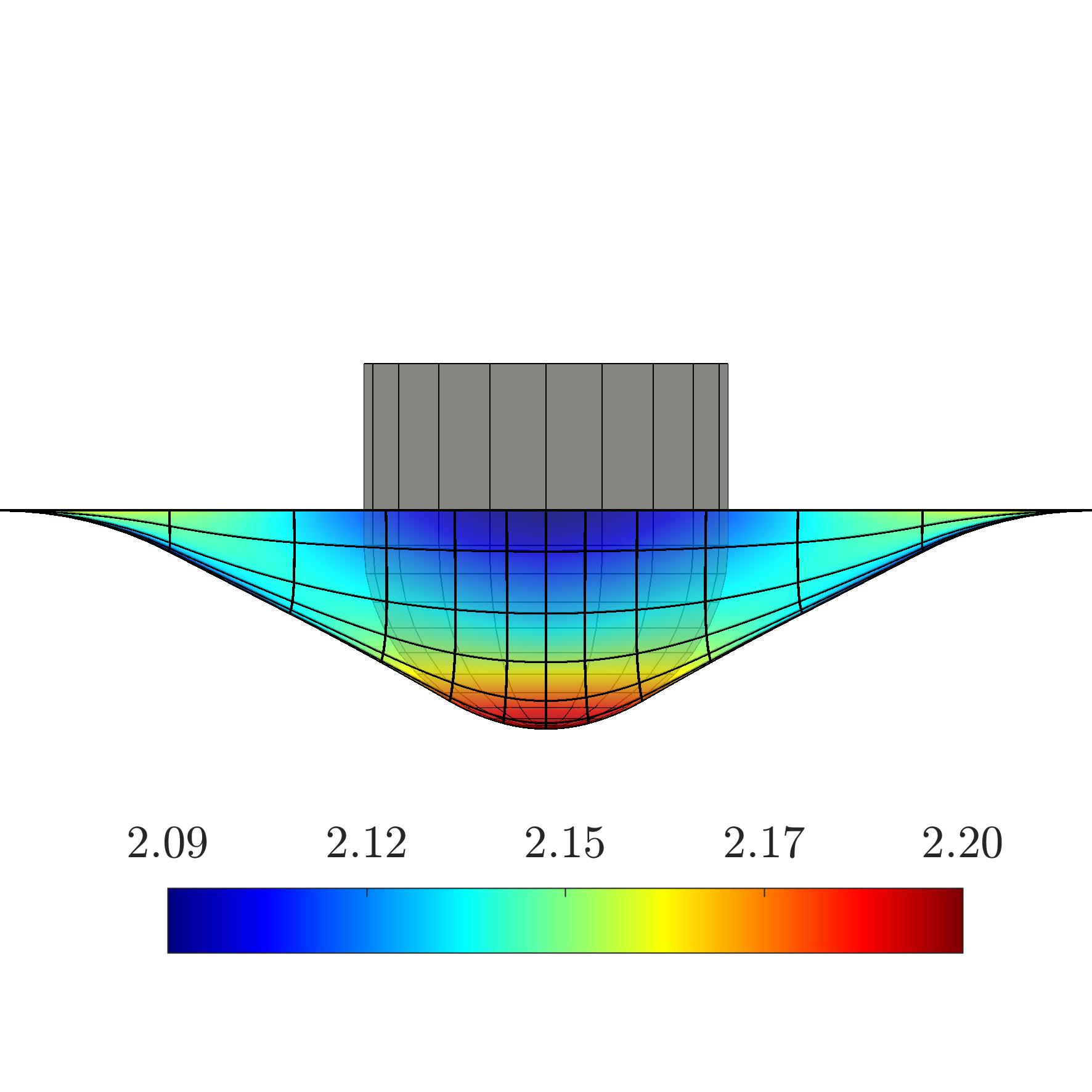}}
\put(0.0,0.0){\includegraphics[height=50mm,trim={350px 300px 0 300px},clip]{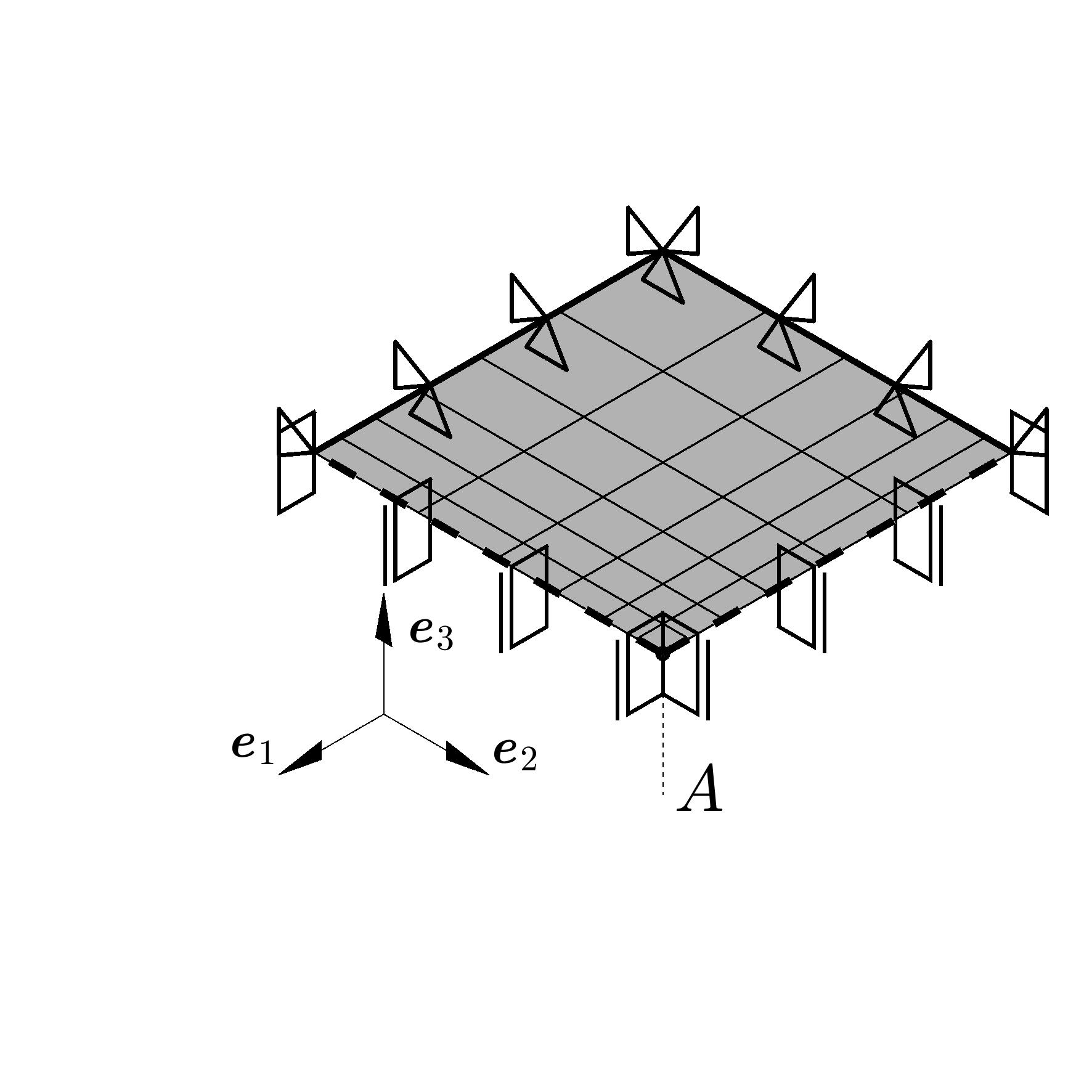}}
\put(6.2,0.0){\includegraphics[width=90mm,trim={0 200px 0 550px},clip]{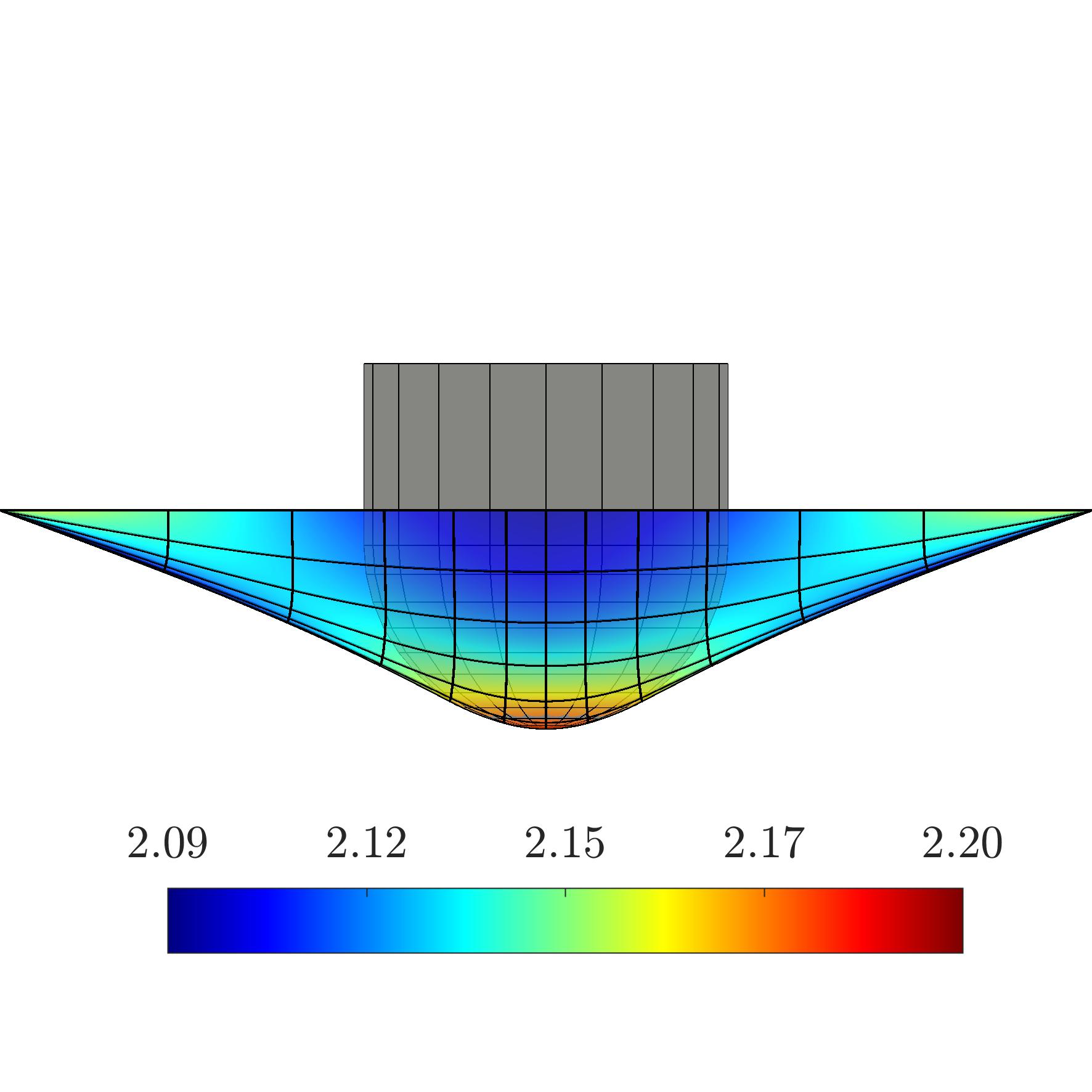}}
\put(-0.5,6.0){(a)}
\put( 6.0,6.0){(c)}
\put(-0.5,1.5){(b)}
\put( 6.0,1.5){(d)}
\end{picture}
\caption{Indentation test: Reference configuration (quarter system) with (a) clamped and (b) fixed outer edges. Deformed configuration (full system) with (c) clamped and (d) fixed outer edges for the GOH model (with the compression/tension switch and $\kappa_i = 0.226$) colored by $I_1 := \tr \bC$.}
\label{f:indent_1}
\end{center}
\end{figure}
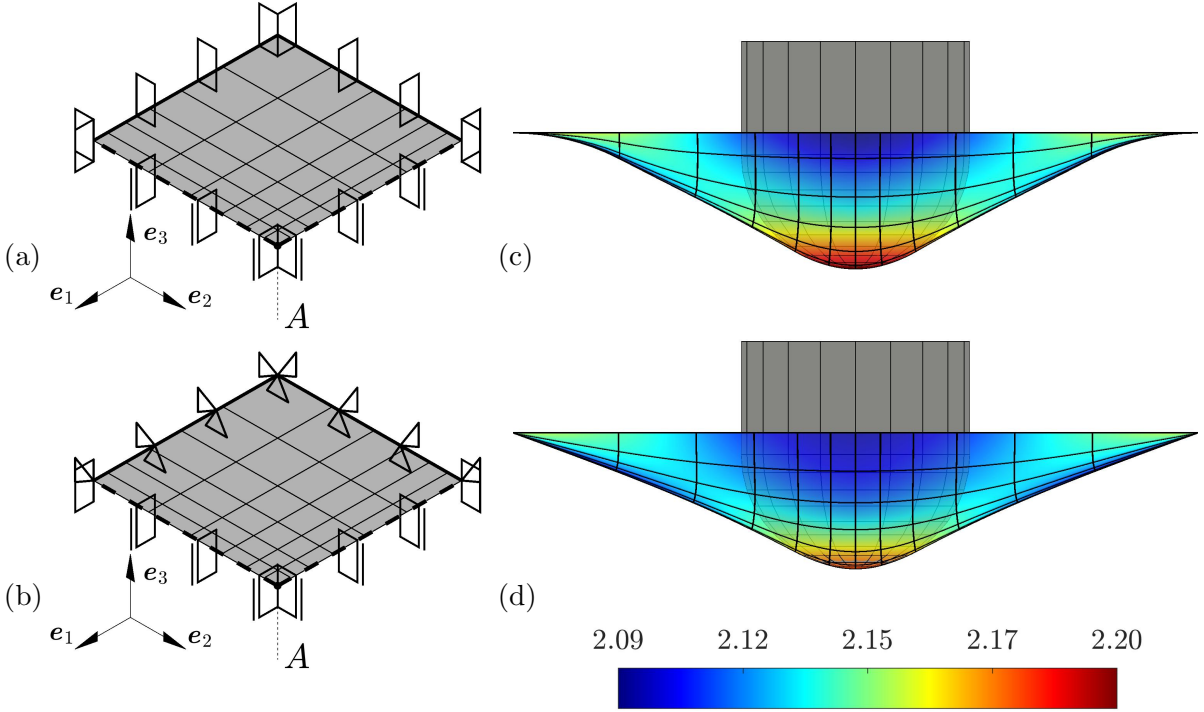

Here, the indentation of a square sheet is simulated. The sheet has the same dimensions and material properties as the plate of Sec.~\ref{s:plate}. As shown in Figs.~\ref{f:indent_1}.a~and~\ref{f:indent_1}.b, two types of boundary conditions are considered, i.e.~the outer edges are either fixed or clamped. The sheet is pressed by an indenter with a rigid spherical cap (see Figs.~\ref{f:indent_1}.c~and~\ref{f:indent_1}.d). The indenter radius is $R=L/6$, where $L$ is the width of sheet. Here, the sheet is meshed by $6 \times 6$ quadratic NURBS-based elements. In the contact area, the mesh is finer. The size of the finest element is $1/4$ of the coarsest one. The sheet constited of the GOH material model with the constants given in Tab.~\ref{t:const}. Following \citet{spbf}, the contact computations is based on an unbiased  penalty formulation applied at the quadrature points of the isogeometric finite elements. The penalty parameter is set to $\epsilon_\mrc = 10^8\,E\,T$, where $E = 3\,\tmu$.   

\begin{figure}[H]
\begin{center} \unitlength1cm
\begin{picture}(15.0,4.2)
\put(2.6,-0.2){\includegraphics[height=42mm]{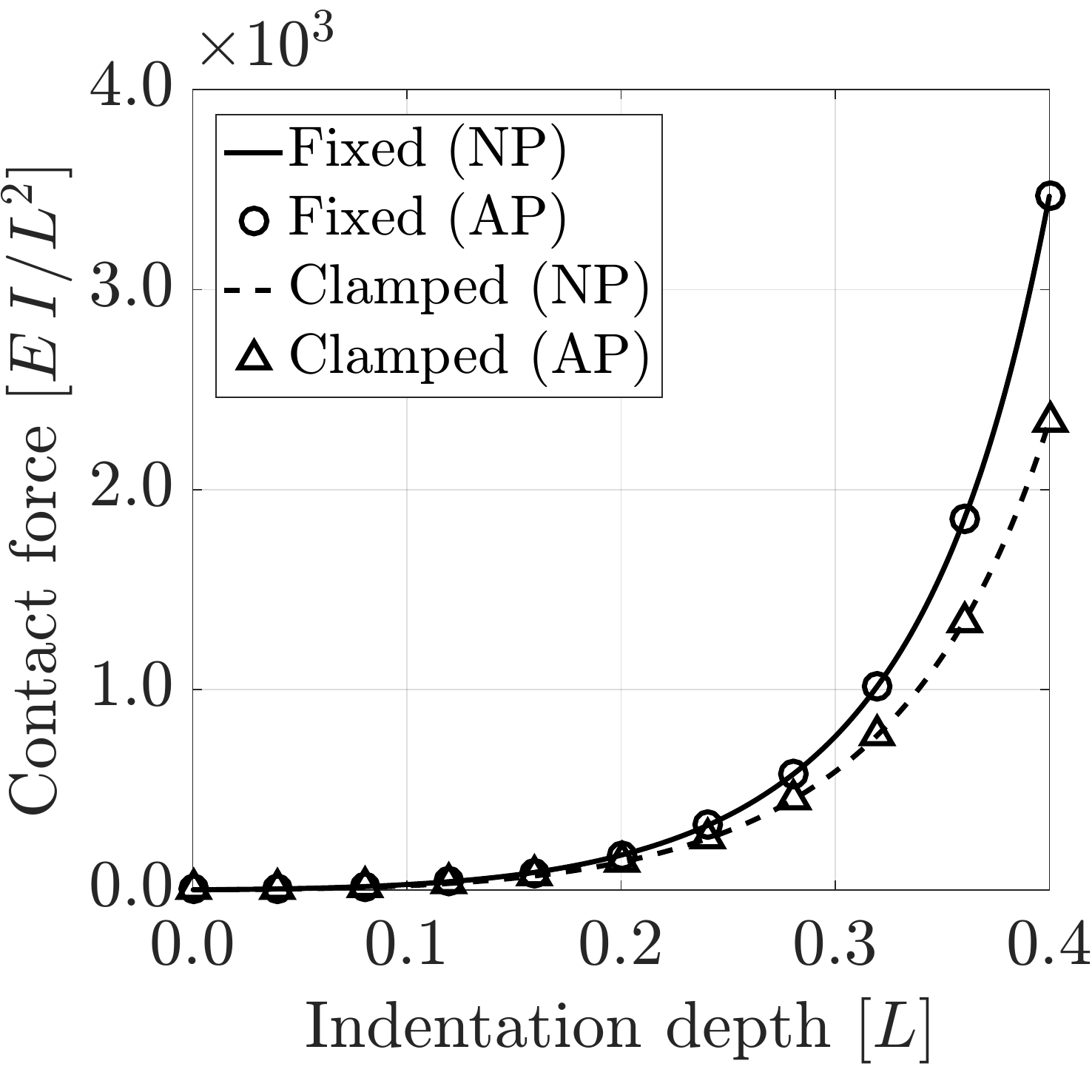}}
\put(2.1,0.3){(a)}
\put(8.0,-0.2){\includegraphics[height=42mm]{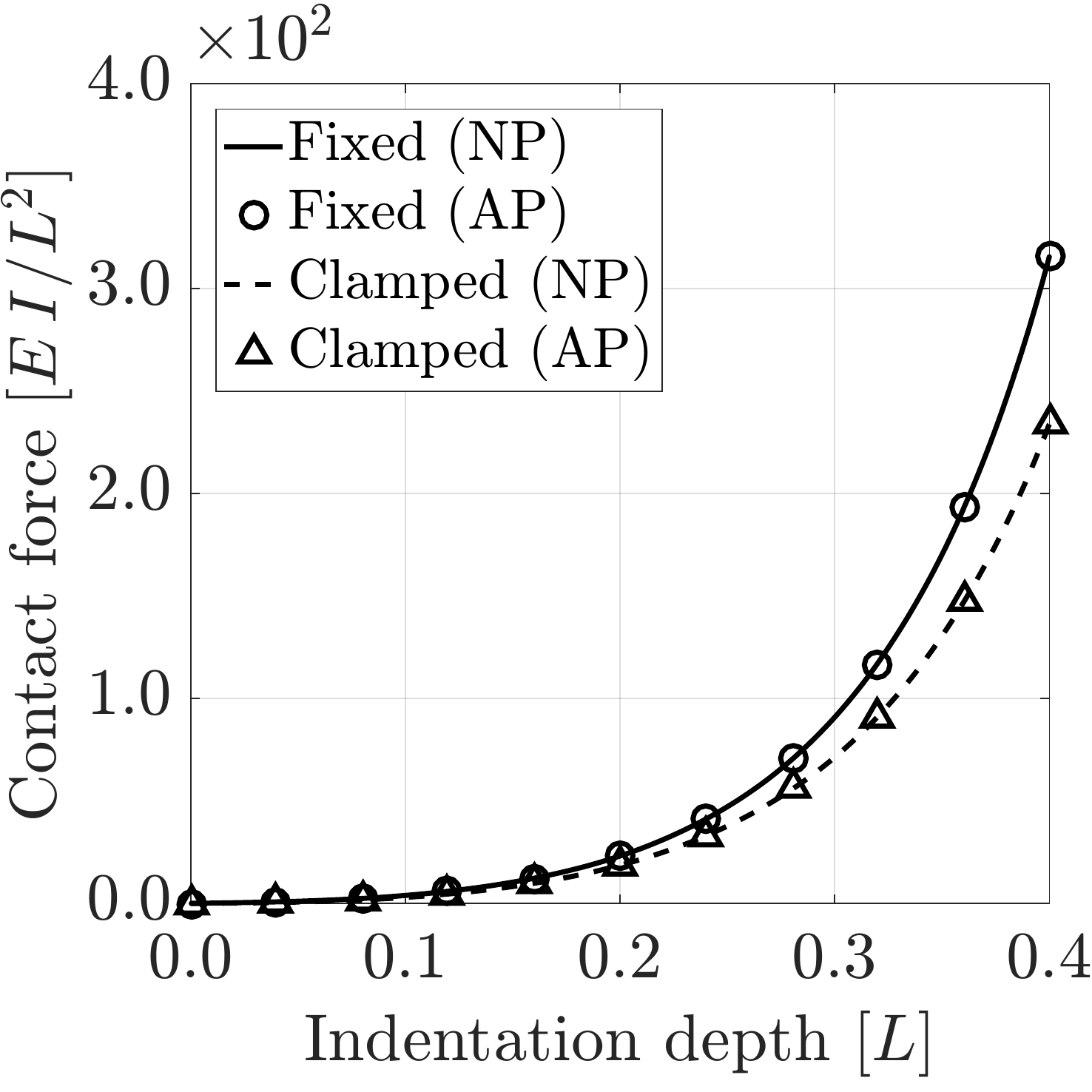}}
\put(7.5,0.3){(b)}
\end{picture}
\caption{Indentation test - contact force vs.~the indentation depth: GOH model with the compression/tension switch and (a) $\kappa_i = 0.226$ and (b) $\kappa_i = 1/3$.}
\label{f:indent_2}
\end{center}
\end{figure}

In Fig.~\ref{f:indent_2}, the vertical component of the total contact force is plotted against the indentation depth for different indenter radius $R$. As expected, both the AP and NP shell models perform similarly. 

\subsection{Angioplasty}\label{s:angio}
Balloon angioplasty is the typical treatment to widen obstructed arteries or veins \citep{humphrey13cardio}. This procedure has been studied computationally by many scholars \citep[e.g.][]{holzapfel96,rogers99,holzapfel02,gasser07,gervaso08,pant12} in order to optimize the internal pressure, mechanical properties and location of the balloon. 

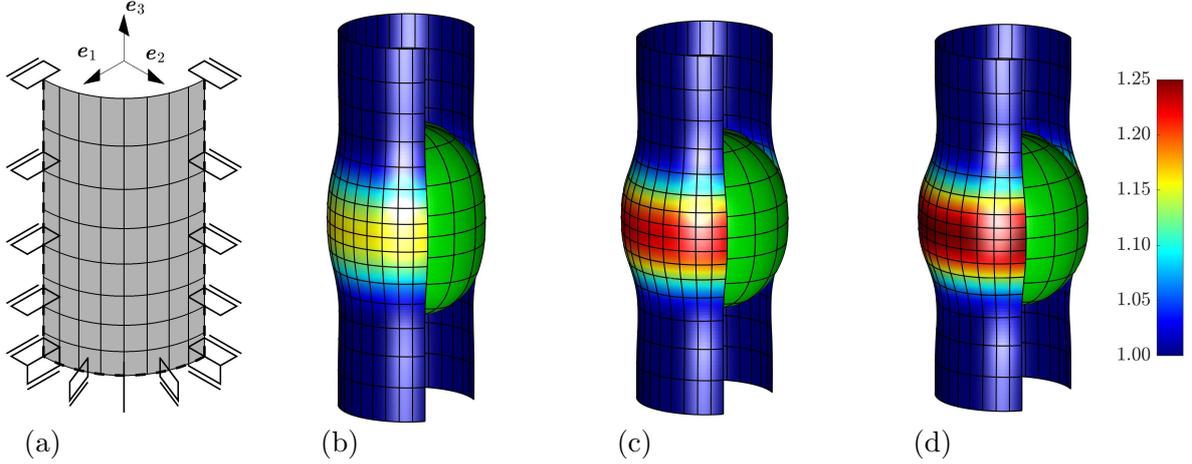
\begin{figure}[ht]
\begin{center} \unitlength1cm
\begin{picture}(15.0,6)
\put(-0.5,0.0){\includegraphics[height=60mm]{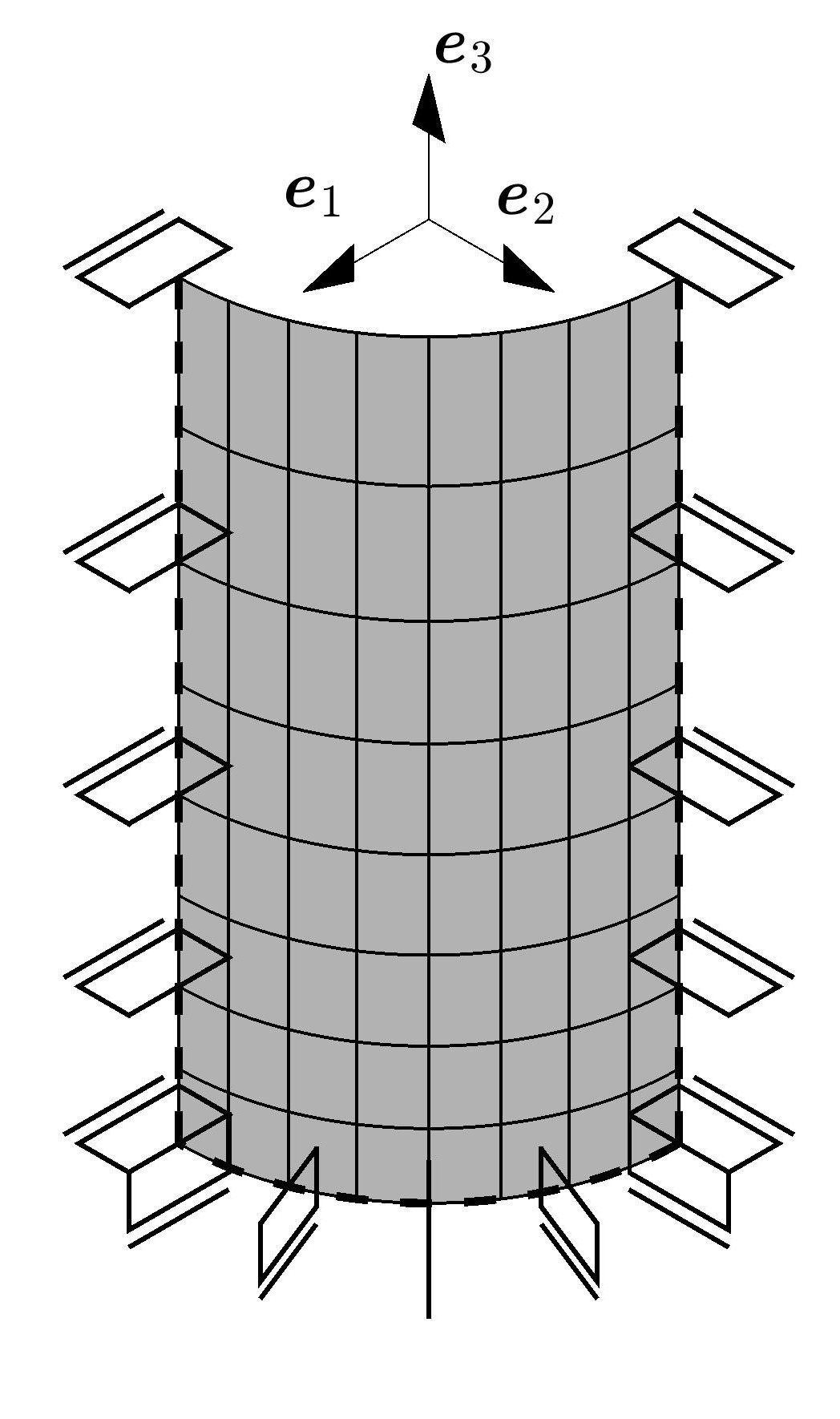}}
\put(0.0,-0.1){(a)}
\put(3.7,0.0){\includegraphics[height=60mm]{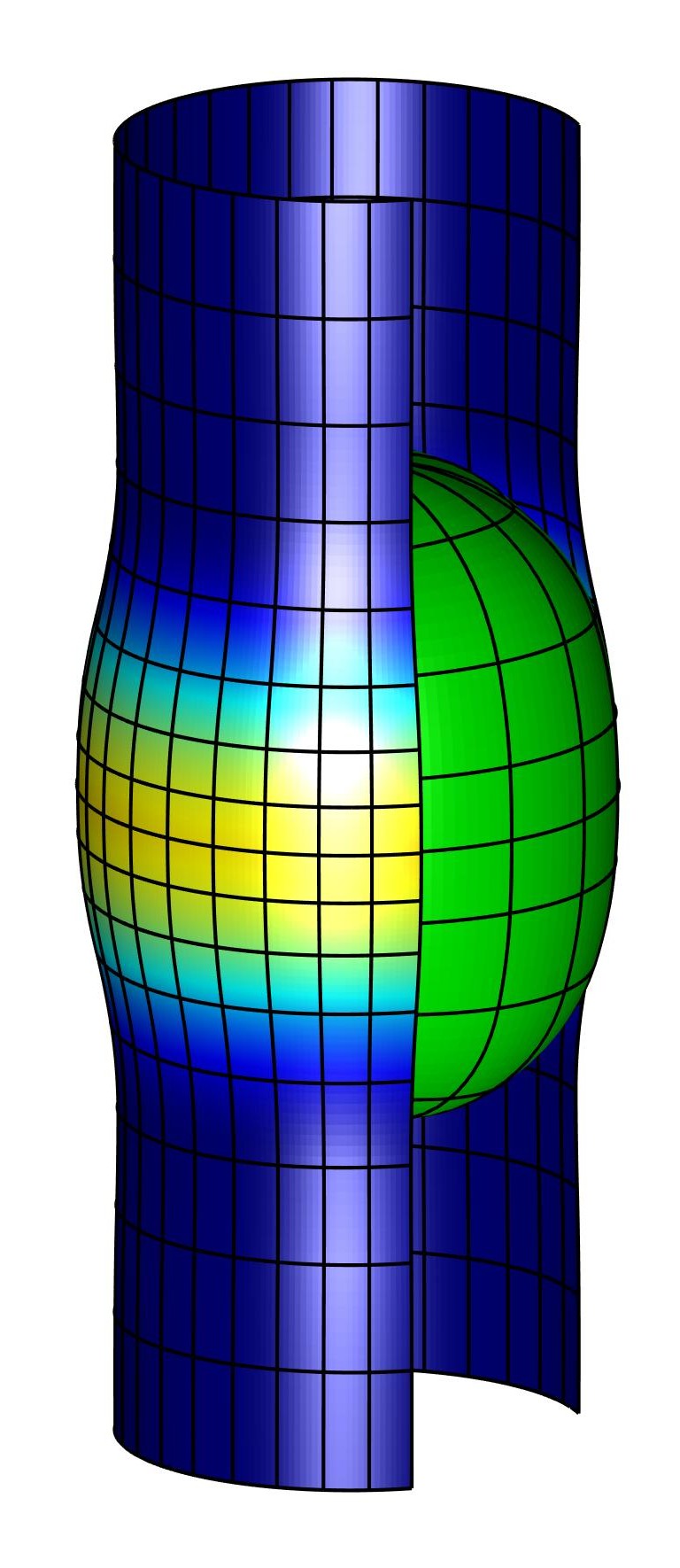}}
\put(3.9,-0.1){(b)}
\put(7.6,0.0){\includegraphics[height=60mm]{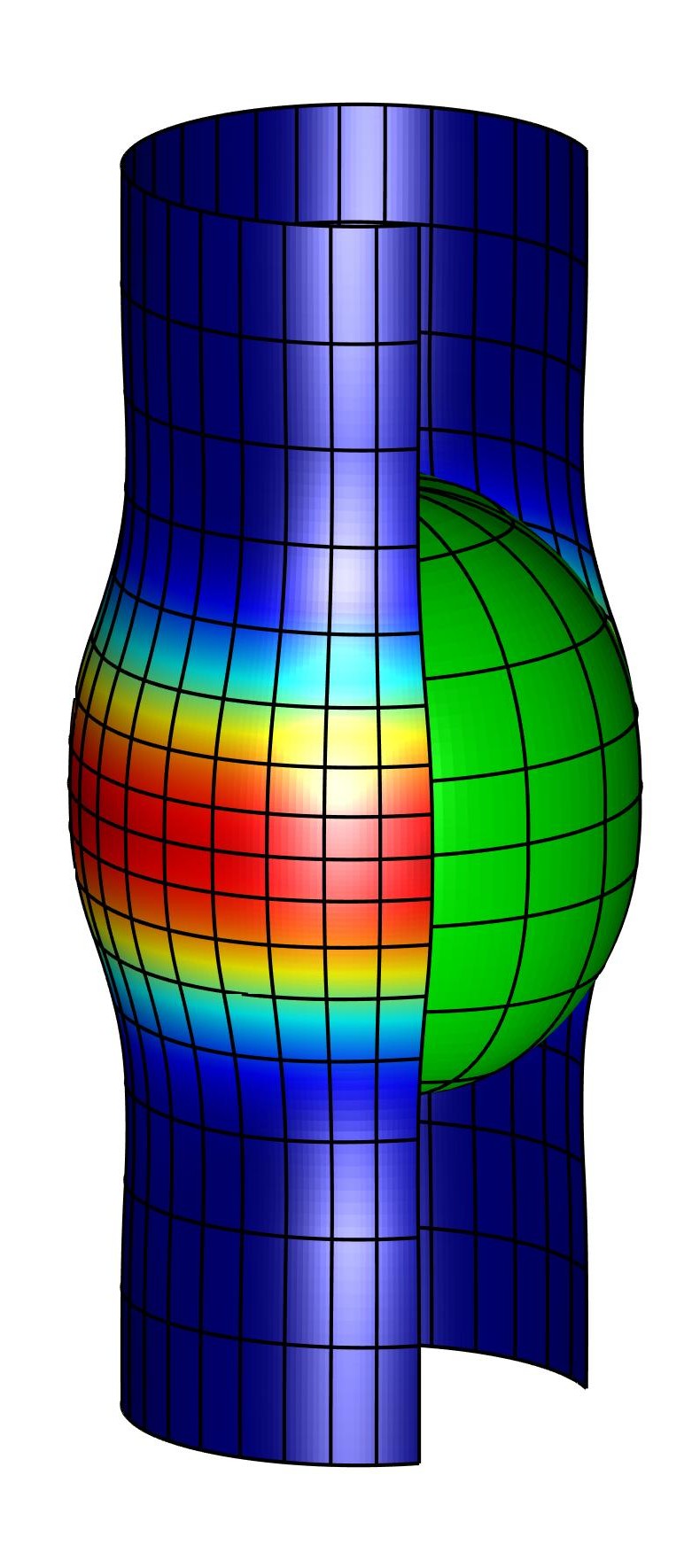}}
\put(7.8,-0.1){(c)}
\put(11.5,0.0){\includegraphics[height=60mm]{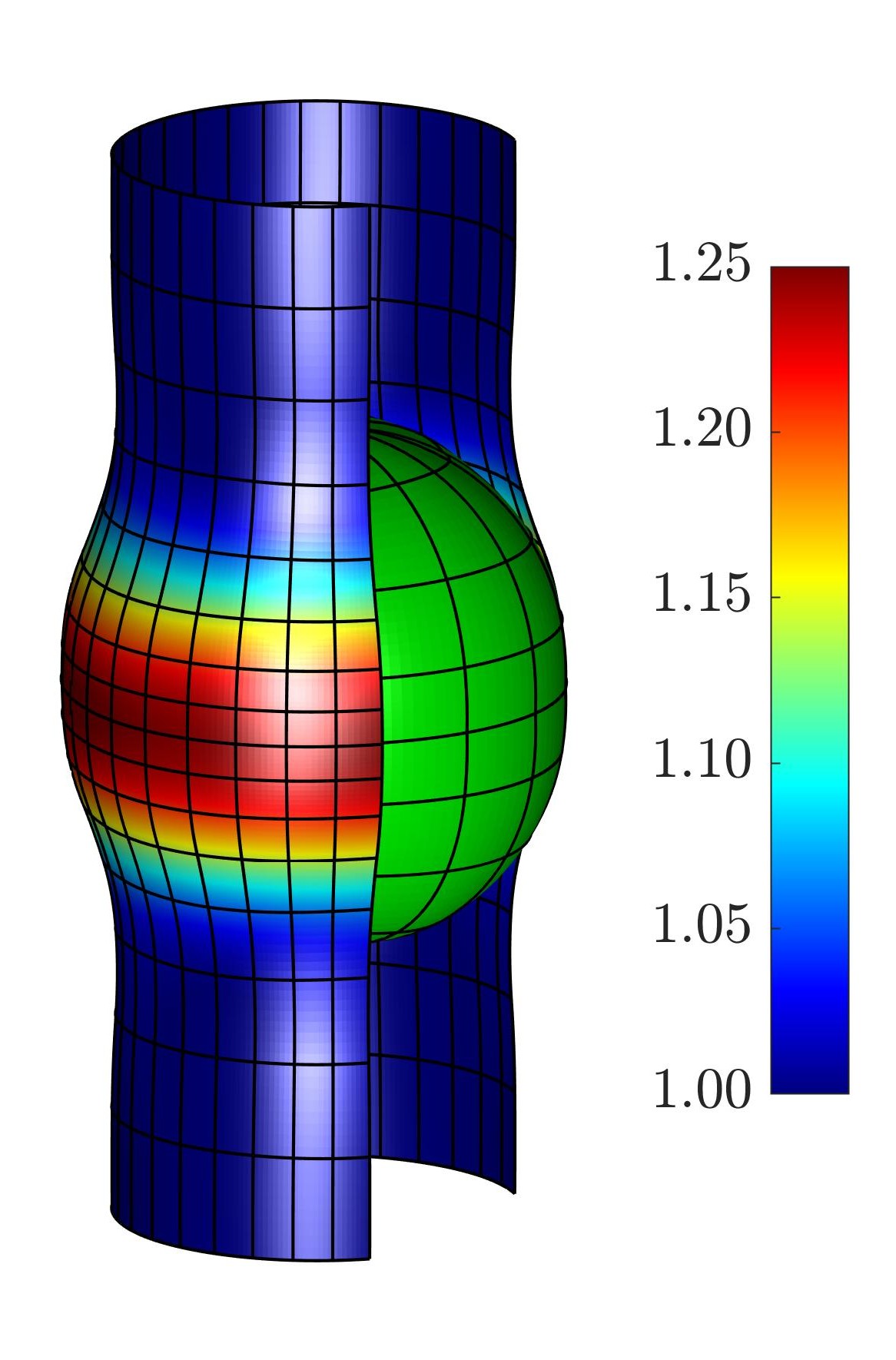}}
\put(11.7,-0.1){(d)}
\end{picture}
\caption{Balloon angioplasty: (a) Reference configuration (1/8 system) with boundary conditions and deformed configuration colored by the circumferential stretch $\lambda_\theta$ for (a) $\tmu_\mrb = 2\,\tmu_\mra$, (b) $\tmu_\mrb = 10\,\tmu_\mra$ and (c) $\tmu_\mrb = 20\,\tmu_\mra$.}
\label{f:angio_1}
\end{center}
\end{figure}  

Here, the angioplasty procedure is simulated as shown in Fig.~\ref{f:angio_1}. A portion of an artery with the dimensions $T_\mra \times R_\mra \times L = 0.5 \times 5 \times 30~[\mathrm{mm}^3]$ is inflated by a balloon with initial radius $ R_\mrb = 0.9\,R_\mra$ and thickness $T_\mrb = 0.1\,R_\mrb$. The balloon is initially pre-stretched by $\lambda_\mrp = 1.1$. The artery is modeled by the GOH material model with $\kappa_i = 0$. The material constants are taken from Tab.~\ref{t:const}, which is similar to the properties of the adventitia of an artery \citep{gasser06}. Two families of fibers are considered with
\eqb{l}
\tbL_i = \sin\theta_i\,\sin\psi\,\be_1 + \sin\theta_i\,\cos\psi\,\be_2 + \cos\theta_i\,\be_3,\quad(i=1,2)~,
\eqe
where $\theta_i \pm 45^\circ$ and $\psi$ is the angular coordinate around $\be_3$ axis. The NP and AP shell models are used for the artery and the compression/tension switch is included. For the NP shell model, 5 Gaussian quadrature points are used for the numerical integration through the thickness. The balloon is modeled by an incompressible Neo-Hookean membrane \citep{membrane} with the shear moduli $\tmu_\mrb = 2,\,10$ and $20\,\tmu_\mra$, where $\tmu_\mra$ is the shear modulus of the anisotropic part of the artery material model (see Tab.~\ref{t:const}). 
The balloon is inflated up to $V=3\,V_0$, where $V_0$ is the initial volume of the balloon. The contact constraint is enforced following the penalty formulation of \citet{spbf}. The penalty parameter is set to $\epsilon_\mrc = 10^7\,E\,T_\mra$, where $E = 3\,\tmu_\mra$. 

\begin{figure}[ht]
\begin{center} \unitlength1cm
\begin{picture}(15.0,4.2)
\put(2.2,-0.2){\includegraphics[height=42mm]{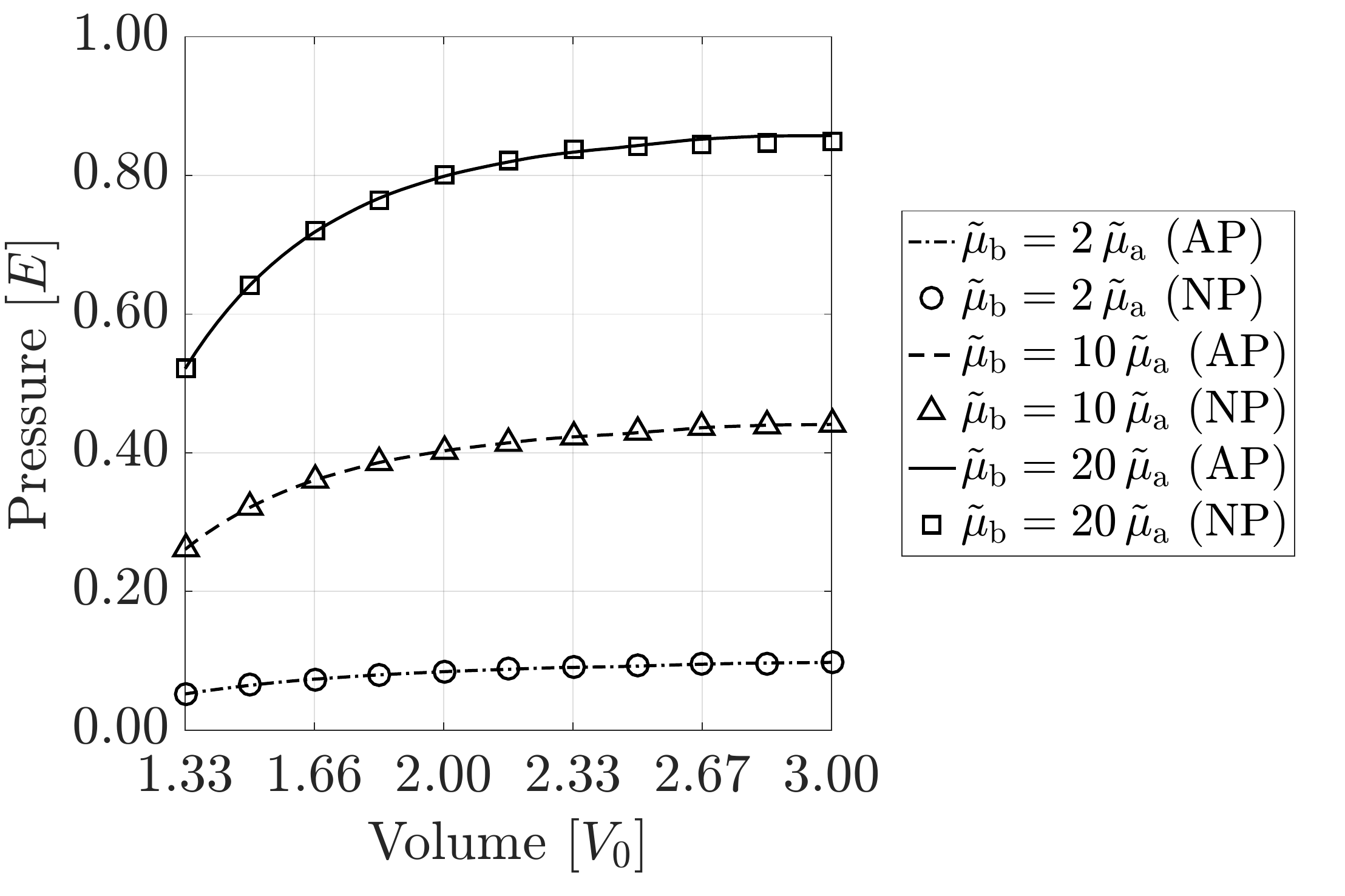}}
\put(1.7,0.3){(a)}
\put(8.5,-0.2){\includegraphics[height=42mm,trim={0 0 220px 0},clip]{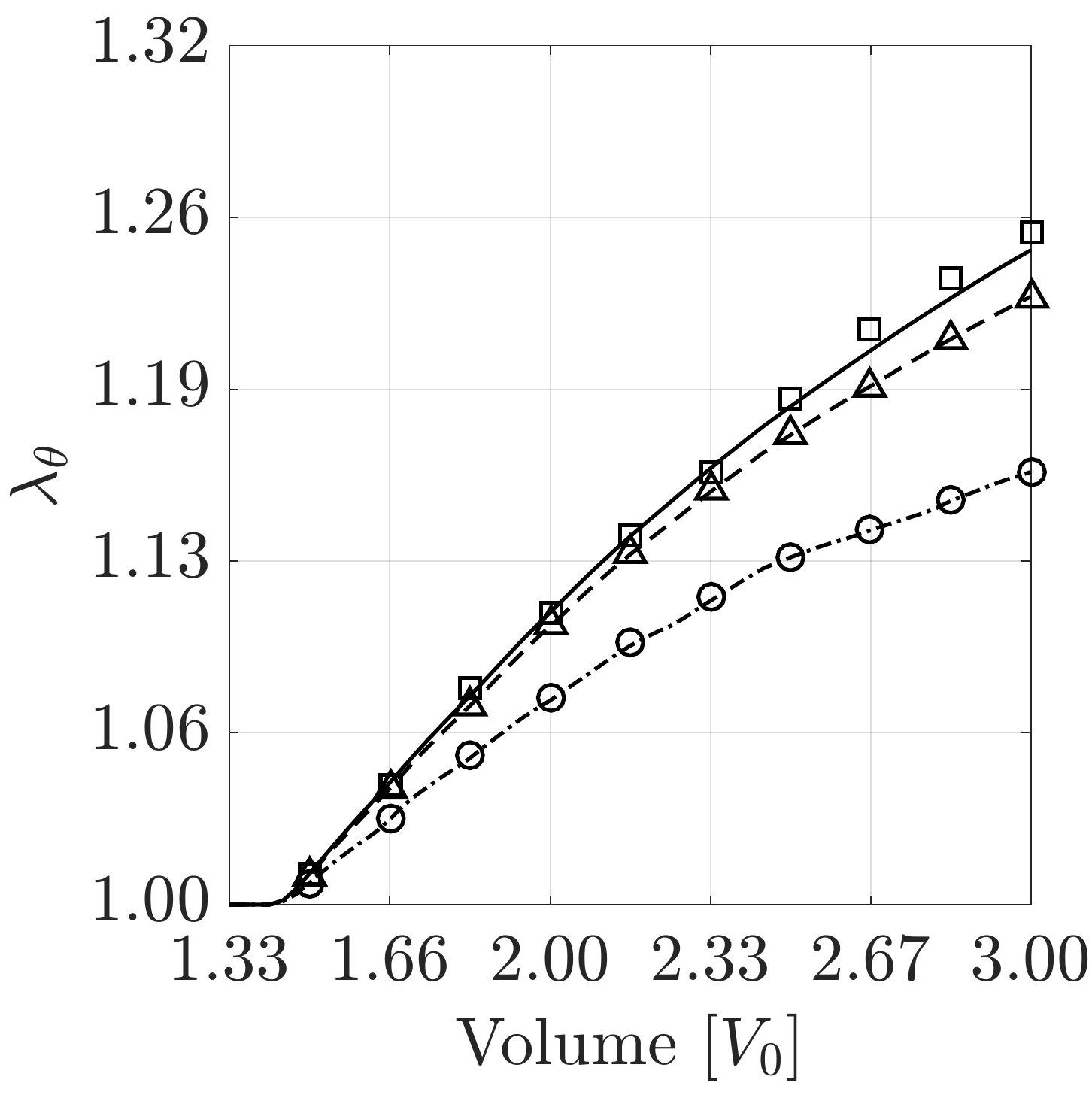}}
\put(8.0,0.3){(b)}
\end{picture}
\caption{Balloon angioplasty: (a) The internal pressure and (b) the average circumferential stretch, $\lambda_\theta$, in the middle of the artery vs.~the volume of the inflated balloon.}
\label{f:angio_2}
\end{center}
\end{figure}

In Fig.~\ref{f:angio_2}.a the internal pressure of the balloon is plotted against its volume for different values of $\tmu_\mrb$. Fig.~\ref{f:angio_2}.b shows the average circumferential stretch,  $\lambda_\theta$, computed in the middle of the artery (see the circumferential dashed line in Fig.~\ref{f:angio_1}.a) against the volume of the inflated balloon. As expected, the results of the AP and NP shell models are very close even though the artery is quite thick ($T/R=0.1$).
%
%
\section{Conclusion}\label{s:con}
This paper presents different rotation-free shell formulations to model thin structures composed of soft biological materials. The formulation is designed for large deformations and allows for geometrical and material nonlinearities, which makes it very suitable for the modeling of soft tissues. The formulation is based on the Kirchhoff--Love hypothesis; thus, it needs only displacement degrees of freedom. Following an isogeometric approach, NURBS-based finite elements are used for the FE discretization and the FE solution, which satisfies the necessary $C^1$-continuity of solution for rotation-free shells.   

Three different approaches to model thin shells are introduced: The \emph{numerically-projected} (NP) shell model, which uses numerical integration through the shell thickness, and the \emph{analytically-projected} (AP) and \emph{directly-decoupled} (DD) shell models, which do not need any numerical through-the-thickness integration. The NP shell model is the most general approach; however, it can be computationally expensive e.g.~for anisotropic constitutive laws like the Gasser--Ogden--Holzapfel material model \citep{gasser06}. For such materials, one may need many quadrature points across the shell thickness to capture discontinuities of the stress across the thickness. This has motivated us to develop the AP shell model, which is computationally more feasible. If the shell thickness is considerably smaller than the in-plane dimensions, for an initially-planar shell, or the radii of curvature, for an initially-curved shell, the NP and AP shell models perform similarly. Furthermore, the DD shell model is presented, which is directly defined on a 2D manifold. This formulation assumes that the material properties and the constitutive law are symmetric w.r.t.~the shell mid-surface. Apart from this restriction, the DD shell model is the most efficient approach. 

Furthermore, the exclusion of compressed fibers is considered for each type of the three shell models. As shown by different examples, the introduced compression/tension switch works very well for both the NP and AP shell models. The DD shell model, however, cannot capture the effect of the switch if the bending forces are dominant.    

Altogether, the presented formulations can be characterized by increased computational efficiency and algorithmic complexity. Accordingly, an appropriate formulation should be chosen by a trade-off between efficiency and complexity for any specific application, structure and constituent.

The introduced shell models are specifically derived for different isotropic and anisotropic material models, which are commonly used for soft biological materials. For both the isotropic and anisotropic models, two types of strain energy density function are examined: Polynomial forms (i.e.~Neo--Hooke, Mooney--Rivlin, and anisotropic Mooney--Rivlin material models) and exponential forms (i.e.~Fung and Gasser--Ogden--Holzapfel material models). The procedure can be easily applied to other material models. Furthermore, the robustness and accuracy of the presented shell models is demonstrated by different examples, which examine pure membrane modes (see the uniaxial tension test), pure bending modes (see the cantilever bending test) and mixed modes (see the pressured clamped plate) of the shell deformation. Moreover, the applicability of the shell models is demonstrated by two examples: The indentation of a sheet under a rigid spherical indenter and an angioplasty example that involves contact between two deformable bodies.

%
\vspace{5mm}
{\bf Acknowledgement}\\
Financial support from the German Research Foundation (DFG) through grant GSC 111 is gratefully acknowledged. The authors also thank Reza Ghaffari for checking the manuscript
carefully.
%
%
\appendix 
\section{Variation of kinematic variables}\label{s:var}
Following the approach of \citet{shelltheo}, the variation of kinematic variables are expressed in terms of the metric tensor $a_{\alpha\beta}$, which captures the stretching deformations, and the curvature tensor $b_{\alpha\beta}$, which captures the bending deformations. These variations are then used to derive the stresses and linearize the governing equations. 

From Eq.~\eqref{e:g_ab}, the variation of $g_{\alpha\beta}$ is
\eqb{l}
\delta g_{\alpha\beta} = \delta a_{\alpha\beta} - 2\,\xi\,\delta b_{\alpha\beta} 
\label{e:vg_ab}
\eqe
and since $[g^{\alpha\beta}]=[g_{\alpha\beta}]^{-1}$,
\eqb{l}
\delta g^{\alpha\beta} = g^{\alpha\beta\gamma\delta}\,\delta g_{\gamma\delta}~, 
\label{e:vgab}
\eqe
with
\eqb{l}
g^{\alpha\beta\gamma\delta}:=\ds\pa{g^{\alpha\beta}}{g_{\gamma\delta}}=-\ds\frac{1}{2}\left(g^{\alpha\gamma}g^{\beta\delta} + g^{\alpha\delta}g^{\beta\gamma}\right) ~.
\eqe

As shown by \citet{shelltheo}, $\delta a^{\alpha\beta} = a^{\alpha\beta\gamma\delta}\,\delta a_{\gamma\delta}$, where
\eqb{l}
a^{\alpha\beta\gamma\delta}:=\ds\pa{a^{\alpha\beta}}{a_{\gamma\delta}}=-\ds\frac{1}{2}\left(a^{\alpha\gamma}a^{\beta\delta} + a^{\alpha\delta}a^{\beta\gamma}\right) 
\label{e:aabgd}\eqe
and $\delta b^{\alpha\beta} = b^{\alpha\beta\gamma\delta}\,\delta a_{\gamma\delta} - a^{\alpha\beta\gamma\delta}\,\delta b_{\gamma\delta}$, where
\eqb{l}
b^{\alpha\beta\gamma\delta}:=\ds\pa{b^{\alpha\beta}}{a_{\gamma\delta}}= 2 H \left(a^{\alpha\beta}a^{\gamma\delta} + a^{\alpha\beta\gamma\delta}\right) - \left(a^{\alpha\beta}b^{\gamma\delta} + b^{\alpha\beta}a^{\gamma\delta} \right)~.
\eqe
Besides, it can be proven that 
\eqb{l}
\delta J = \ds\frac{J}{2}\,a^{\alpha\beta}\,\delta a_{\alpha\beta}~,\quad\quad
\delta \aJ = \ds\frac{\aJ}{2}\,g^{\alpha\beta}\,\delta g_{\alpha\beta}~.
\label{e:vtJ}\eqe

The variation of the first invariant of the right Cauchy--Green tensor is
\eqb{l}
\delta\tI_1 = \delta I_1 + 2\,\lam\,\delta\lam~,\quad\quad \delta\tI_1 = \delta \aI_1 + \tmod{\delta g_{33}}~,
\label{e:vtI1}\eqe
where the in-plane components are
\eqb{l}
\delta I_1 = A^{\alpha\beta}\,\delta a_{\alpha\beta} ~, \quad\quad
\delta \aI_1 = G^{\alpha\beta}\, \delta g_{\alpha\beta} ~.
\eqe

Similarly, the variation of the other invariants can be found as
\eqb{lll}
\delta\tI_2 \is \lam^2 \, \delta I_1 + 2 J\,\delta J + 2\,I_1\,\lam\,\delta\lam~, \\[3mm]
\delta\tI_2 \is  \tmod{g_{33}} \, \delta \aI_1 + 2 \aJ\,\delta\aJ + \tmod{\aI_1\,\delta g_{33}}
\label{e:vtI2}
\eqe
and
\eqb{l}
\delta\tilde{I_3} = 2\,\lam^2 \, J\,\delta J + 2\,J^2\lam\,\delta\lam ~, \quad\quad
\delta\tilde{I_3} =  2\,\tmod{g_{33}} \, \aJ\,\delta \aJ + \tmod{\aJ^2\,\delta g_{33}} ~.
\label{e:vtI3}
\eqe	

\section{Linearization of the external virtual work}\label{s:DGext}
If the in-plane components of the body force, i.e.~$f^\alpha\,\vaua$, are neglected, the linearized external virtual work contribution is \citep{shelltheo} 
\eqb{l}
\Delta G_\mathrm{ext} = \Delta G_{\mathrm{ext}p} + \Delta G_{\mathrm{ext}t} +  \Delta G_{\mathrm{ext}m}~,
\label{e:dgext}\eqe
where \citep{membrane}
\eqb{l}
\Delta G_{\mathrm{ext}p} = \ds\int_{\sS}p\,\delta\bx\cdot(\bn \otimes \ba^\alpha - \ba^\alpha \otimes \bn)\,\Delta\ba_\alpha\,\dif a~.
\label{e:dgextp}\eqe
Denoting the convective coordinate along the curve $\partial_m \sS$ as $\xi^\epsilon$, where $\epsilon = 1$ or $\epsilon = 2$, the co-variant base vector at $\bx \in \partial_m \sS$ is $\ba_\epsilon := \partial\bx/\partial\xi^\epsilon$. The corresponding contra-variant base vector is then $\ba^\epsilon := a^{\epsilon\alpha}\ba_\alpha~(\alpha = 1,2)$. This gives \citep{solidshell}
\eqb{lll}
\Delta G_{\mathrm{ext}m} \is \ds\int_{\partial_m \sS}\,\delta\ba_\alpha\cdot\big(\nu^\beta\,\bn\otimes\ba^\alpha + \nu^\alpha\,\ba^\beta\otimes\bn\big)\,\Delta\ba_\beta\,\dif s \\[4mm]
\mi \ds\int_{\partial_m \sS}\,\dfrac{1}{\norm{\ba_\epsilon}^2}\,m_\tau\,\nu^\alpha\, \delta\ba_\alpha\cdot\big(\bn\otimes\ba^\epsilon\big)\,\Delta\ba_\epsilon\,\dif s
\label{e:dgextm}\eqe
and
\eqb{l}
\Delta G_{\mathrm{ext}t} = \ds\int_{\partial_t\sS} \delta\bx\cdot\Delta\bt\,\dif s +
\ds\int_{\partial_t \sS}\,\dfrac{1}{\norm{\ba_\epsilon}^2}\,\delta\bx\cdot\big(\bt\otimes\ba^\epsilon\big)\, \Delta\ba_\epsilon\,\dif s~,
\label{e:dgextt}\eqe
where $\norm{\ba_\epsilon} = \sqrt{\ba_\epsilon\cdot\ba_\epsilon}$. If the traction $\bt$ is not a follower load (of the displacement), $\Delta\bt = \boldsymbol{0}$. 

\section{Out-of-plane linearization of kinematic variables}\label{s:vxi}
In this section, on the shell mid-surface, the kinematical variables are linearized w.r.t. $\xi$. From Eq.~\eqref{e:g_ab}, we have
\eqb{l}
g_{\alpha\beta,3} := \ds\pa{g_{\alpha\beta}}{\xi} = -2\,b_{\alpha\beta} ~.
\label{e:g_ab_xi}\eqe
Combining Eqs.~\eqref{e:vgab}~and~\eqref{e:g_ab_xi} gives
\eqb{l}
g^{\alpha\beta}_{,3} := \ds\pa{g^{\alpha\beta}}{\xi} = -2\,g^{\alpha\beta\gamma\delta}\,b_{\gamma\delta} ~.
\eqe
It can be shown that $g^{\alpha\beta\gamma\delta}\,b_{\gamma\delta} = -b^{\alpha\beta}$ if $\xi = 0$. Thus, on the shell mid-surface,
\eqb{l}
\hg^{\alpha\beta}_{,3} := \Big(g^{\alpha\beta}_{,3}\Big)_{\xi=0} = 2\,b^{\alpha\beta} ~.
\label{e:gab_xi}\eqe
Similar quantities can be derived for $G_{\alpha\beta}$ and $G^{\alpha\beta}$ in the reference configuration. Plugging Eq.~\eqref{e:g_ab_xi} into  Eq.~\eqref{e:vtJ}, we have
\eqb{l}
\aJ_{,3} := \ds\pa{\aJ}{\xi} = \aJ\left(G^{\alpha\beta}\,B_{\alpha\beta} - g^{\alpha\beta}\,b_{\alpha\beta}\right) ~,
\eqe
which gives
\eqb{l}
\hJ_{,3} := \Big(\aJ_{,3}\Big)_{\xi=0} = 2\,J\left(H_0 - H\right)
\label{e:J_xi}\eqe
on the shell mid-surface. In a similar fashion, the first invariant of the right Cauchy--Green tensor is linearized as
\eqb{l}
\hI_{1,3} := \Big(\ds\pa{\aI_1}{\xi}\Big)_{\xi=0} = 2\,\big(a_{\alpha\beta}\,B^{\alpha\beta} - b_{\alpha\beta}\,A^{\alpha\beta}\big)~.
\label{e:I1_xi}\eqe

\section{First-order compression/tension switch for fibers}\label{s:switch_lin}
For the principal directions of anisotropic materials, it can be shown that 
\eqb{l}
\hL^{\alpha\beta}_{i,3} := \Big(\ds\pa{\aL^{\alpha\beta}_i}{\xi}\Big)_{\xi=0} = 2\,L_i^\alpha\,L_{i,3}^\beta ~,
\label{e:Lab_xi}\eqe 
where $L_{i,3}^\alpha := B^{\alpha\beta}\,L_\beta^{i}$ and $L_\alpha^{i} := \bL_i\cdot\bA_\alpha$. Thus, on the mid-surface, we have
\eqb{l}
\hI^{i}_{4,3} := \Big(\ds\pa{\aI^{i}_4}{\xi}\Big)_{\xi=0} = -2\,b_{\alpha\beta}\,L^{\alpha\beta}_i + a_{\alpha\beta}\,\hL^{\alpha\beta}_{i,3} ~.
\label{e:hI4_xi}\eqe 

Using a first order Taylor expansion, on a shell layer at $\xi$, $ \aI^{i}_4 $ can be related to the similar quantity $I^{i}_4$ on the mid-surface as
\eqb{l}
\aI^{i}_4 = I^{i}_4 + \xi\,\hI^{i}_{4,3} ~,
\label{e:tI4_taylor}\eqe 
which gives $\xi^{i}_0 = \xi^{i}_0(\bx)$, where $\tI^{i}_4 = \aI^{i}_4 = 1$\footnote{Here, it is assumed that $\aL^{33}_i=L^{33}_i=0$ (see Remark~\ref{r:fibers})}, as
\eqb{l}
\xi^{i}_0 := \ds\frac{1 - I^{i}_4}{\hI^{i}_{4,3}} ~.
\label{e:xi0}\eqe 

Then, $T^{i}_1$ and $T^{i}_2$ are defined according to the algorithm shown in Tab.~\ref{t:T1T2}. 
Furthermore, for the material tangents of Sec.~\ref{s:parsh}, one needs to linearize $T_1^{i}$ and $T_2^{i}$ as   
\eqb{l}
U_{1i}^{\alpha\beta} := \ds\pa{T_1^{i}}{a_{\alpha\beta}}~,\quad\quad U_{2i}^{\alpha\beta} := \ds\pa{T_2^{i}}{a_{\alpha\beta}}~, \quad\quad
V_{1i}^{\alpha\beta} := \ds\pa{T_1^{i}}{b_{\alpha\beta}}~,\quad\quad V_{2i}^{\alpha\beta} := \ds\pa{T_2^{i}}{b_{\alpha\beta}}~,
\eqe
which depend on
\eqb{lllll}
Y^{\alpha\beta}_i \dis \ds\pa{\xi^{i}_0}{\auab} \is -\left(\hI^{i}_{4,3}\right)^{-2}\left[\hI^{i}_{4,3}\,L^{\alpha\beta}_i + \left(1 - I_4^{i}\right)\hL^{\alpha\beta}_{i,3}\right] ~, \\[2mm]
Z^{\alpha\beta}_i \dis \ds\pa{\xi^{i}_0}{\buab} \is 2\,\left(\hI^{i}_{4,3}\right)^{-2}\,\left(1 - I_4^{i}\right)L^{\alpha\beta}_i 
\label{e:xi0_ab}\eqe 
as shown in Tab.~\ref{t:T1T2_ab}.

\begin{table}[H]
\tabulinesep = 5pt
\centering
\begin{tabu} to 0.9\textwidth { X[-1.5,c,m] | X[c,m] | X[c,m] | X[c,m] }
\toprule  
{} & $\xi_0 < -\ds\frac{T}{2}$ & $-\ds\frac{T}{2} \leq \xi_0 \leq \ds\frac{T}{2}$ & $\ds\frac{T}{2} < \xi_0$ \\
\cline{2-4}
$\hat{I}^{i}_{4,3} > 0$ & $T_1^{i} = -\ds\frac{T}{2},~ T_2^{i} = \ds\frac{T}{2}$ & $T_1^{i} = \xi_0^{i},~ T_2^{i} = \ds\frac{T}{2} $ & N.A. \\
\hline  
$\hat{I}^{i}_{4,3} < 0$ & N.A. & $T_1^{i} = -\ds\frac{T}{2},~ T_2^{i} = \xi_0^{i} $ & $T_1^{i} = -\ds\frac{T}{2},~ T_2^{i} = \ds\frac{T}{2} $\\
\hline 
\end{tabu}\par\vskip0.4pt

\begin{tabu} to 0.9\textwidth { X[-1.5,c,b] | X[c,m] | X[c,m] }
\hline
{} & $ I_4^{i}>1 $ & $ I_4^{i}\leq1 $ \\ \cline{2-3}
$\hI^{i}_{4,3} = 0$ & $ T_1^{i} = -\ds\frac{T}{2},~ T_2^{i} = \ds\frac{T}{2}$ & N.A. \\
\bottomrule
\end{tabu}
\caption{Algorithm to find $[T^{i}_1,T^{i}_2]$, where $\tI^{i}_4 > 1$.}
\label{t:T1T2}
\end{table}

\begin{table}[H]
\tabulinesep = 5pt
\centering
\begin{tabu} to 0.9\textwidth { X[-1.5,c,m] | X[c,m] | X[1.2,c,m] | X[c,m] }
\toprule  
{} & $\xi_0 < -\ds\frac{T}{2}$ & $-\ds\frac{T}{2} \leq \xi_0 \leq \ds\frac{T}{2}$ & $\ds\frac{T}{2} < \xi_0$ \\
\cline{2-4}
\multirow{2}{*}{$\hat{I}^{i}_{4,3} > 0$} & $U^{\alpha\beta}_{1i} = V^{\alpha\beta}_{1i} = 0 $ & $U^{\alpha\beta}_{1i} = Y^{\alpha\beta}_i,~ V^{\alpha\beta}_{1i} = Z^{\alpha\beta}_i $ & \multirow{2}{*}{N.A.} \\
{} & $U^{\alpha\beta}_{2i} = V^{\alpha\beta}_{2i} = 0$ & $U^{\alpha\beta}_{2i} = V^{\alpha\beta}_{2i} = 0$ & {}\\
\hline  
\multirow{2}{*}[-4pt]{$\hat{I}^{i}_{4,3} < 0$} & \multirow{2}{*}[-4pt]{N.A.} & $U^{\alpha\beta}_{1i} = V^{\alpha\beta}_{1i} = 0 $ & $U_{1i} = V_{1i} = 0 $ \\
{} & {} & $U^{\alpha\beta}_{2i} = Y^{\alpha\beta}_i,~ V^{\alpha\beta}_{2i} = Z^{\alpha\beta}_i $ & $U^{\alpha\beta}_{2i} = V^{\alpha\beta}_{2i} = 0$ \\
\hline 
\end{tabu}\par\vskip0.4pt

\begin{tabu} to 0.9\textwidth { X[-1.5,c,b] | X[c,m] | X[c,m] }
\hline
{} & $ I_4^{i}>1 $ & $ I_4^{i}\leq1 $ \\ \cline{2-3}
$\hI^{i}_{4,3} = 0$ & $U^{\alpha\beta}_{1i} = V^{\alpha\beta}_{1i} = U^{\alpha\beta}_{2i} = V^{\alpha\beta}_{2i} = 0$ & N.A. \\
\bottomrule
\end{tabu}
\caption{Algorithm to linearize $T_1^{i}$ and $T_2^{i}$.}
\label{t:T1T2_ab}
\end{table}

%
\vspace{5mm}
{\bf Conflict of Interest}\\
The authors declare that they have no conflict of interest.
%
%
\bibliographystyle{apalike}
\bibliography{ShellBib,BioShellBib,bibliography}

\end{document}